\documentclass[11pt,titlepage]{article}

\usepackage{graphicx}
\usepackage{float}

\usepackage{mathptmx}
\usepackage{euscript}
\usepackage{latexsym}
\usepackage{amsmath}
\usepackage{amsfonts}
\usepackage{amssymb}
\usepackage{ulem}

\newcommand{\al}{\alpha}
\newcommand{\om}{\omega}
\newcommand{\Om}{\Omega}
\newcommand{\ep}{\epsilon}

\newcommand{\lam}{\lambda}
\newcommand{\sig}{\sigma}
\newcommand{\bmbeta}{\pmb{\beta}}
\newcommand{\bbR}{\mathbb{R}}

\newcommand{\cA}{\EuScript{A}}
\newcommand{\bz}{\mathbf{z}}

\newcommand{\eq}{\begin{eqnarray*}}
\newcommand{\eqq}{\end{eqnarray*}}
\newcommand{\eqn}{\begin{eqnarray}}
\newcommand{\eqqn}{\end{eqnarray}}

\newcommand{\IID}{\mathrm{IID}}

\DeclareMathSymbol{,}{\mathpunct}{operators}{"2C}

\begin{document}

\title{Quantile-Frequency Analysis and Spectral Measures for Diagnostic Checks of Time Series With Nonlinear Dynamics}

\author{Ta-Hsin Li\footnote{IBM T.\ J.\ Watson Research Center, Yorktown Heights, NY 10598-0218, USA (thl@us.ibm.com)}}

\date{October 9, 2020}

\maketitle

\begin{abstract}

Nonlinear dynamic volatility has been observed in many financial time series. The recently proposed quantile periodogram offers an alternative way to examine this phenomena in the frequency domain. The quantile periodogram is constructed from trigonometric quantile regression of time series data at different frequencies and quantile levels, enabling the quantile-frequency analysis (QFA) of nonlinear serial dependence. This paper introduces some spectral measures based on the quantile periodogram for diagnostic checks of financial time series models and for model-based discriminant analysis. A simulation-based parametric bootstrapping technique is employed to compute the $p$-values of the spectral measures. The usefulness of the proposed method is demonstrated by a simulation study and a motivating application using the daily log returns of the S\&P 500 index together with GARCH-type models. The results show that the QFA method is able to provide additional insights into the goodness of fit of these financial time series models that may have been missed by conventional tests. The results also show that the QFA method offers a more informative way of discriminant analysis for detecting regime changes in financial time series.

\vspace{0.3in}

\noindent
{\it Abbreviated Title}: Quantile-Frequency Analysis for Time Series With Nonlinear Dynamics

\end{abstract}

\section{Introduction}

\begin{figure}[t]
\centering
\includegraphics[width=1.55in,angle=-90]{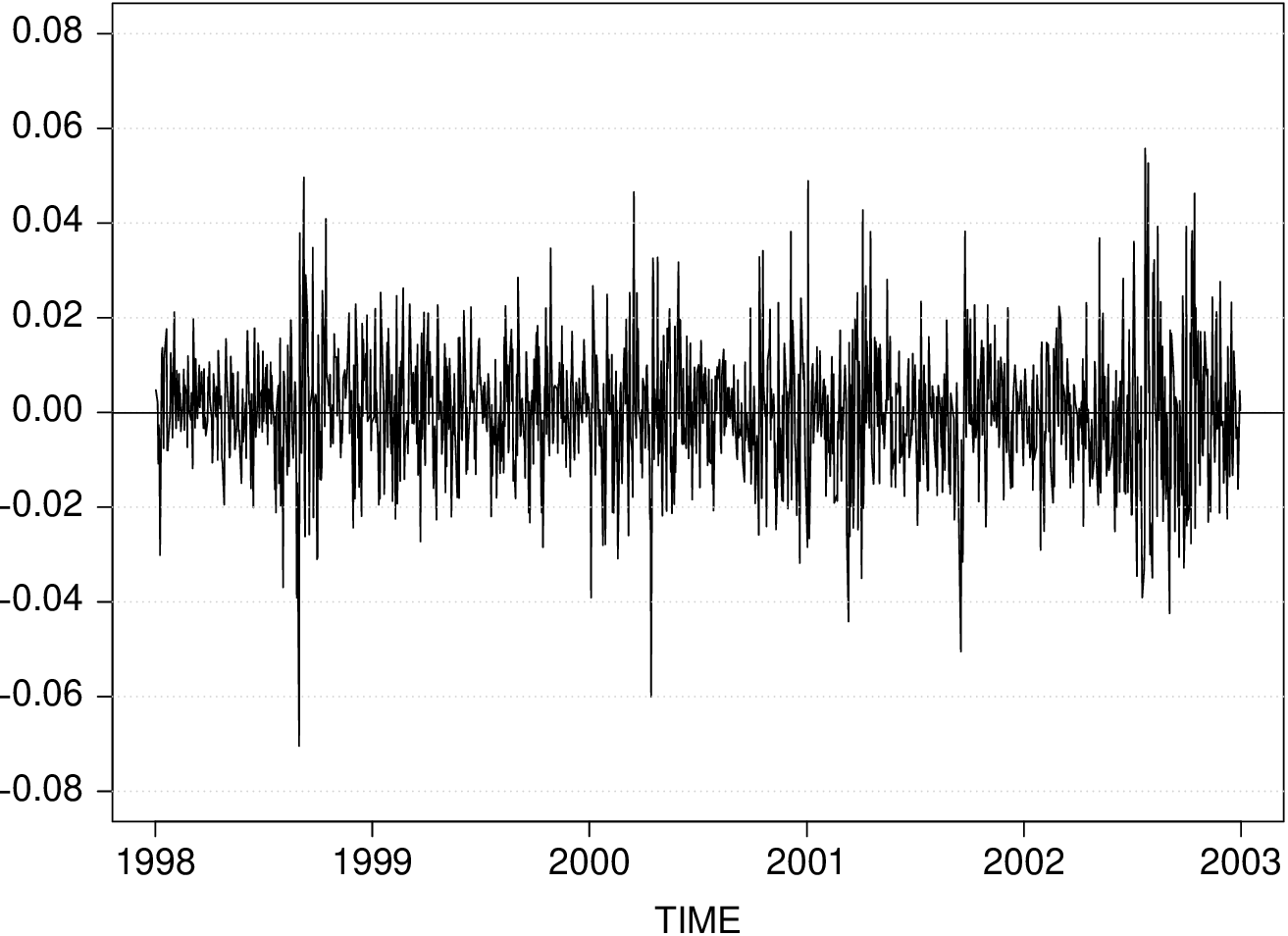}  \hfill
\includegraphics[width=1.55in,angle=-90]{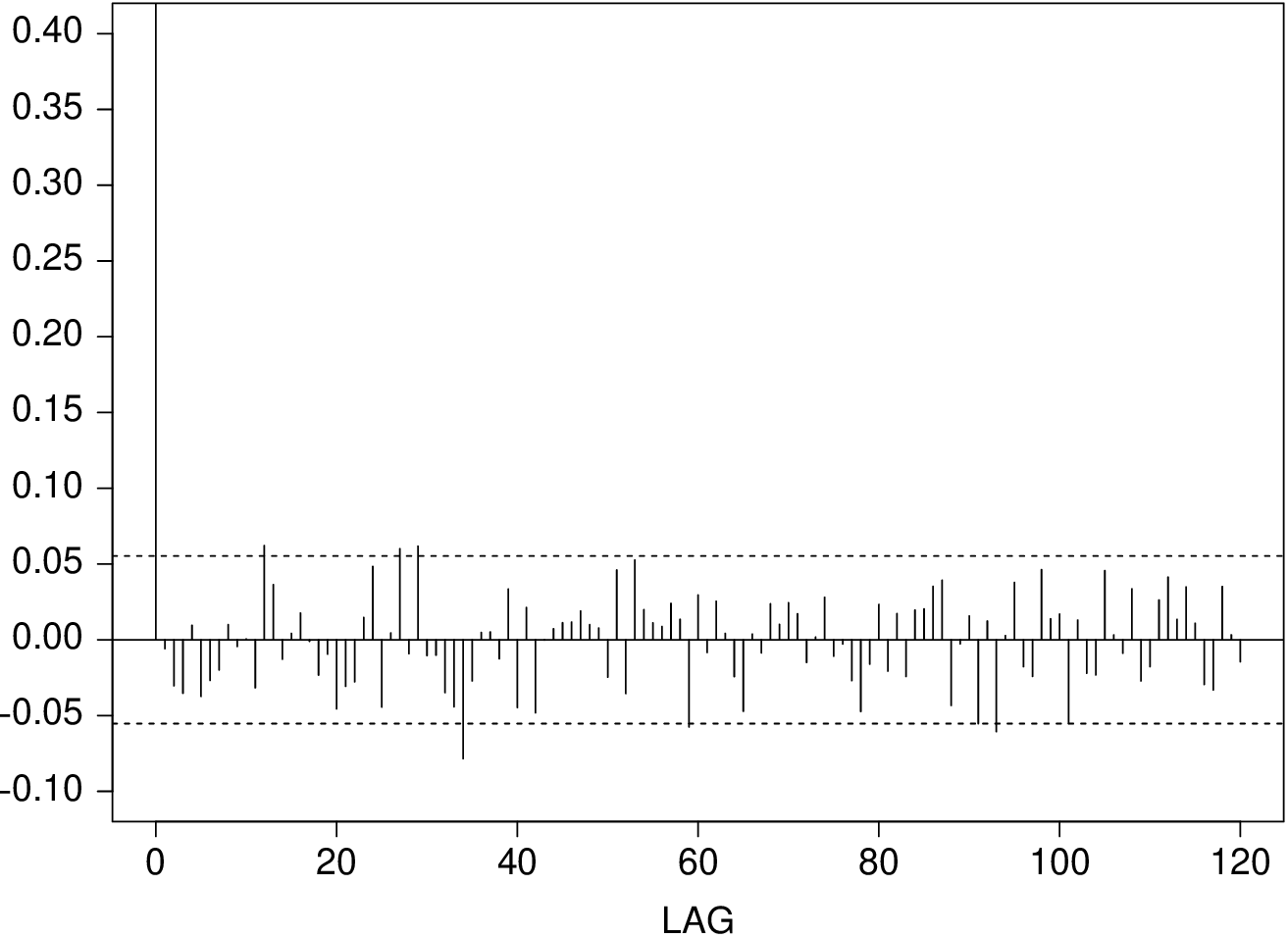}  \hfill
\includegraphics[width=1.55in,angle=-90]{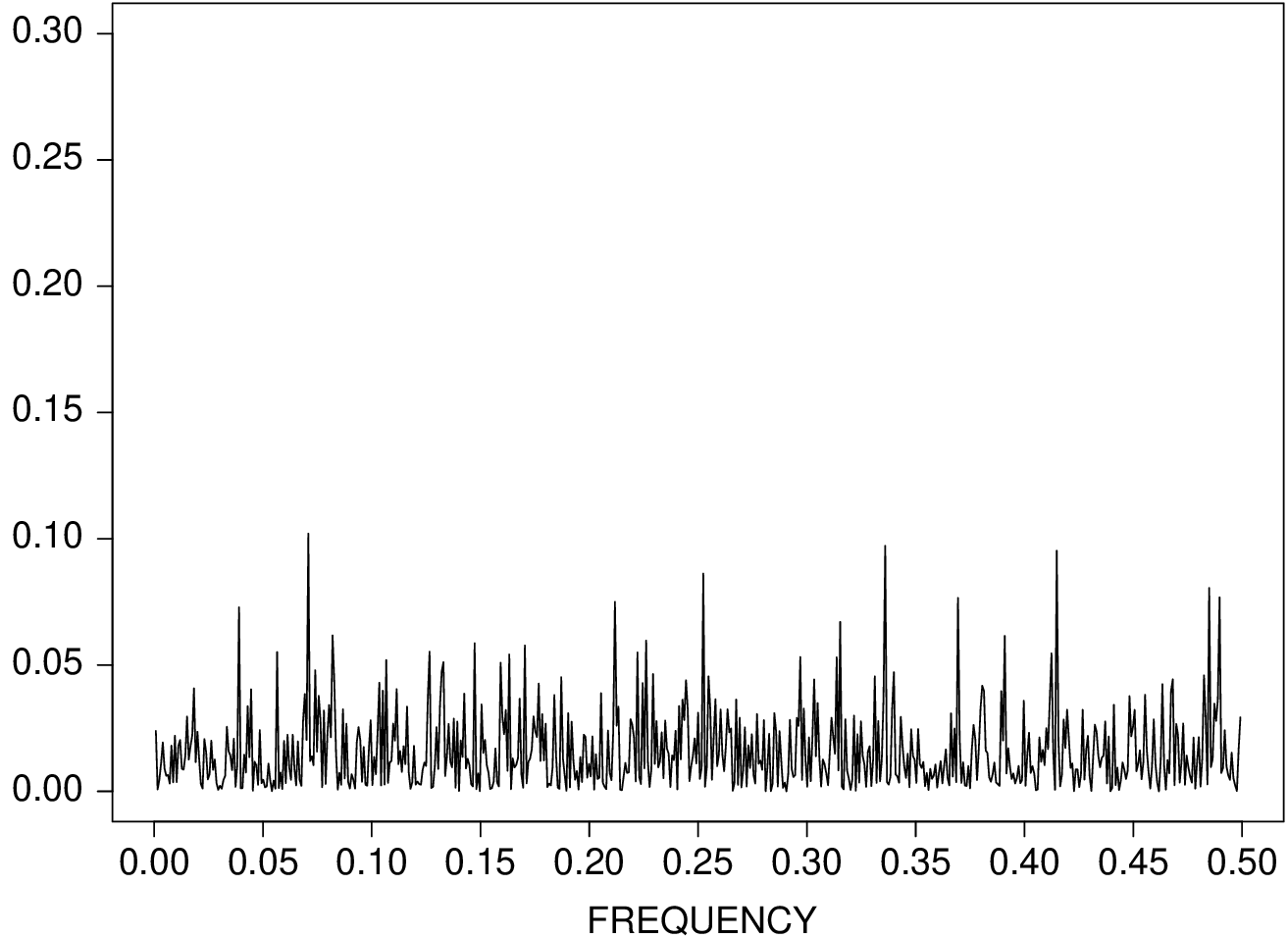} \\
\includegraphics[width=1.55in,angle=-90]{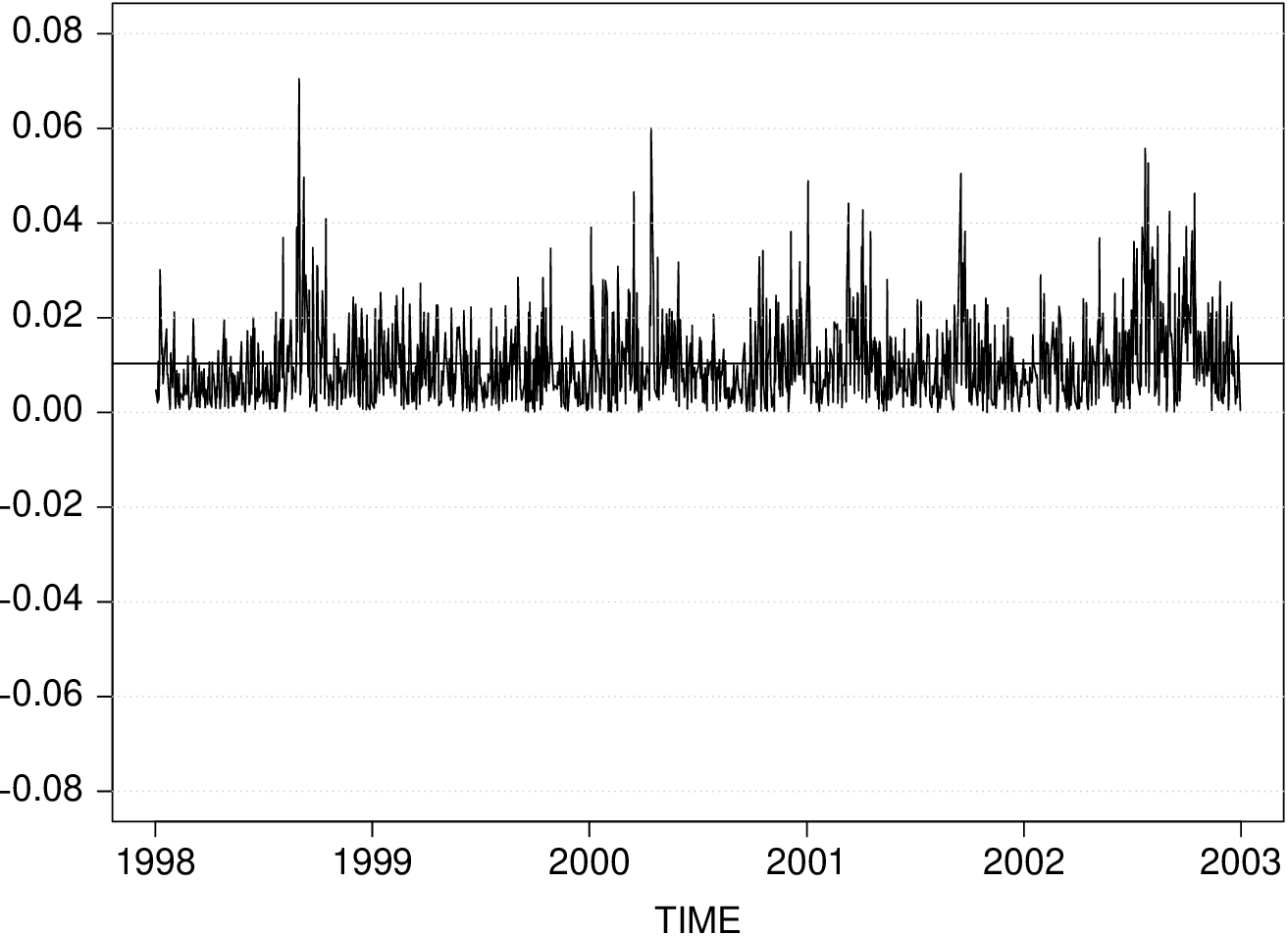} \hfill
\includegraphics[width=1.55in,angle=-90]{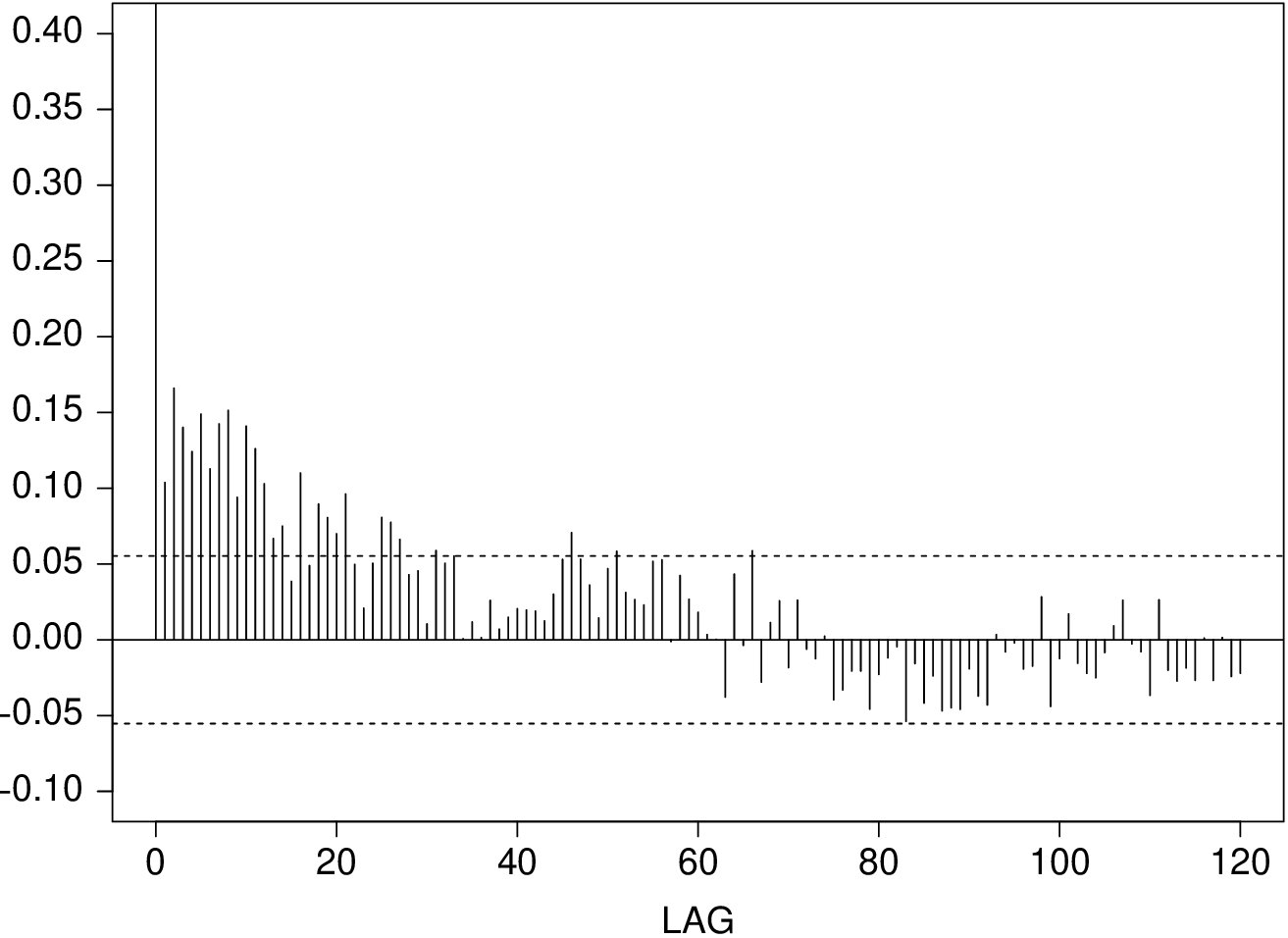} \hfill
\includegraphics[width=1.55in,angle=-90]{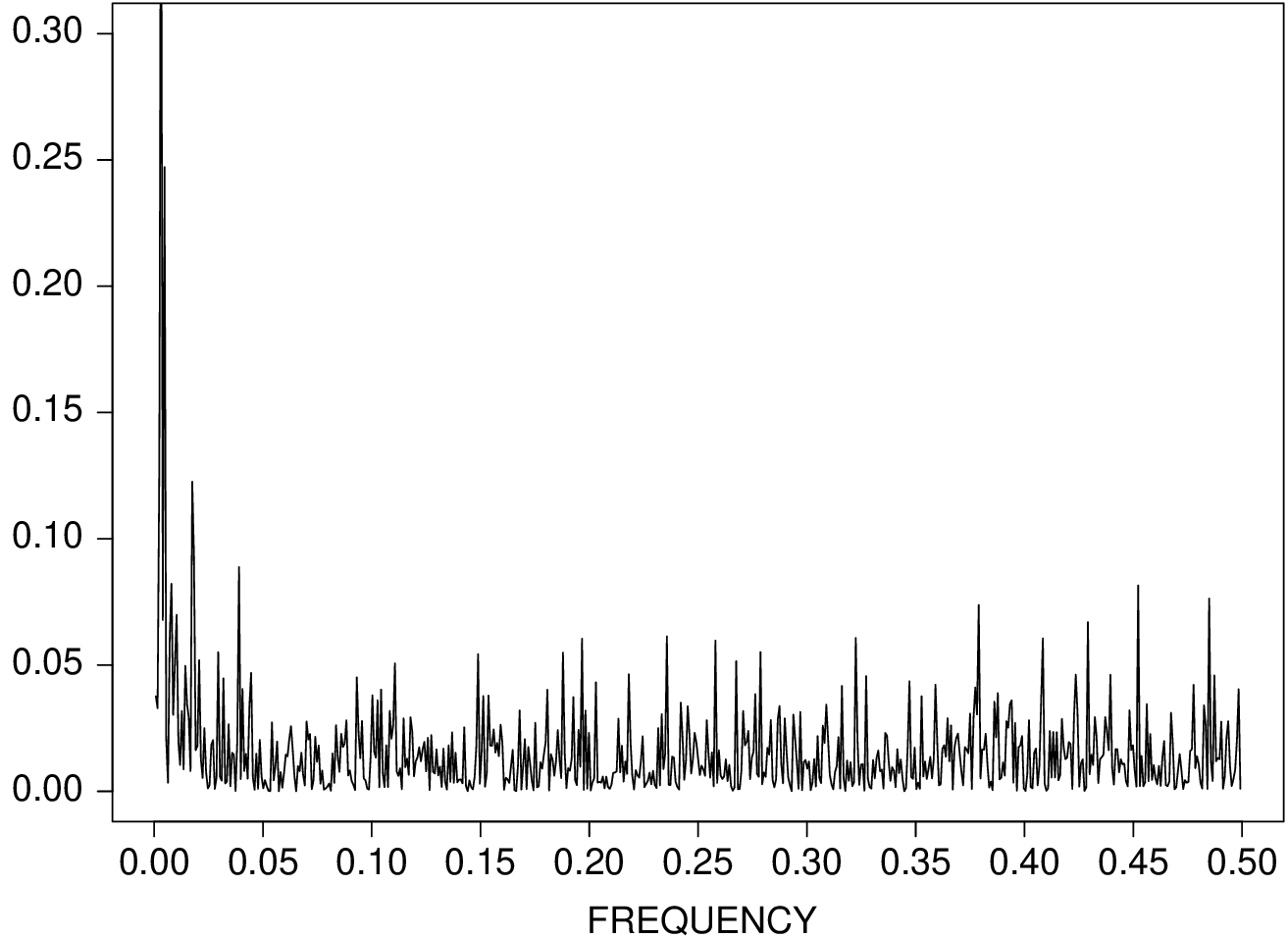} 
\caption{Top row: time series of daily log  returns of the S\&P 500 index for the 5-year 
period of 1998--2002 (left) with its autocorrelation function (middle) and normalized periodogram (right).
Bottom row: absolute-value transform of the series (left) with its
autocorrelation function (middle) and normalized periodogram (right). 
Solid horizontal line in the time series plots represents the mean of the series. All periodograms  are plotted here and thereafter as functions 
of the regular frequency variable $\om/(2\pi) \in (0,0.5)$ 
and their amplitude is rescaled by a factor of 10. }
\label{fig:spx}
\end{figure}

Many financial time series exhibit complicated nonlinear dynamics that cannot be adequately handled by conventional diagnostic tools that are based on the second-order statistics (Fama, 1965; Taylor, 1986). For example, consider the time series of daily log returns of the S\&P 500 index (SPX) 
shown in Figure~\ref{fig:spx}.  These series are almost indistinguishable from white noise when examined by the autocorrelation function in the time domain and the periodogram in the frequency domain. Their serial dependence is revealed by these conventional techniques only after certain nonlinear transformations, such as square and absolute value, are applied to the original data, as illustrated in the bottom panel of Figure~\ref{fig:spx}. In effect, the nonlinear transformations ``fold'' the original series along the zero line and turn the negative values into positive ones. As a result, the autocorrelation function and the periodogram no longer reflect the serial dependence in the excesses of the original series above or below its mean which is very close to zero; instead, they reflect the serial dependence in the excesses of the folded series above or below a value much higher than zero.

As an example, consider the absolute-value transformation, and let $X_t$ denote the original series.
Then, by definition, the event that $|X_t|$ and $|X_{t+\tau}|$ are on the same side of $\mu_{\rm abs} := E(|X_t|)$ contributes positively to the lag-$\tau$ autocorrelation of the transformed series, and the event that these values are on the opposite side of $\mu_{\rm abs}$ makes a negative contribution. Note that $|X_t|$ greater than $\mu_{\rm abs}$ means $X_t$ above $\mu_{\rm abs}$ or below $- \mu_{\rm abs}$, and that $|X_t|$ less than  $\mu_{\rm abs}$ means $X_t$ above $- \mu_{\rm abs}$ and below $\mu_{\rm abs}$. Therefore, in essence, the autocorrelation function and the periodogram of the transformed data represent  the oscillatory behavior of the original series around the horizontal lines $-\mu_{\rm abs}$ and $\mu_{\rm abs}$. A similar explanation applies to the square transformation if $\mu_{\rm abs}$ is replaced by $\sqrt{\mu_{\rm sq}}$, where $\mu_{\rm sq} := E(X_t^2)$. In comparison, the autocorrelation function and the periodogram of the original series describe its oscillatory behavior 
around $\mu_{\rm o} := E(X_t)$. Because $\mu_{\rm o}$ is near the center of the marginal distribution whereas $\pm \mu_{\rm abs}$ and $\pm \sqrt{\mu_{\rm sq}}$ are far away from the center, it is reasonable to conclude that the ability of the conventional tools to reveal nonlinear serial dependence when applied to the transformed data can be largely attributed to the shift of focus from the center to a higher or lower level.

Owing to its simplicity, the coupling of a folding transformation with conventional 
diagnostic techniques has become a standard way of testing nonlinear serial dependence. 
For example, as part of the diagnostic checks for GARCH-type models, 
the R package {\tt fGarch} (Wuertz, 2017) employs the Ljung-Box test (Ljung and Box, 1978) 
and the Lagrange multiplier ARCH test (Engle, 1982) which are based respectively on the autocorrelation function of the squared residuals and the autoregressive $R^2$-statistic of the squared residuals. Though popular and useful, this approach has some limitations: it depends on the choice of the transformation (e.g., square versus absolute value) which may result in different assessment of serial dependence (Taylor, 1986); it is unable to distinguish excursions of the original series above a target level (e.g., $\mu_{\rm abs}$ in the absolute-value example) from those below the opposite level (e.g., $-\mu_{\rm abs}$) and therefore blind to the asymmetric characteristics of serial dependence observed in certain financial time series. 

The recently introduced quantile periodogram (Li, 2008; 2012a) is well suited to overcome these limitations. 
Derived by applying the quantile regression method (Koenker and Bassett, 1978) with a trigonometric regressor directly to the original untransformed series, the quantile periodogram offers a capability of examining serial dependence at any quantile level of the marginal distribution. By varying the quantile level as well as the trigonometric frequency parameter, one obtains a two-dimensional function that can be used to diagnose nonlinear serial dependence. We call this method  quantile-frequency analysis (QFA).

Among diagnostic tools for serial dependence, some are more narrowly focused but easier to use; others are more comprehensive but harder to compute, visualize, and interpret. 
The former include conventional techniques such as the autocorrelation function and the periodogram (Priestley, 1981) and more recent ones such as that discussed by Hecke, Volgushev and Dette (2018). The latter include the techniques based on the bivariate distribution function $P(X_t \le  x, X_{t+\tau} \le y)$ or the corresponding bivariate characteristic function (Skaug and Tj{\o}stheim, 1993; Hong, 2000; Lee and Subba Rao, 2011; Dette, Hallin, Kley and Volgushev, 2015). The QFA method discussed in this paper represents a trade-off between these two groups: it is more comprehensive than the conventional tools because it explores 
the variability around all quantile levels rather than just the mean; it is less difficult to use because it retains some essential properties of conventional spectral analysis based on the periodogram. Related works in recent literature include Hagemann (2013), Lim and Oh (2015), Jordanger and Tj{\o}stheim (2017), and Fajardo, Reisen, L\'{e}vy-Leduc and Taqqu (2018).

The success of ARCH and GARCH models (Engle, 1982; Bollerslev, 1986) for financial time series such as the SPX daily returns owes largely to their ability to capture the serial dependence of volatility. The original ARCH and GARCH models have been extended in various ways to handle more complicated behaviors including asymmetry and nonlinearity. Examples are the GJR-GARCH models (Glosten, Jagannathan and Runkle, 1993) and the more general APARCH models (Ding, Granger and Engle, 1993) which encompass several others as special cases 
(Taylor, 1986; Schwert, 1990; Higgins and Bera, 1992; Zakoian, 1994).  The R package {\tt fGarch}  (Wuertz, 2017) enables parameter estimation, goodness of fit testing, and simulation of these models.
It will be used in our experiments.

In this paper, we introduce some spectral measures (or divergence metrics) 
based on the quantile periodogram and 
demonstrate their potential usefulness for goodness of fit testing of time series models and for model-based discriminant analysis.   To compute the $p$-values of the QFA-based spectral measures, we employ the well-known technique called parametric bootstrapping (Efron and Tibshirani, 1993). 
Our motivating application using the SPX data  shows that the QFA method is able to identify certain lack-of-fit problems that may have been overlooked by standard tests, especially regarding the asymmetry in serial dependence of large negative returns versus that of large positive returns. 
The QFA method demonstrates similar diagnostic capabilities in the application for model-based discriminant analysis to detect regime changes over time.

The remainder of this paper is organized as follows. In Section 2, we review the technique of quantile periodogram and introduce the idea of QFA. In Section 3, we define the QFA-based spectral measures and discuss their applications in diagnostic checking of time series models 
and discriminant analysis using the parametric bootstrapping technique. 
In Section 4, we provide the results of a simulation study that helps illustrate the
detection power of the spectral measures. In Section 5, 
we provide the results of an application using the SPX data and some popular GARCH-type models. 

\section{Quantile Periodogram and Quantile-Frequency Analysis}

For a time series $\{X_t: t=1,\dots,n\}$ of length $n$, 
and for given angular frequency $\om \in (0,\pi)$ and 
quantile level $\al \in (0,1)$, consider the trigonometric
quantile regression problem
\eqn
\{\hat{\lam}_n(\om,\al), \hat{A}_{n}(\om,\al),\hat{B}_{n}(\om,\al)\}
:=  \arg\min_{\lam, \, A, \, B \in \bbR} 
\sum_{t=1}^n \rho_\al(X_t - \lam - A \cos(\om t) - B \sin(\om t)),
\label{qr}
\eqqn
where $\rho_\al(x)$ 
is defined as $\al x$ for $x \ge 0$ and $-(1-\al) x$ for $x<0$. It can be solved efficiently by
linear programming techniques (Portnoy and Koenker, 1997; Koenker, 2005). 
There are two methods to define quantile periodograms based on (\ref{qr}) (Li, 2012a). 
Method I focuses on the squared magnitude of the regression coefficients and 
defines the quantile periodogram (of the first kind) as
\eqn
q_{n}(\om,\al) := \frac{n}{4} (\hat{A}_n^2(\om,\al) 
+ \hat{B}_n^2(\om,\al)).
\label{qper}
\eqqn
Method II examines the contribution of the trigonometric regressors
in reducing the cost function and defines the quantile periodogram 
(of the second kind) as
\eqn
q_{n}(\om,\al) := \sum_{t=1}^n \rho_\al(X_t - \hat{\lam}_n(\al)) - \sum_{t=1}^n
\rho_\al(X_t - \hat{\lam}_n(\om,\al) - \hat{A}_n(\om,\al) \cos(\om t) - \hat{B}_n(\om,\al) \sin(\om t)),
\label{qper2}
\eqqn
where $\hat{\lam}_n(\al) :=  \arg\min_{\lam\in \bbR} \sum_{t=1}^n \rho_\al(X_t - \lam)$
is just the $\al$-quantile of $\{ X_t: t=1,\dots,n \}$.

The quantile periodogram defined by (\ref{qper}) has the advantage of being easily generalizable to quantile cross-periodograms for multiple time series and biquantile cross-periodograms for single time series  (Li, 2013). The quantile periodogram defined by (\ref{qper2}) has a smoother sample path and enables more accurate estimation of spectral peaks (Li,  2012b). Both are natural extensions of the ordinary periodogram defined by $I_n(\om) := n^{-1} |\sum_{t=1}^n X_t \exp(-i\om t)|^2$
with $i := \sqrt{-1}$, in the sense that at the Fourier frequencies $\om_k := 2\pi k /n \in (0,\pi)$, with $k$ being integer, $I_n(\om_k)$ can be derived according to (\ref{qper}) and (\ref{qper2}) by replacing the quantile regression coefficients and the quantile objective function $\rho_\al(\cdot)$ with the least-squares regression coefficients and the least-squares objective function $\frac{1}{2} (\cdot)^2$.

Note that  in defining the quantile periodograms by (\ref{qper}) and (\ref{qper2}) we exclude 
the cases of $\om=0$ and $\pi$. This is sufficient for most practical purposes. However, 
there are situations such as one discussed in Li (2019) where it is necessary to include these values
in the definition
as follows: for $\om=0$, it is natural to set $q_n(0,\al) := 0$; for $\om = \pi$, it suffices to consider the 2-parameter quantile regression problem $\{\hat{\lam}_n(\pi,\al), \hat{A}_{n}(\pi,\al)  \}
:=  \arg\min_{\lam , \, A \in \bbR} \sum_{t=1}^n \rho_\al(X_t - \lam - A \cos(\pi t))$
while setting $\hat{B}_n(\pi,\al)  := 0$.

\begin{figure}[t]
\centering
\centerline{\footnotesize \hspace{0.2in}1992--1996\hspace{1.6in}1998--2002\hspace{1.7in}2008--2012\vspace{-0.3in}}
\includegraphics[width=1.55in,angle=-90]{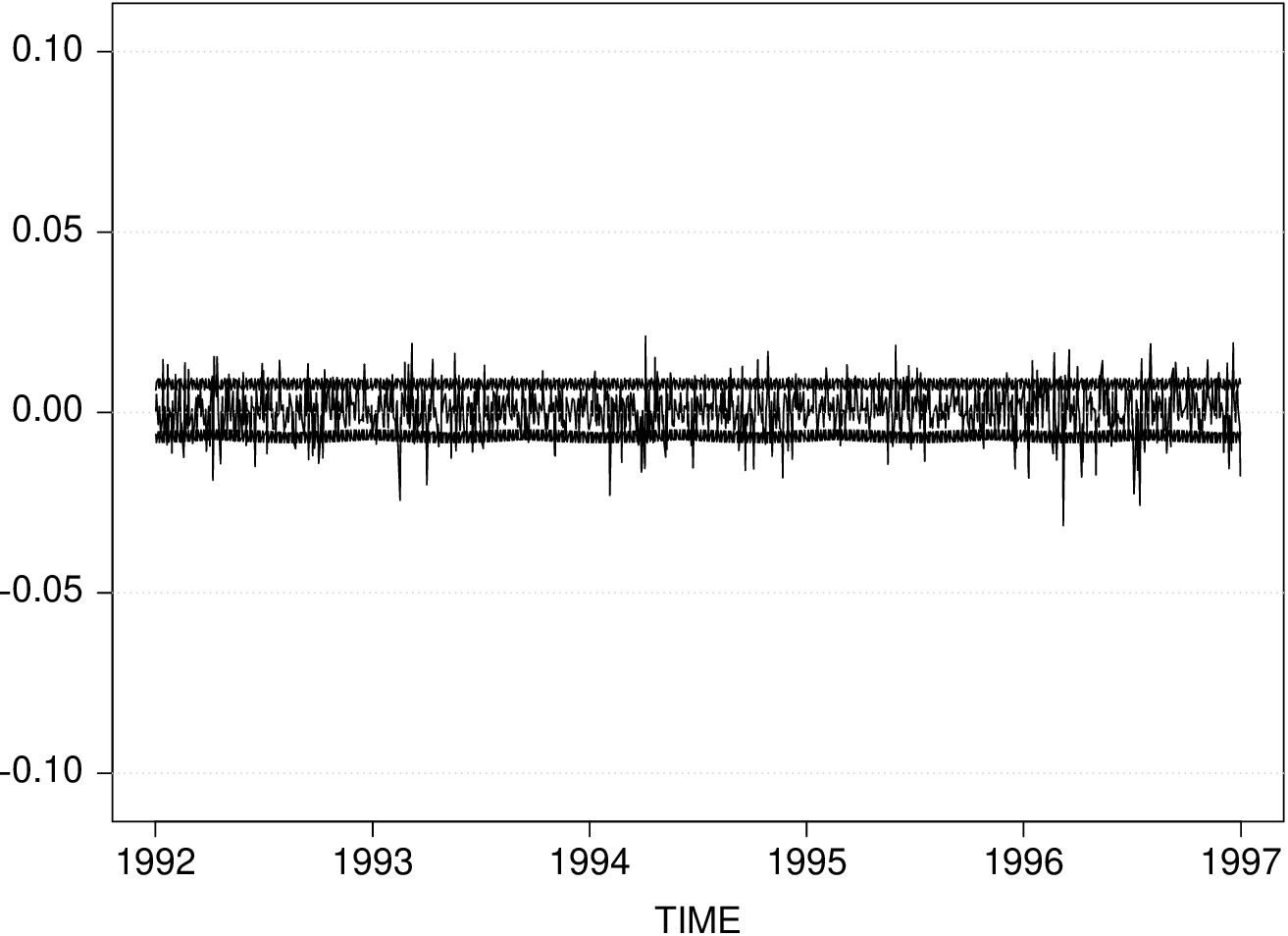} \hfill
\includegraphics[width=1.55in,angle=-90]{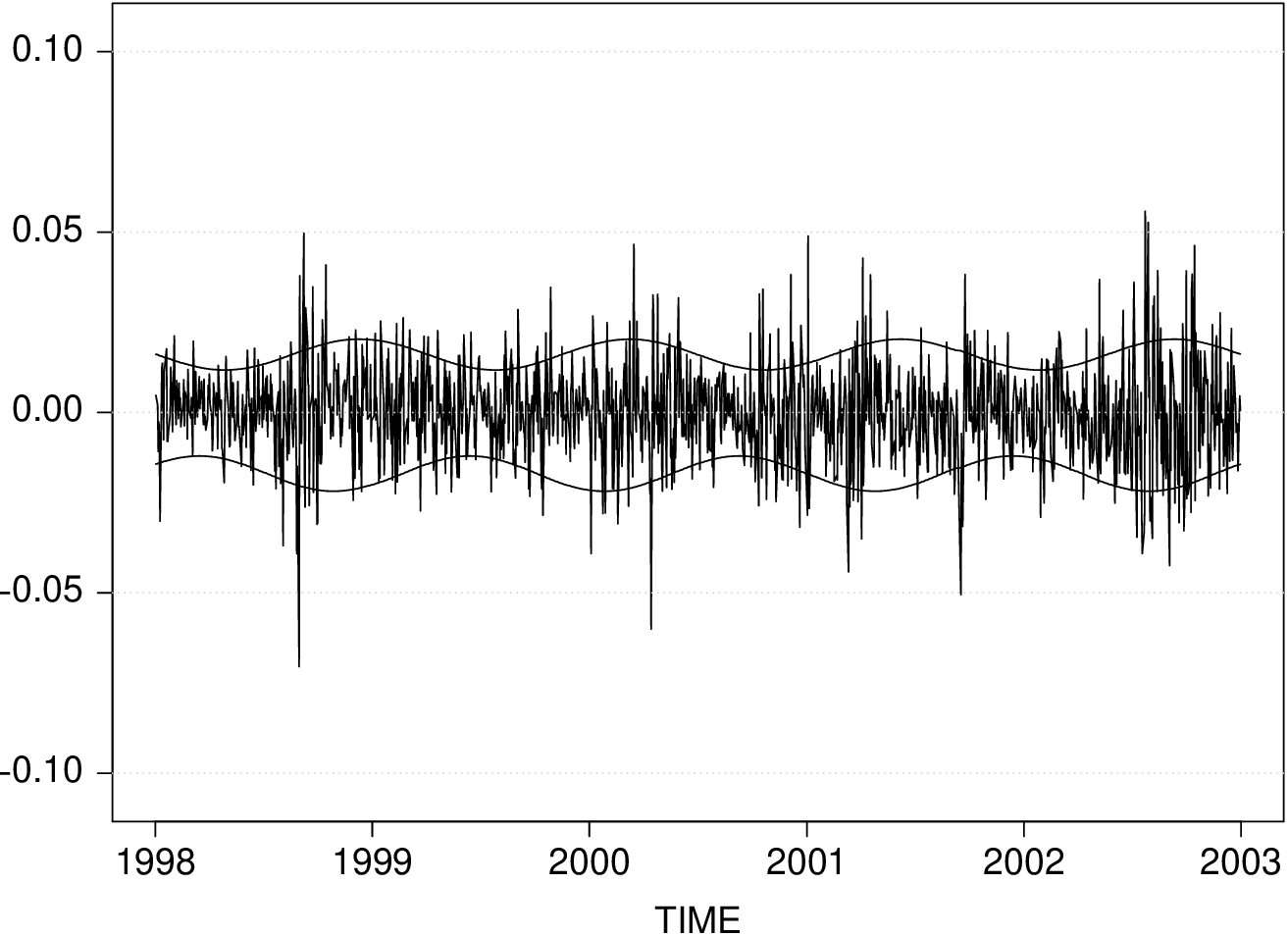} \hfill
\includegraphics[width=1.55in,angle=-90]{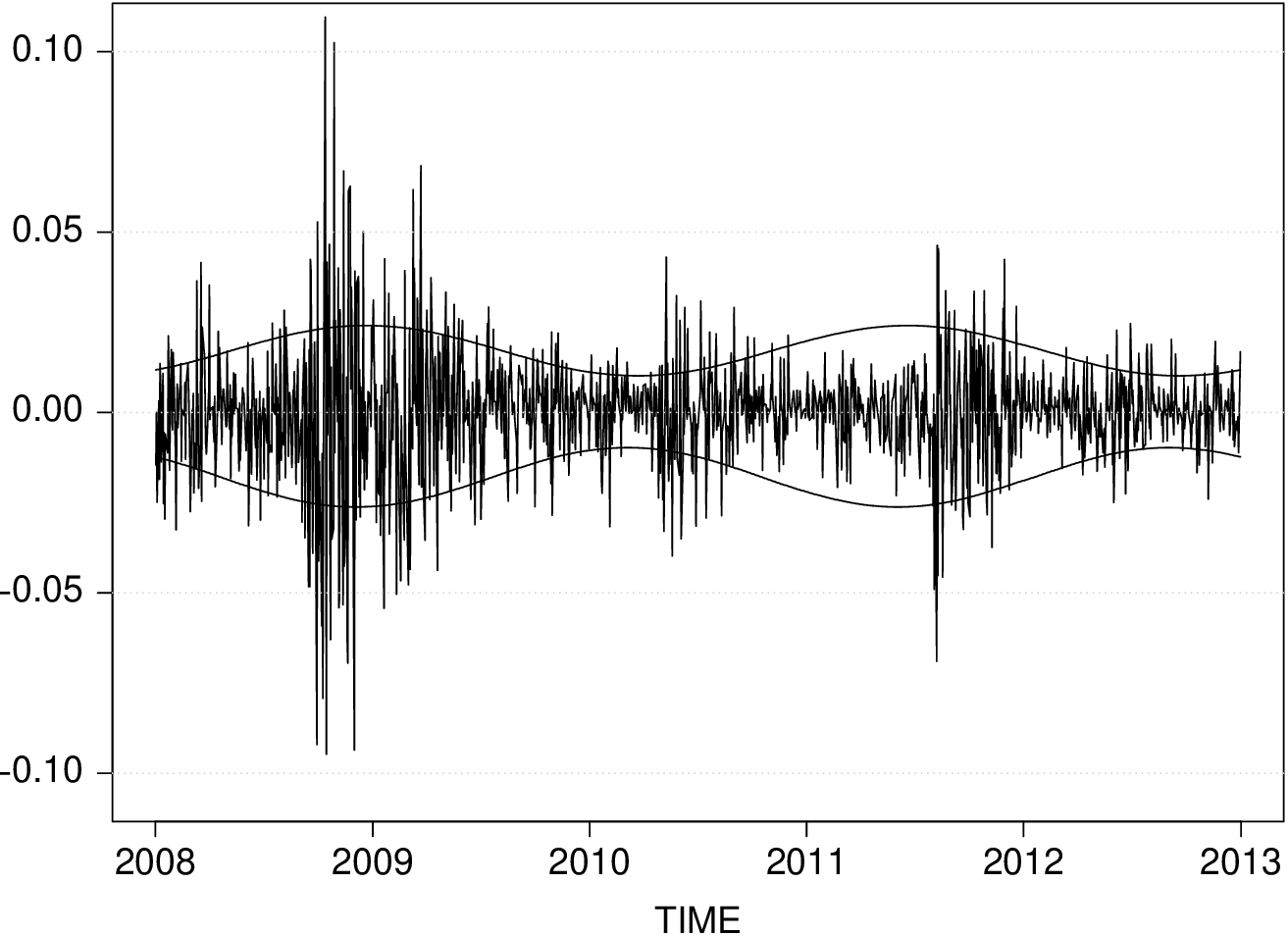} \\
\vspace{0.15in}
\centerline{\footnotesize \hspace{-1in}$\al = 0.1$ \hspace{1.75in}$\al=0.1$\hspace{1.8in}$\al=0.1$ \vspace{-0.55in}}
\includegraphics[width=1.55in,angle=-90]{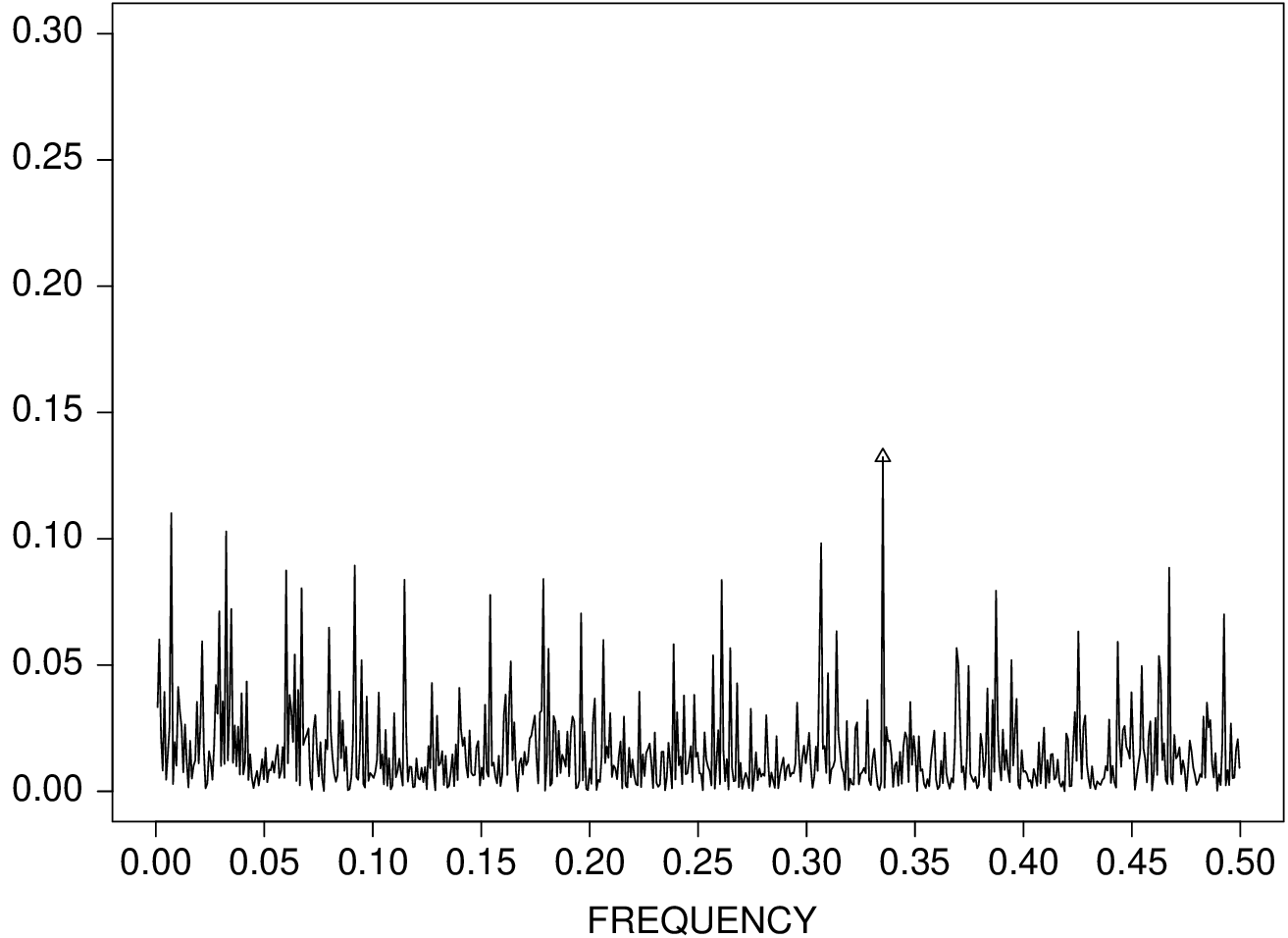} \hfill
\includegraphics[width=1.55in,angle=-90]{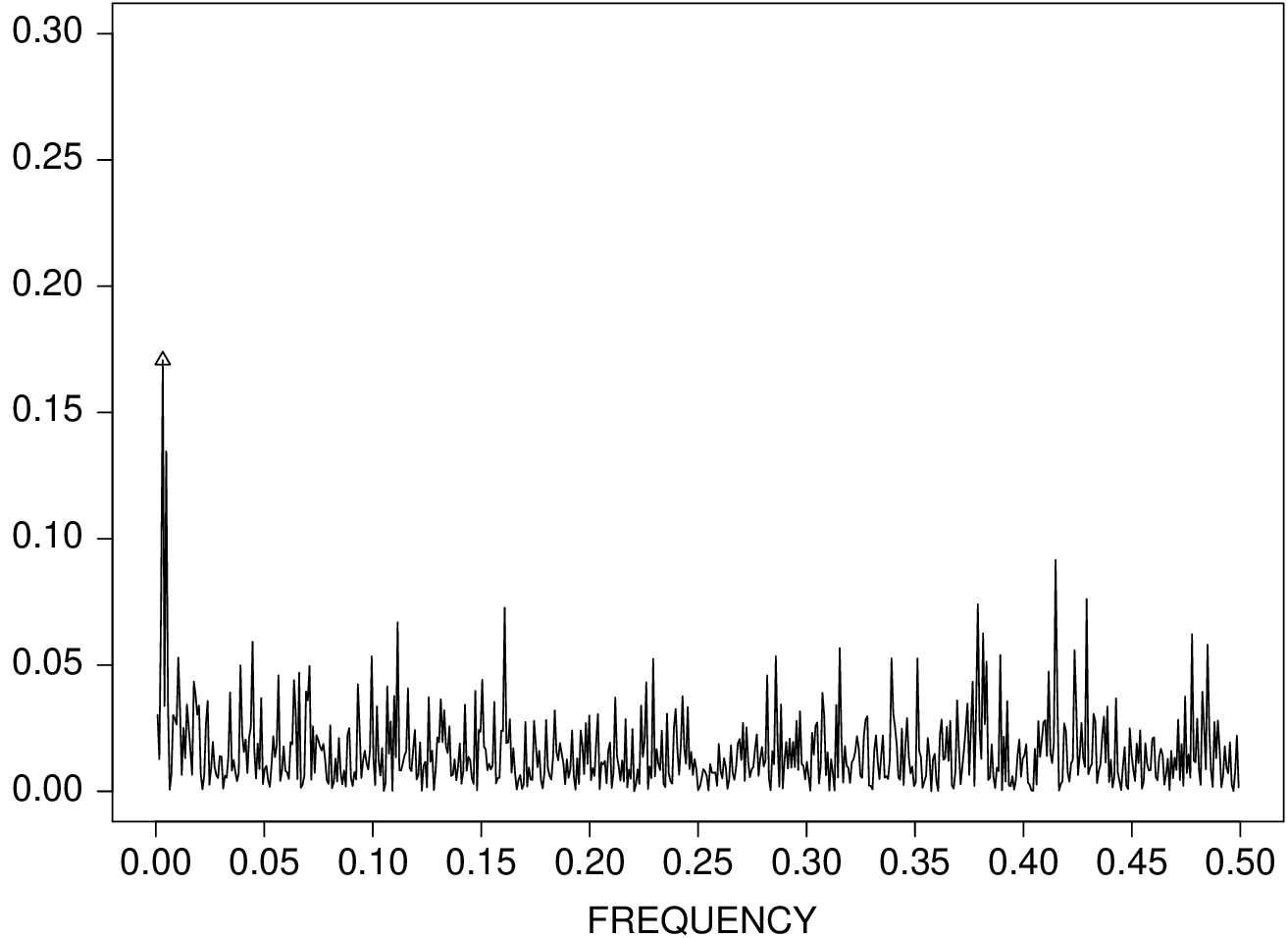} \hfill
\includegraphics[width=1.55in,angle=-90]{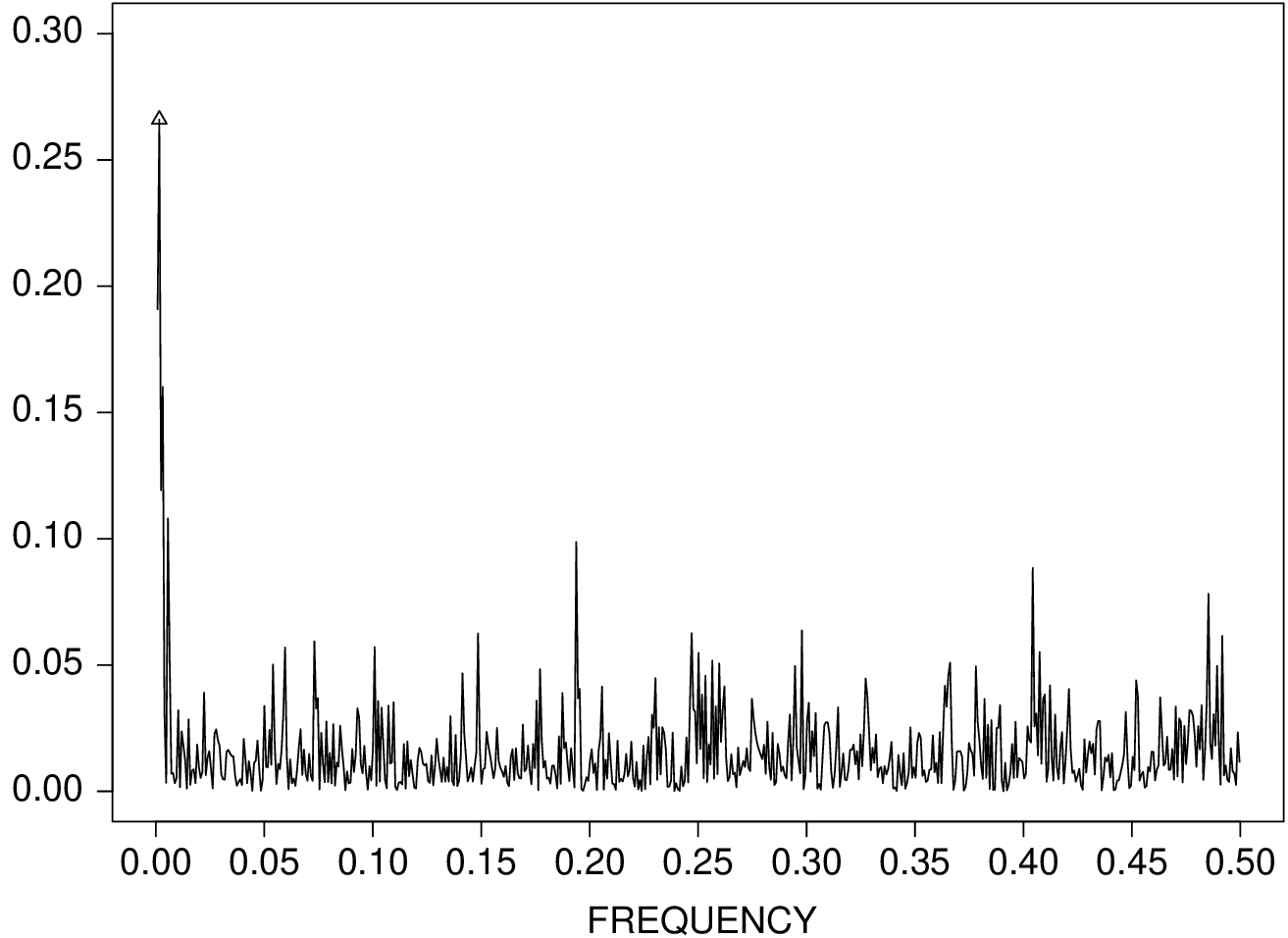} \\
\vspace{0.15in}
\centerline{\footnotesize \hspace{-1in}$\al = 0.9$ \hspace{1.75in}$\al=0.9$\hspace{1.8in}$\al=0.9$ \vspace{-0.55in}}
\includegraphics[width=1.55in,angle=-90]{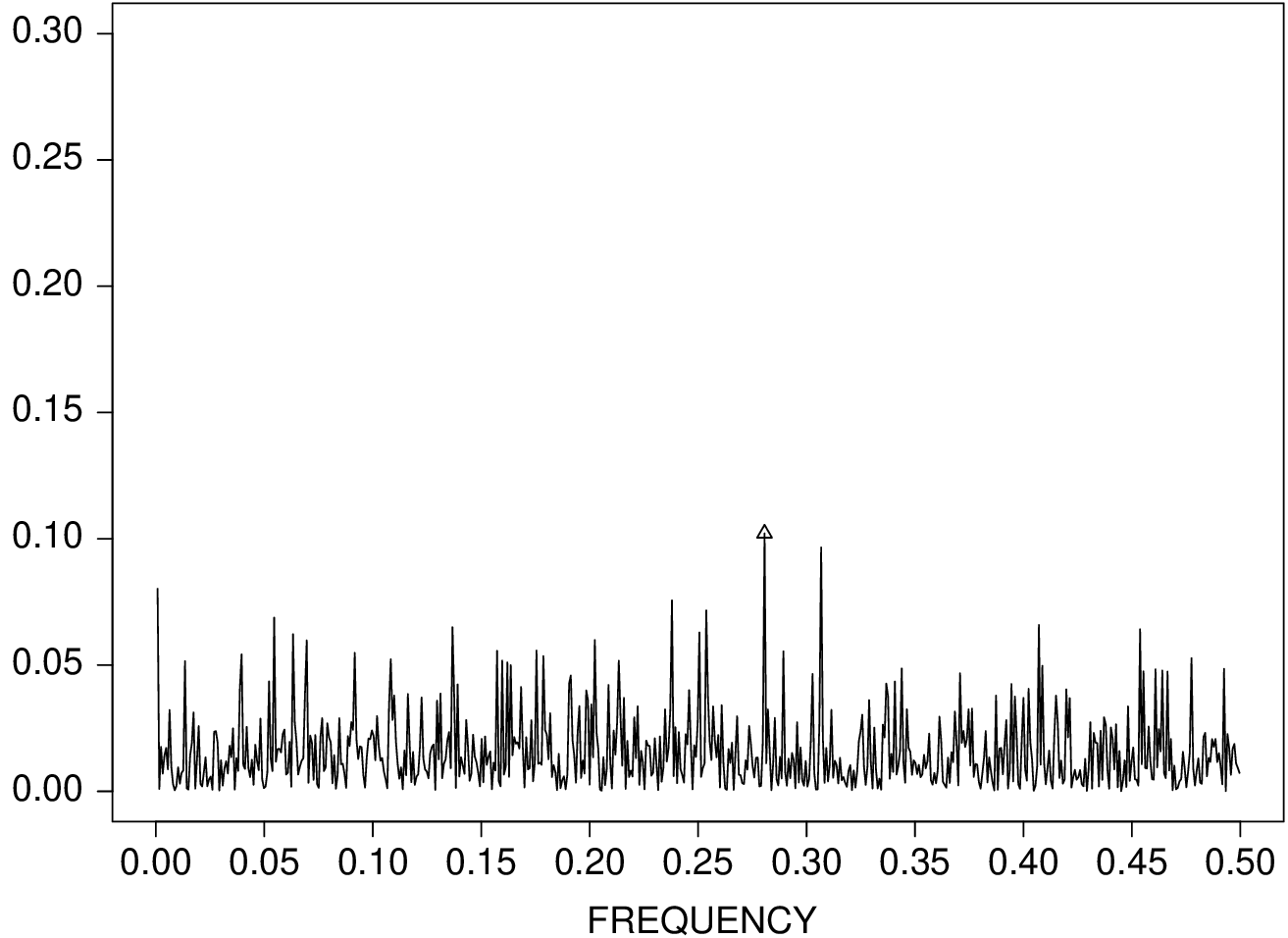} \hfill
\includegraphics[width=1.55in,angle=-90]{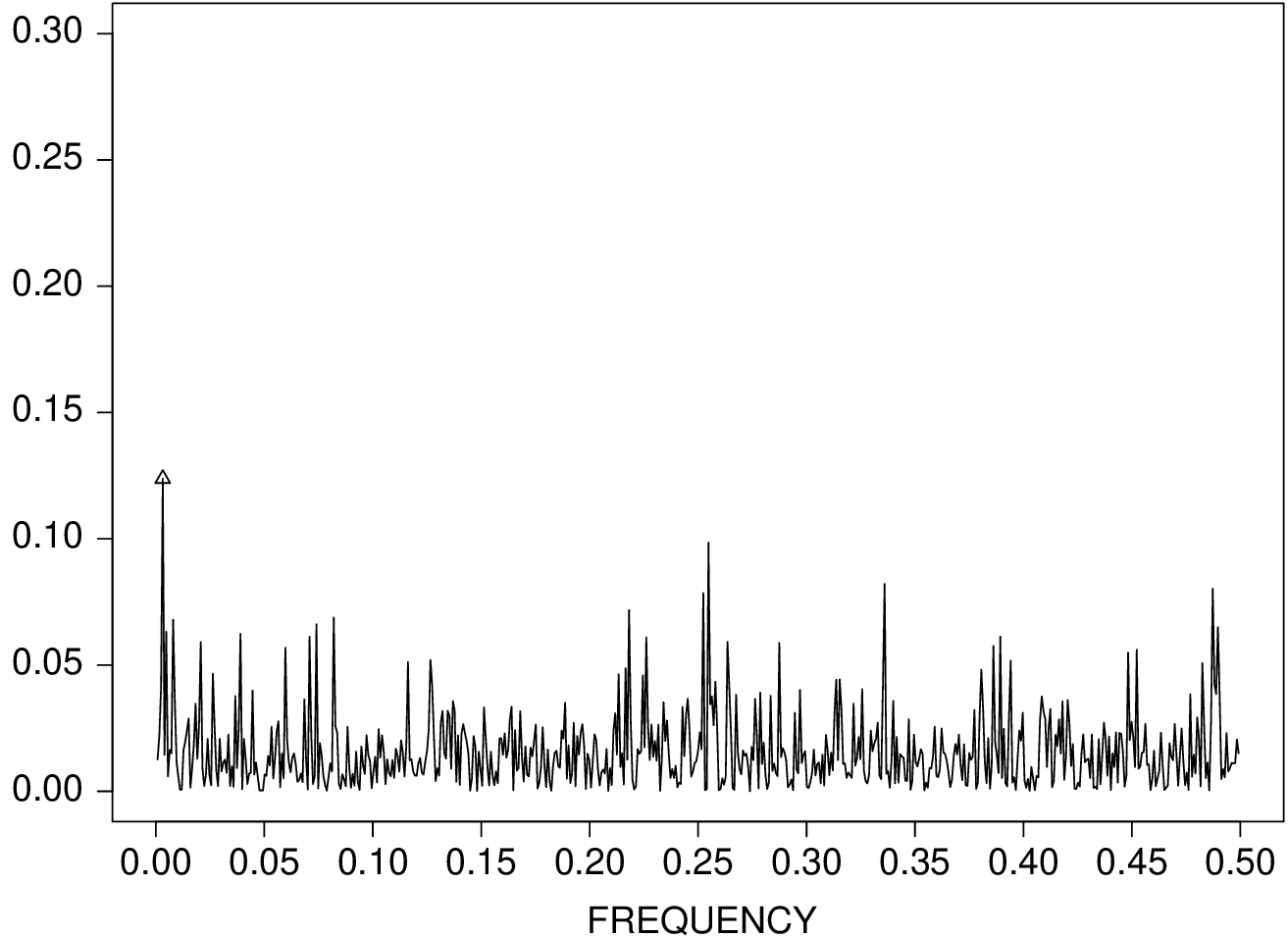} \hfill
\includegraphics[width=1.55in,angle=-90]{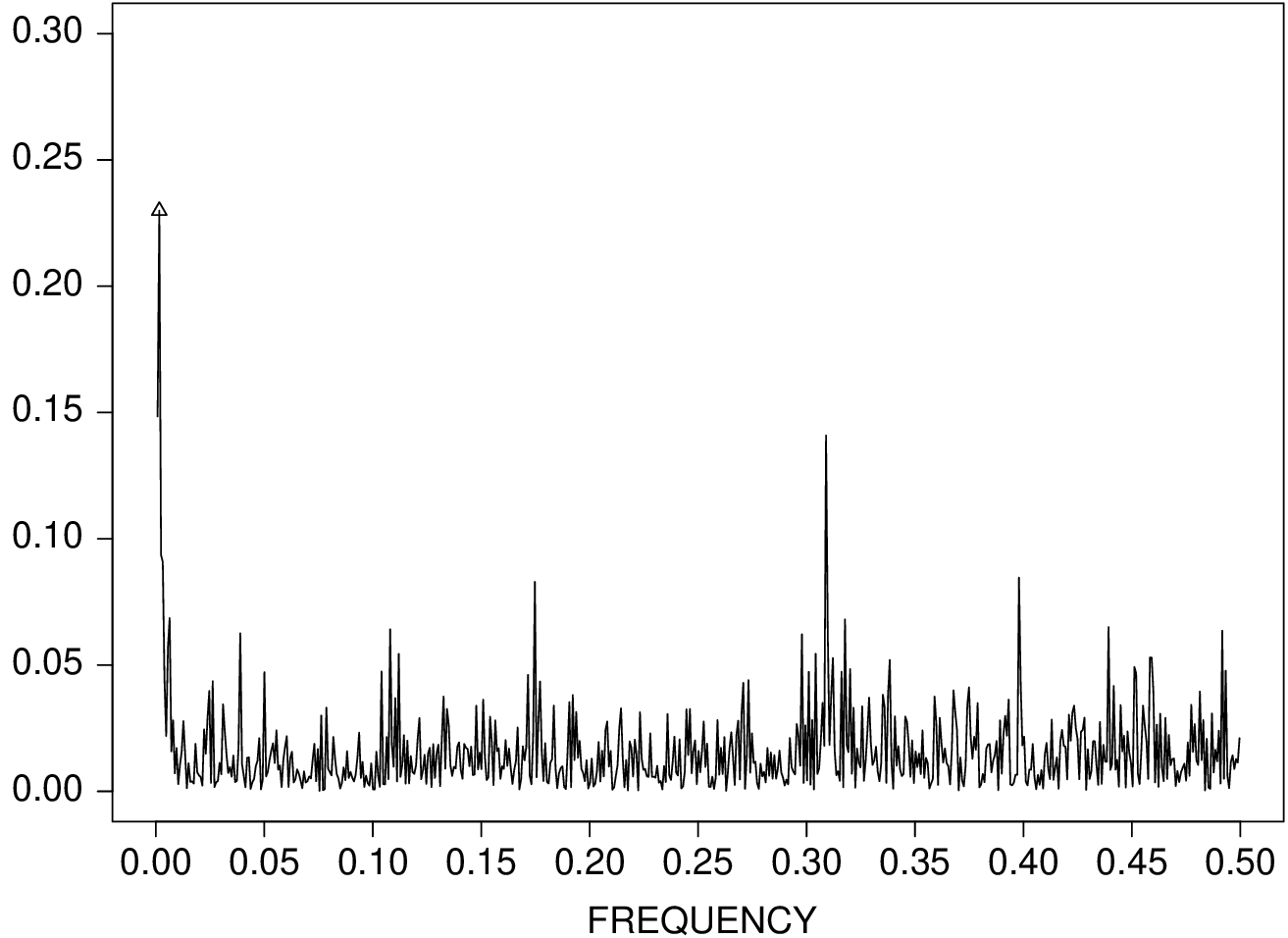} 
\caption{SPX daily log return series (top row) in the preriod 1992--1998  (left), 1998--2002 (middle),
and 2008--2012 (right),  superimposed with the trigonometric quantile regression fits at level $0.1$ and $0.9$ using the peak frequency of the corresponding quantile periodograms (middle and bottom rows) indicated by triangle. }
\label{fig:fit}
\end{figure}

The quantile periodograms  are representations of the oscillatory behavior of series $\{X_t \}$ around its $\al$-quantile level. To illustrate, consider the SPX daily returns in three periods of five years: 1992--1996, 1998--2002, and 2008--2012. Figure~\ref{fig:fit} shows these series and their quantile periodograms defined by (\ref{qper2}) at level $0.1$ and $0.9$. It also shows the quantile regression fits corresponding to the largest periodogram ordinates, which  represent the dominant patterns in these series at the respective quantile levels. As can be seen, the fits of series 1998--2002 and 2008--2012 at level 0.9 exhibit 
low-frequency dynamics that follow quite closely the broad clusters of large positive returns; the fits at level 0.1 behave similarly but for large negative returns. A lack of such clusters in series 1992--1996 results in the dominant patterns reflecting short-term dynamics at both levels. Furthermore, the peak values in the quantile periodograms of series 1998--2002 and 2008--2012 are greater at level 0.1 than at level 0.9, suggesting a stronger clustering effect in large negative returns than in large positive returns. This asymmetric behavior cannot be revealed by the ordinary periodogram of absolute or squared returns.

The theoretical underpinning of the quantile periodograms 
is discussed in Li (2012a). To summarize, let us assume that $\{ X_t \}$ is a strictly stationary process with continuous marginal distribution function $F(\cdot)$; let  $\lam_\al := F^{-1}(\al)$ denote the $\al$-quantile of the marginal distribution
and  $\{ r_\al(\tau): \tau = 0, \pm 1,\dots\}$ denote the autocorrelation function of the so-called level-crossing process $\{ X_t(\lam_\al)\}$, where $X_t(\lam_\al) := 1$ if $X_t \ge \lam_\al$ and $X_t(\lam_\al) := -1$ otherwise; let us further assume that $\{ r_\al(\tau) \}$
is absolutely summable and the corresponding (normalized) spectral density function 
$f_\al(\om) := \sum_{\tau=-\infty}^\infty r_\al(\tau) \exp(i\om\tau)$, known as the
level-crossing spectrum, is strictly positive. Then, under some additional technical conditions (Li, 2012a),  it can be shown that for fixed $\om \in (0,\pi)$ and $\al \in (0,1)$ the quantile periodograms defined by (\ref{qper}) and (\ref{qper2}) have an asymptotic exponential distribution 
with mean 
\eqn
q(\al,\om) := \eta_\al^2 f_\al(\om), \label{qspec}
\eqqn
called quantile spectrum. This property is similar to that of the ordinary periodogram. Indeed, under suitable conditions  (Priestley, 1981, p.\ 425),
the ordinary periodogram $I_n(\om)$ also has an asymptotic exponential distribution, 
but the mean is the power spectrum 
\eqn
g(\om) := \sig^2 f(\om), 
\label{gper}
\eqqn
with $\sig^2$ being the variance of $\{ X_t\}$, 
$f(\om) := \sum_{\tau=-\infty}^\infty r(\tau) \exp(i\om\tau)$ being 
the (normalized) spectral density function of $\{ X_t\}$, and 
$\{ r(\tau)\}$ denoting its autocorrelation function. 

The level-crossing spectrum $f_\al(\om)$ in (\ref{qspec}) is independent of the marginal distribution $F(\cdot)$ and hence invariant to scale; it serves as a spectral representation of the serial dependence of $\{ X_t \}$ in terms of the diagonal bivariate cumulative probabilities $F_\tau(\lam_\al,\lam_\al) := P(X_t \le \lam_\al,X_{t+\tau} \le \lam_\al)$ or, equivalently, the diagonal bivariate copulas $C_\tau(\al,\al) := F_\tau(F^{-1}(\al),F^{-1}(\al))$. This extends the capability of the ordinary
spectrum $f(\om)$ which merely represents the second-order moments
of the bivariate distributions. The scaling factor $\eta_\al^2$ in (\ref{qspec}) takes the form $\al(1-\al)/[F'(\lam_\al)]^2$ or $\al(1-\al)/F'(\lam_\al)$, depending on whether the quantile periodogram is defined by (\ref{qper}) or (\ref{qper2}). This quantity carries the information of the marginal distribution.
It plays the role that $\sig^2$ does in the power spectrum (\ref{gper}).

For simplicity, we will use the quantile periodogram defined by (\ref{qper2}) exclusively in the remainder 
of this paper to demonstrate the proposed method.
We will evaluate the quantile periodogram $q_{n}(\om,\al)$ at the Fourier frequencies 
$\om_k := 2\pi k/n \in (0,\pi)$, with $k = 1, \dots, \lfloor (n-1)/2 \rfloor$, and at equally-spaced 
quantile levels $\al_\ell := \ell/(m+1)$, with $\ell = 1,\dots,m$ for some $m \ge 2$ (e.g., $m=99$).

To focus on the serial dependence rather than the marginal distribution, one may 
consider the normalized quantile periodogram
\eqn
\tilde{q}_n(\om_k,\al_\ell) := \frac{q_n(\om_k,\al_\ell)} { \sum_{k'}
q_n(\om_{k'},\al_\ell)},
\label{nqper}
\eqqn
which satisfies $\sum_{k} \tilde{q}_n(\om_{k},\al_\ell) = 1$ for all $\al_\ell$ $(\ell=1,\dots,m)$. 
One may also consider the cumulative quantile periodogram
\eqn
Q_n(\om_k,\al_\ell) := \sum_{k' \le k} \tilde{q}_n(\om_{k'},\al_\ell).
\label{Q}
\eqqn
Both $\{\tilde{q}_n(\om_k,\al_\ell)\}$ and $\{ Q_n(\om_k,\al_\ell) \}$
can be displayed graphically as images.
Investigating the patterns of these two-dimensional arrays visually or numerically for the underlying serial dependence properties they represent constitutes what we call the quantile-frequency analysis, or QFA. 

The R package {\tt quantreg} by Koenker (2005) can be used to compute the quantile regression coefficients in (\ref{qr}) needed to construct these arrays. In particular, for fixed frequency $\om_k$, a single call of the {\tt rq} function with the supply of all quantile levels is able to produce the regression coefficients for all $\al_\ell$. Parallelization with respect to the frequencies $\om_k$ using the {\tt foreach} package (Weston, 2019) further  speeds up the computation. An R code for computing the quantile periodograms is available at https://github.com/IBM/qfa.

\begin{figure}[t]
\centering
\centerline{\footnotesize \hspace{0.0in}1992--1996\hspace{1.6in}1998--2002\hspace{1.6in}2008--2012\vspace{-0.3in}}
\hspace{0.25in}
\includegraphics[width=1.4in,angle=-90]{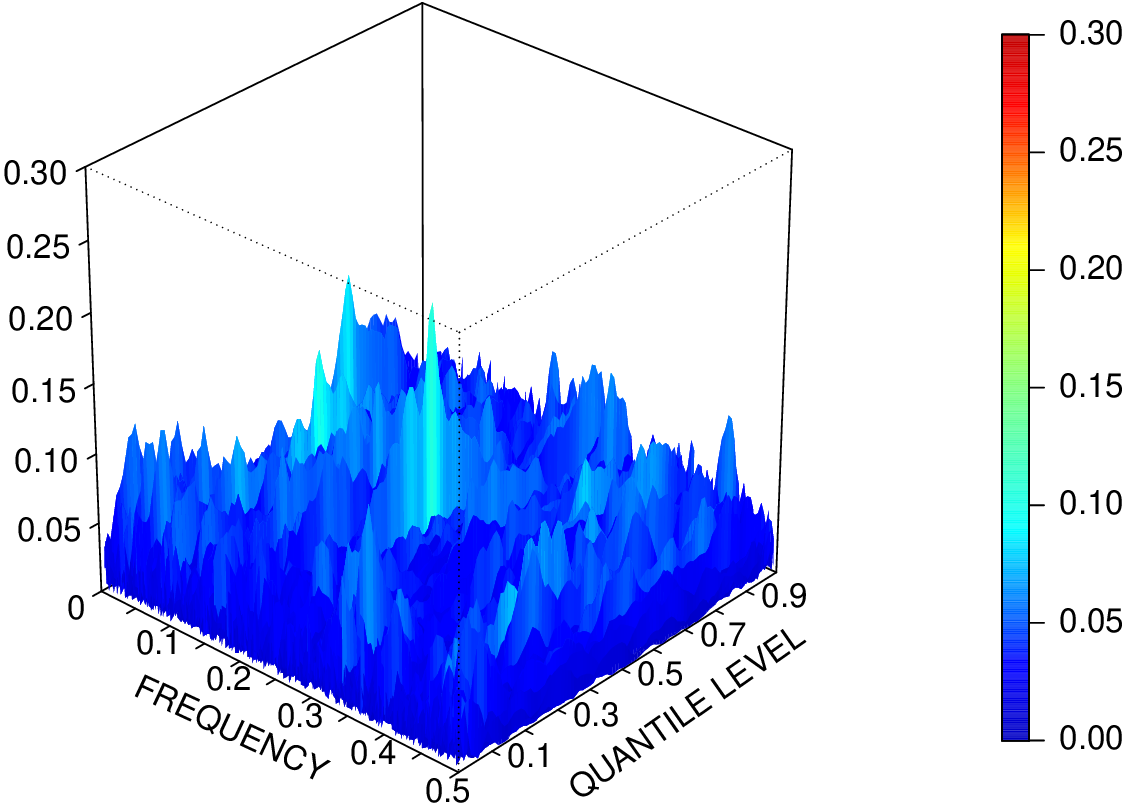}  \hfill
\includegraphics[width=1.4in,angle=-90]{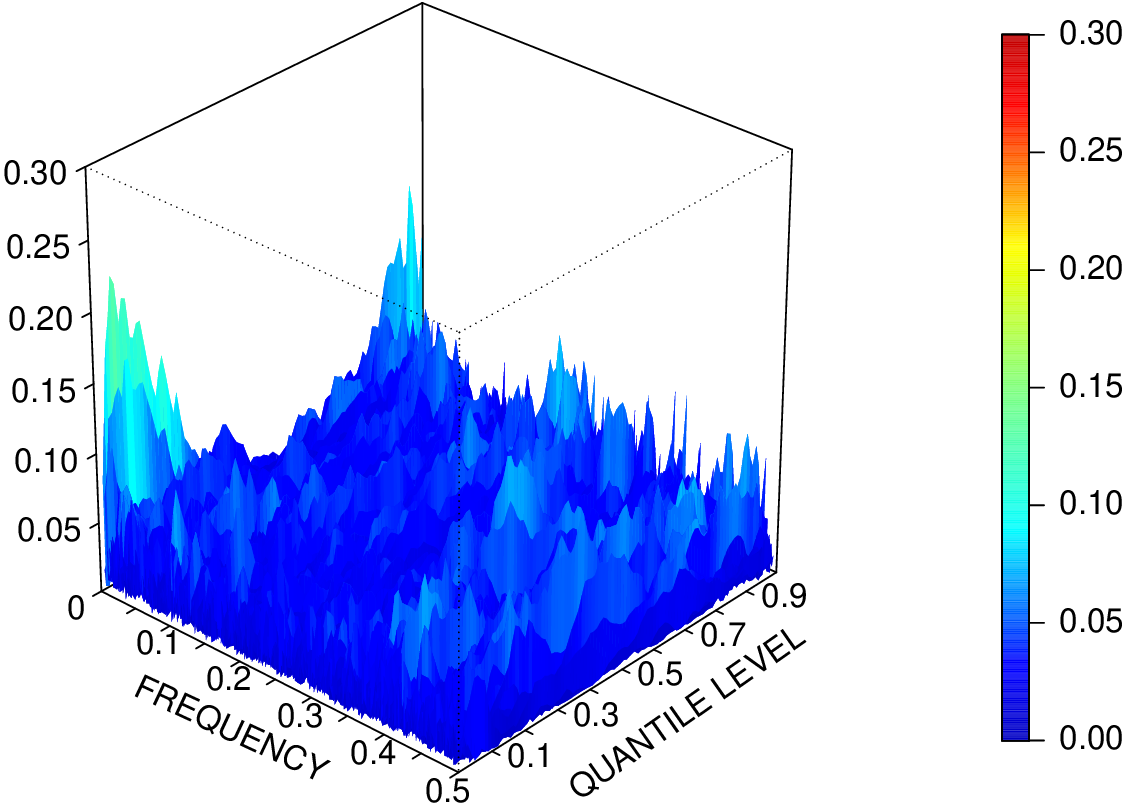}  \hfill
\includegraphics[width=1.4in,angle=-90]{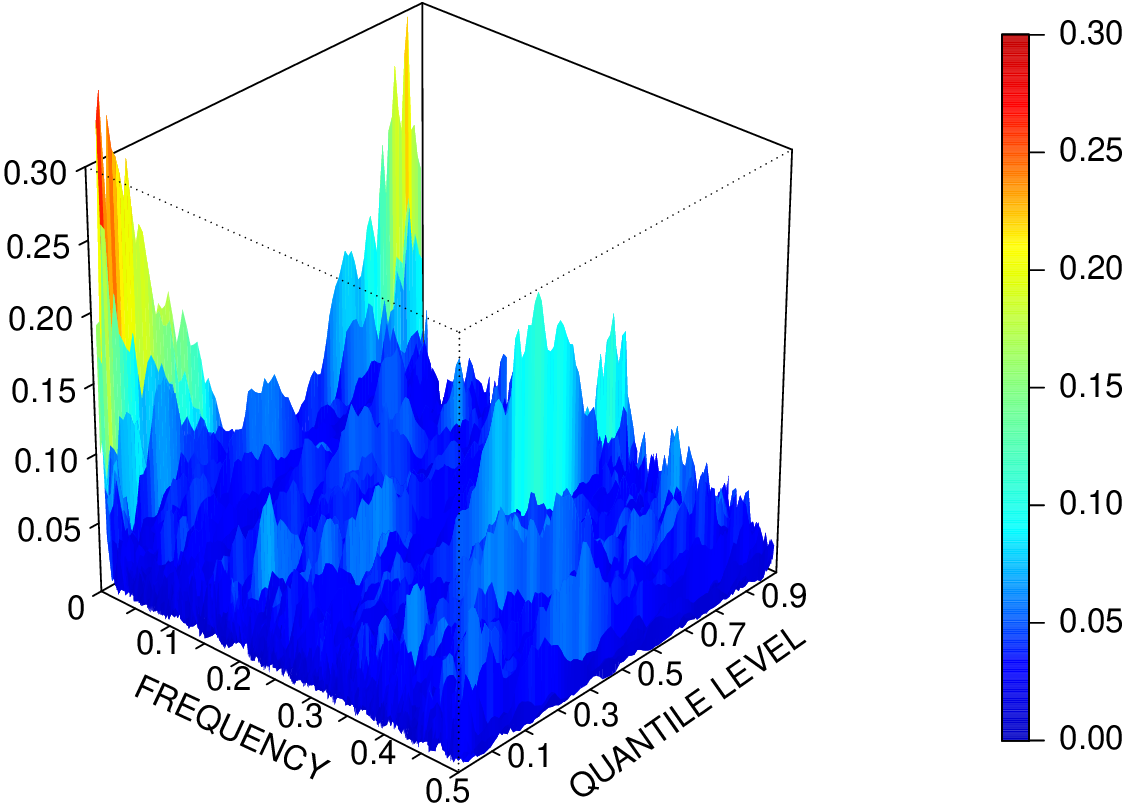}  \\
\includegraphics[width=1.55in,angle=-90]{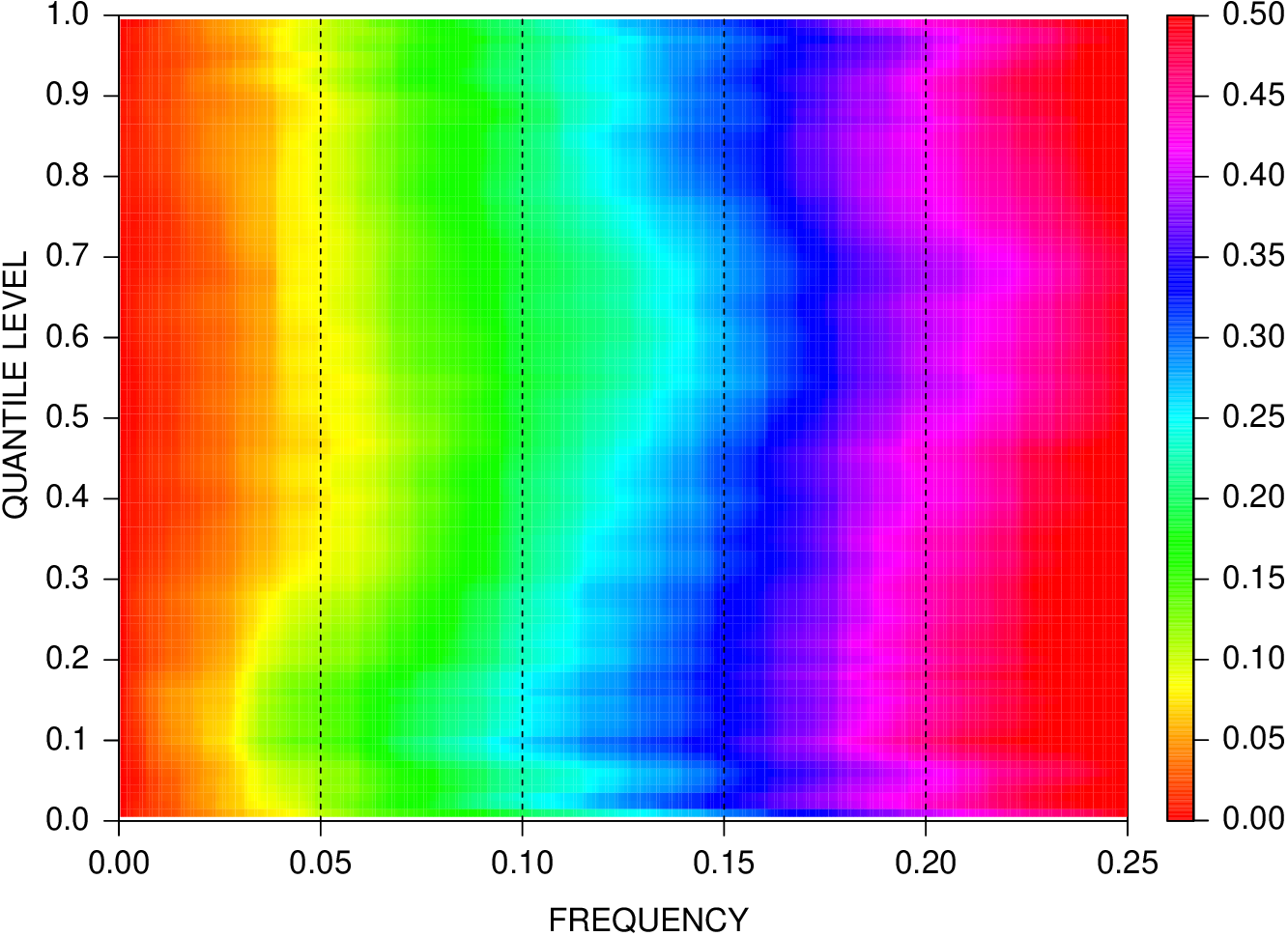} \hfill
\includegraphics[width=1.55in,angle=-90]{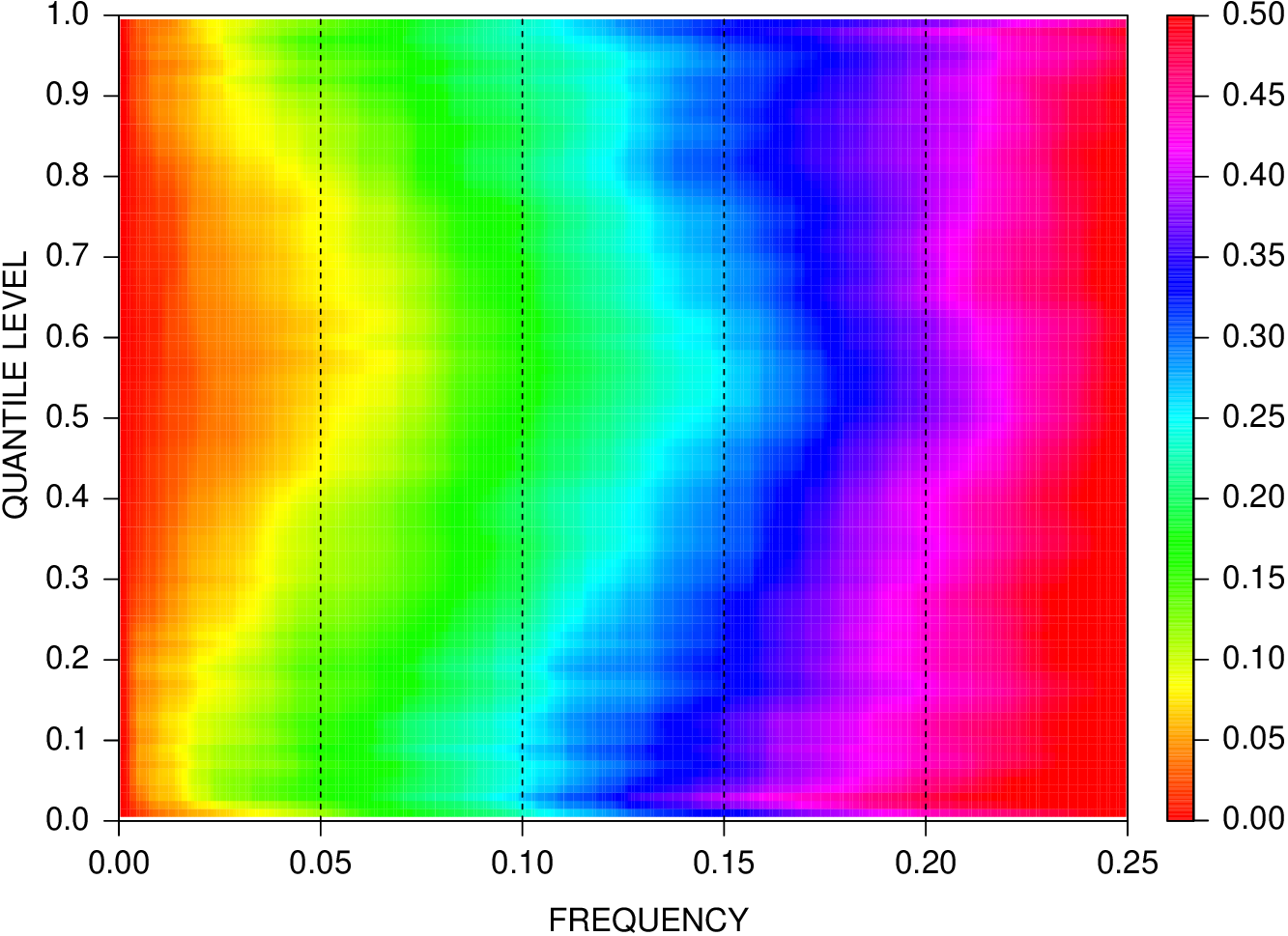} \hfill
\includegraphics[width=1.55in,angle=-90]{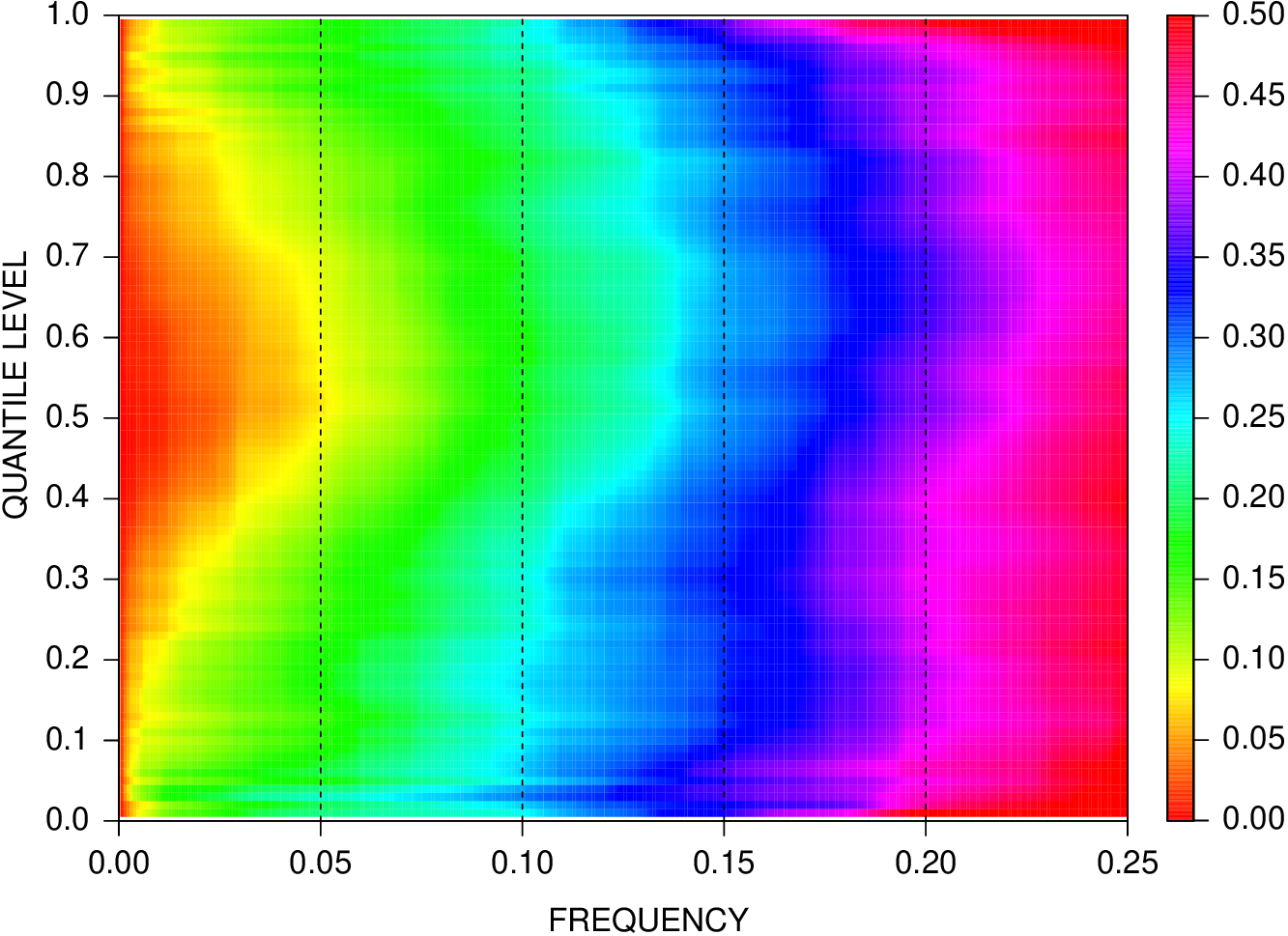}
\caption{Normalized quantile periodograms (top row) and cumulative quantiles periodograms (bottom row) of series 1992--1996 (left), 
series 1998--2002 (middle), and series 2008--2012 (right). All cumulative quantile periodograms  here and thereafter are plotted only for the lower half of the frequencies where the most interesting features reside.}
\label{fig:QFA}
\end{figure}

Figure~\ref{fig:QFA} depicts the arrays of normalized quantile periodogram and cumulative quantile periodogram for the three SPX series in Figure~\ref{fig:fit}. Strong low-frequency
activities appear in the lower and upper quantile regions for series 1998--2002 and 2008--2012, whereas a lack of such is noticed for series 1992--1996. An asymmetry between lower quantiles and upper quantiles
is shown for all series, especially 1998--2002 and 2008--2012. Large values 
at middle quantiles and higher frequencies are also observed for series  1992--1996
and 2008--2012, reflecting some short-term behaviors of small returns.

\section{Spectral Measures and Applications}

Given the arrays of quantile periodograms in (\ref{nqper}) 
and cumulative quantile periodograms in (\ref{Q}), one can borrow traditional spectral analysis techniques to create QFA-based spectral measures to quantify the devation of the observed behavior of serial dependence from the expected one.

For example, motivated by the Kolmogorov-Smirnov statistics for the ordinary periodogram (Priestley, 1981, 
p.\ 481), we define
\eqn
KS_{\rm max} & := & \max_{\al_\ell \in \cA} \bigg\{
\sqrt{|\Om|} \max_{\om_k \in \Om}  |Q_n(\om_k,\al_\ell) - Q(\om_k,\al_\ell)| \bigg\}, \label{KSmax} \\
KS_{\rm mean}  & := & \underset{\al_\ell \in \cA}{\rm mean} \bigg\{
\sqrt{|\Om|} \max_{\om_k \in \Om}  |Q_n(\om_k,\al_\ell) - Q(\om_k,\al_\ell)| \bigg\}. \label{KSmean}
\eqqn
In these expressions, $Q(\om,\al)$ denotes the expected cumulative quantile spectrum of a model in question, $\cA \times \Om$ defines the quantile-frequency region to focus on, and $|\Om|$ is the cardinality of $\Om$.
The two metrics are distinguished by how the Kolmogorov-Smirnov statistics are aggregated over the quantile levels. It is expected that $KS_{\max}$ be more sensitive
when deviations are concentrated on a few quantiles and that $KS_{\rm mean}$  be more sensitive when deviations are spread out across many quantiles.

In addition,  motivated by the so-called Whittle likelihood for the ordinary periodogram (Whittle, 1962),
we can also define
\eqn
WL_{\rm max} & := & \max_{\al_\ell \in \cA} \bigg\{
\frac{1}{\sqrt{ |\Om| }} \sum_{\om_k \in \Om}  \
d(\tilde{q}_n(\om_k,\al_\ell)/\tilde{q}(\om_k,\al_\ell)) \bigg\}, \label{WLmax} \\
WL_{\rm mean}  & := & \underset{\al_\ell \in \cA}{\rm mean} \bigg\{
\frac{1}{\sqrt{ |\Om| }} \sum_{\om_k \in \Om} 
d(\tilde{q}_n(\om_k,\al_\ell)/\tilde{q}(\om_k,\al_\ell))   \bigg\}, \label{WLmean}
\eqqn
where $\tilde{q}(\om,\al)$ denotes the expected (normalized) quantile spectrum and
$d(x) := x - \log(x) - 1$ for $x > 0$ is a nonnegative convex function
with a unique minimum zero at $x=1$. These metrics examine the quantile periodogram rather than the cumulative quantile periodogram, and they are closely related to the Kullback-Leibler divergence (Kullback and Leibler, 1951). Note that it is also possible to define the WL metrics using the unnormalized quantile periodogram and spectrum to take the marginal distribution into account.

Similarly to the conventional spectral measures of the ordinary periodogrm, the QFA-based spectral measures in (\ref{KSmax})--(\ref{WLmean}) can be used for diagnostic checking of the goodness of fit of time series models. We consider two approaches, which we call the residual approach and the direct approach, respectively.

In the residual approach, the metrics are applied to the residuals of a model to check whether and where the white noise assumption may be violated. To that end, it suffices to set $\tilde{q}(\om_k,\al_\ell) := 1/\lfloor (n-1)/2 \rfloor$ and $Q(\om_k,\al_\ell) := k/\lfloor (n-1)/2 \rfloor$.
This approach complements the standard portmanteau tests, such as the Ljung-Box (LB) test based on the autocorrelations of the squared residuals, and the Lagrange multiplier (LM) ARCH test based on the autoregressive $R^2$-statistics of the squared residuals. An advantage of the QFA-based tests is that they avoid the need to specify certain sensitive parameters such as the number of autocorrelations in the LB test and the order of autoregression in the LM test.

In the direct approach, the spectral measures are applied directly to the original time series rather than the residuals, and the targets $\tilde{q}(\om,\al)$ and $Q(\om,\al)$ are replaced by the expected quantile spectrum and cumulative quantile spectrum under the assumed model. Because mathematical formulas of the expected spectra are not generally available, we resort to a Monte Carlo simulation procedure which comprises (a) generating a large number of independent realizations from the model and (b)  calculating the ensemble average of the quantile periodograms and cumulative quantile periodograms of the simulated series. 

Another application of the spectral measures is model-based discriminant analysis which determines whether or not two observed time series can be regarded as having similar serial dependence properties. In this application, we first fit a suitable model to one series and then check how well it fits the other series using the spectral measures. We refer to this application as discriminant testing. Because this approach is prone to model misspecification, just like any model-based method in hypothesis testing, it is important that the model fitted to one series does not fail the goodness-of-fit test based on the same spectral measure that detects 
a lack of fit for the other series.

The quantile periogram at a fixed quantile level has similar asymptotic statistical properties to the ordinary periodogram (Li, 2008; 2012a). Therefore, in the special case where $\cA$ contains a single value, the asymptotic theory for the conventional KS and WL statistics of the ordinary periodogram (Huang, Ombao and Stoffer, 2004; Kakizawa, Shumway and Tanaguchi, 1998) can be used to approximate the 
distributions of the QFA-based spectral measures in (\ref{KSmax})--(\ref{WLmean}) under the null hypothesis. When $\cA$ contains multiple values, the situation becomes more complicated and the theoretical distributions remain unknown at this time. In this paper, we  rely on more practical means to demonstrate the QFA method.

More specifically, we employ the well-known technique called parametric bootstrapping  (Efron and Tibshirani, 1993), where a large number of independent realizations are simulated from the fitted model under the null hypothesis and the empirical distributions of the resulting metrics are used to determine the $p$-values of the observed metrics. Although it requires a complete specification of the model (including the distribution of the residuals), the parametric bootstrapping technique does have the advantage of being able to address the finite-sample properties of the metrics that an asymptotic theory cannot. 

Besides parametric bootstrapping, there are other ways to generate bootstrap samples. For example, in goodness of fit testing, the need to specify the residual distribution can be 
avoided by sampling the observed residuals; in discriminant testing, one may use nonparametric techniques to generate bootstrap samples for time series (e.g., Paparoditis and Politis, 1999; B$\ddot{\rm u}$hlmann, 2002).
In this paper, we focus on the fully parametric approach.

\section{Simulation Study}

The spectral measures (\ref{KSmax})--(\ref{WLmean}) are expected to respond  differently to different types of spectral deviations. To shed some light on their sensitivity, we conduct three simulation 
experiments.

The first experiment tests the null hypothesis of Gaussian white noise against the alternative hypothesis
of a GARCH process driven by Gaussian white noise where the parameters are estimated from SPX series 2008--2012 (see next section for details of the model). The second experiment tests
the null hypothesis of the GARCH process in the first experiment against the alternative hypothesis 
where the GARCH process, denoted by $X_{t1}$, is combined with an independent sinusoidal process $X_{t2} := \sqrt{2} \cos(2\pi f t + \phi)$, where $f \in (0,0.5)$ is the frequency parameter and $\phi$ is a uniformly distributed random variable in $[0,2\pi)$. Specifically, the time series under the alternative hypothesis takes the form 
\eq
X_t := (1-w(X_{t1})) X_{t1} + w(X_{t1}) X_{t2},
\eqq
where $w(x) :=  a(\tanh(sx+1)-\tanh(sx-1))/(\tanh(1)-\tanh(-1))$ is a bell-shaped function 
that peaks at $x=0$, with $a \in [0,1]$ and $s>0$ controlling its height and width, respectively. 
In the experiment, we take $f =0.25$, $a=0.4$, and $s=4$. 
The model simulates a periodicity at middle quantiles while retaining 
the GARCH effect at lower and higher quantiles. The third experiment 
tests the null hypothesis of a GJR-GARCH process driven by Gaussian white noise 
with parameters estimated from series 1992--1998 against the alternative hypothesis
of a GJR-GARCH process where the parameters are estimated from series 2008--2012
(see next section for details of the models).
Table~\ref{tab:simu} contains the detection probabilities of the spectral measures 
(\ref{KSmax})--(\ref{WLmean}) for these experiments when the false alarm probability is fixed at 0.05.

It is well known that quantile regression exhibits a greater statistical variability 
when the quantile level becomes very close to 0 or 1 relative to the sample size 
(Koenker, 2005, p.\ 130). Generally, it is conceivable that the inclusion of such ``extreme'' quantiles 
could be beneficial to the lack-of-fit detection  if the signal at these quantiles outweighs the increased variability. On the other hand, if the signal is absent or relatively weak at these quantiles, 
the added variability could reduce the detection power of the resulting tests. 
Table~\ref{tab:simu} contains the results with two choices of the set $\cA$ 
for quantile levels: $\cA = \{ 0.05,0.06,\dots,0.95\}$ or $\cA = \{ 0.01,0.02,\dots,0.99\}$. 
In both cases, $\Om$ is the set of Fourier frequencies in $(0,\pi)$.

The results in Table~\ref{tab:simu} show that the KS metrics are more effective 
in detecting the GARCH effect and the WL metrics are more effective in detecting 
the sinusoidal ``contamination'' in the GARCH model.
The results also show that the tests benefit from the inclusion of quantile levels near 0 and 1
in the first experiment but suffer from it in the second experiment, and its effect  is mixed  in the third
experiment.

\begin{table}[t]
{\footnotesize
\begin{center} 
\caption{Detection Probability of Spectral Measures (\ref{KSmax})--(\ref{WLmean}) } 
\label{tab:simu}
\begin{tabular}{c|cccc|cccc} \hline
&  \multicolumn{4}{c|}{$\cA = \{0.05,0.06,\dots,0.95\}$} 
&  \multicolumn{4}{c}{$\cA = \{0.01,0.02,\dots,0.99\}$} \\[-0.1in]
 \multicolumn{1}{c|}{Experiment} & 
 $KS_{\max}$ & $WL_{\max}$ & $KS_{\rm mean}$ & $WL_{\rm mean}$ &
 $KS_{\max}$ & $WL_{\max}$ & $KS_{\rm mean}$ & $WL_{\rm mean}$ 
  \\  \hline 
1        & 0.923 &  0.165 & 0.835 &  0.462 & 0.949 & 0.177 & 0.913 & 0.530 \\
2        & 0.111 &  0.961 & 0.371 &  0.860 & 0.059 & 0.950 & 0.345 &  0.846 \\
3        & 0.614 & 0.117  & 0.465 &  0.208 & 0.493 & 0.114 & 0.556 & 0.238 \\
\hline
\end{tabular} 
\end{center}
\hspace{0.8in}Note. (a) Results are based on 1000 simulation runs. (b) False alarm probability equals 0.05.
}
\end{table}

\section{Application to S\&P 500 Daily Returns}

In this section we present some results for the financial time series shown in Figure~\ref{fig:fit}. These series are daily log returns derived from the daily closing values of the S\&P 500 index. Specifically, if $Y_t$ and $Y_{t-1}$ denote the closing values in trading days $t$ and $t-1$, respectively, then the log return in day $t$ is defined as $X_t := \log(Y_t/Y_{t-1})$. The three periods are chosen with the intention of representing different social and economic conditions:  the 1992--1996 period has relatively low volatility and no dramatic events; the 1998--2002 period has high volatility and includes the dot-com bubble and the  event of September 11, 2001; and the 2008--2012 period has very high volatility as a result of the financial crisis and its aftermath. Only scale-invariant statistics are employed in this application, because our main interest  is in their serial dependence properties rather than their apparently different scales and marginal distributions.

The application has two diagnostic components as outlined in Section 3: (a) the goodness of fit testing to check whether and where some popular financial time series models may experience a lack of fit using both residual and direct approaches; (b) the discriminant testing to check whether and where these series may differ in the underlying serial dependence properties based on the models. Needless to say,
the goodness of fit testing is an important component of any statistical modeling process, not only for ensuring the integrity of the models in subsequent applications (e.g., inference and forecasting), but also for identifying possible lack-of-fit areas to help improve the models. The discriminant testing enables the  detection of regime changes in financial markets with potentially large consequences in forecasting, investment portfolio optimization, 
and other economic applications (Ang and Timmermann, 2012).

In this application, we consider the two most popular types of models: the GARCH models (Engle, 1982; Bollerslev, 1986) and the GJR-GARCH models (Glosten, Jagannathan and Runkle, 1993). These models belong to the so-called APARCH family  (Ding, Granger and Engle, 1993) that can be expressed as
\eqn
\left\{
\begin{array}{lll}
X_t & = & \mu + \sig_t \ep_t, \quad \{ \ep_t \} \sim \IID(0,1), \\
\sig_t^r & = & {\displaystyle a_0 + \sum_{i=1}^{p_1} a_i (|\ep_{t-i}| - c_i \ep_{t-i})^r
 + \sum_{j=1}^{p_2} b_j \sig_{t-j}^r \quad (r> 0, |c_i| \le 1)}.
 \end{array}
\right.
\label{model}
\eqqn
The GARCH models correspond to the case where $r=2$ and $c_i =0$ for all $i$. The GJR-GARCH models include the additional parameters $c_i$ to allow asymmetric feedback from positive and negative values of previous excitations $\ep_{t-i}$. The possible choice of $r \ne 2$ in (\ref{model}) introduces 
nonlinearity in the volatility.

In the application, we further take $p_1=p_2=1$, which gives the most popular and useful GARCH(1,1) model (Hansen and Lunde 2005).  The remaining parameters are estimated from  data, using the {\tt garchFit} function in the {\tt fGarch} package (Wuertz, 2017). 

For each combination of models and series, 
Table~\ref{tab:fit} summarizes the maximum likelihood estimates of the parameters under the assumption that 
$\{ \ep_t \}$ is Gaussian white noise. Table~\ref{tab:fit} also contains the $p$-values of the standard LB and LM tests
on the basis of squared residuals. The test results suggest that there may be a lack of fit in both models of series 2008--2012.

\begin{table}[t]
{\footnotesize
\begin{center} 
\caption{Estimated Model Parameters and $p$-Vaules of Standard Goodness-of-Fit Tests} 
\label{tab:fit}
\begin{tabular}{l|l|ccccc|ccc} \hline
 \multicolumn{1}{c|}{Model} &
\multicolumn{1}{c|}{Series} &
$\; \ \mu$ & $a_0$ & $a_1$ & $b_1$ & $c_1$  & LB & LM \\  \hline
GARCH                  &  1992--1996 & \; \ 5.29e-4 &  1.42e-6 & 4.33e-2 & 9.19e-1 & & 
 0.405 &  0.381 \\
                            &  1998--2002 & \; \ 2.34e-4 & 7.16e-6 & 9.49e-2 & 8.68e-1 & & 
 0.346 &  0.330  \\
                            &  2008--2012 & \; \ 2.32e-4 & 2.48e-6 & 1.08e-1 & 8.83e-1 & & 
 {\bf 0.001} & {\bf 0.002}    \\
GJR-GARCH           & 1992--1996  & \; \ 4.18e-4 & 2.74e-6 & 2.68e-2 & 8.73e-1 & 1.0 & 
 0.588 & 0.578   \\
                            &  1998--2002 & $-$3.58e-4 & 6.66e-6 & 4.85e-2 & 8.71e-1 & 1.0 &
0.151 & 0.119  \\
                            &  2008--2012 & \; \ 1.77e-4 & 2.65e-6 & 4.33e-2 & 8.97e-1 & 1.0 &
{\bf 0.007}  & {\bf 0.008}   \\
\hline
\end{tabular} 
\end{center}
Note: (a) The LB test employs the first 10 autocorrelations of the squared residuals and 
the LM test is based on the order-10 autoregression of the squared residuals.
(b) Unlike the {\tt fGarch} implementation, the first 10 values 
in the residual series are excluded for the LB and LM tests to mitigate the boundary effect. 
(c) Bold-face font highlights the cases where $p$-values are less than or equal to 0.05.
}
\end{table}

In the remainder of this section, we consider the QFA-based 
spectral measures (\ref{KSmax})--(\ref{WLmean}) for goodness of fit testing and
discriminant testing.  Unless noted otherwise, we take $\Om$ in  (\ref{KSmax})--(\ref{WLmean}) to be the set of Fourier frequencies in $(0,\pi)$ 
and $\cA$ to be the set of grid-points $0.05,0.06,\dots,0.95$.

\subsection{Goodness of Fit Testing}

\begin{table}[t]
{\footnotesize
\begin{center} 
\caption{$p$-Vaules of QFA-Based Goodness of Fit Tests by Residual Approach} 
\label{tab:QFA}
\begin{tabular}{l|l|ccccc} \hline
 \multicolumn{1}{c|}{Model} & \multicolumn{1}{c|}{Series} &
$KS_{\max}$ & $WL_{\max}$ & $KS_{\rm mean}$ & $WL_{\rm mean}$  \\  \hline 
GARCH          &  1992--1996 & 0.380 &  0.360 & 0.105 &  0.134  
\\
                    &  1998--2002 & 0.149 &  0.472 & 0.067 &  0.085 
  \\
                    &  2008--2012 & {\bf 0.001}& 0.546 & {\bf 0.013} &  0.354
  \\
GJR-GARCH   & 1992--1996 & 0.572 & 0.457 &  0.184 &  0.124 
 \\
                    &  1998--2002 & 0.662 & 0.790 &  0.224 & 0.180 
 \\
                    &  2008--2012 & {\bf 0.003} &  0.757 &  {\bf 0.018}  & 0.413 
 \\
\hline
\end{tabular} 
\end{center}
Note. (a) Results are based on 1000 simulation runs of Gaussian white noise.
(b) Bold-face font here and thereafter highlights the estimated $p$-values of the QFA-based tests 
for which the upper limit of the one-sided 95\% confidence interval, given by $p+1.64 \sqrt{p(1-p)/1000}$,
is less than or equal to 0.05 (equivalent to $p \le 0.039$).
}
\end{table}

\begin{figure}[p]
\centering
\includegraphics[width=1.55in,angle=-90]{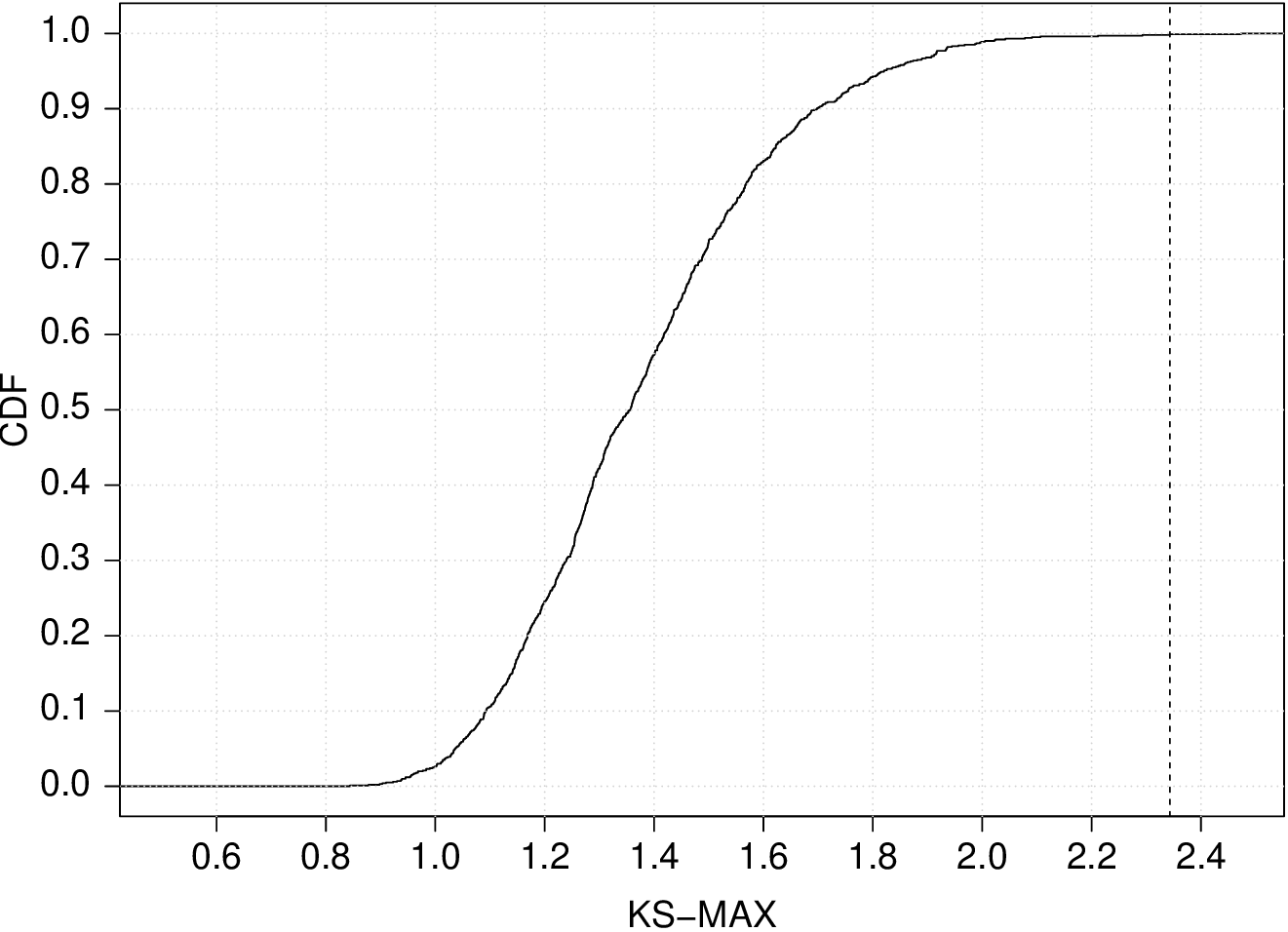}  \hspace{0.1in}
\includegraphics[width=1.55in,angle=-90]{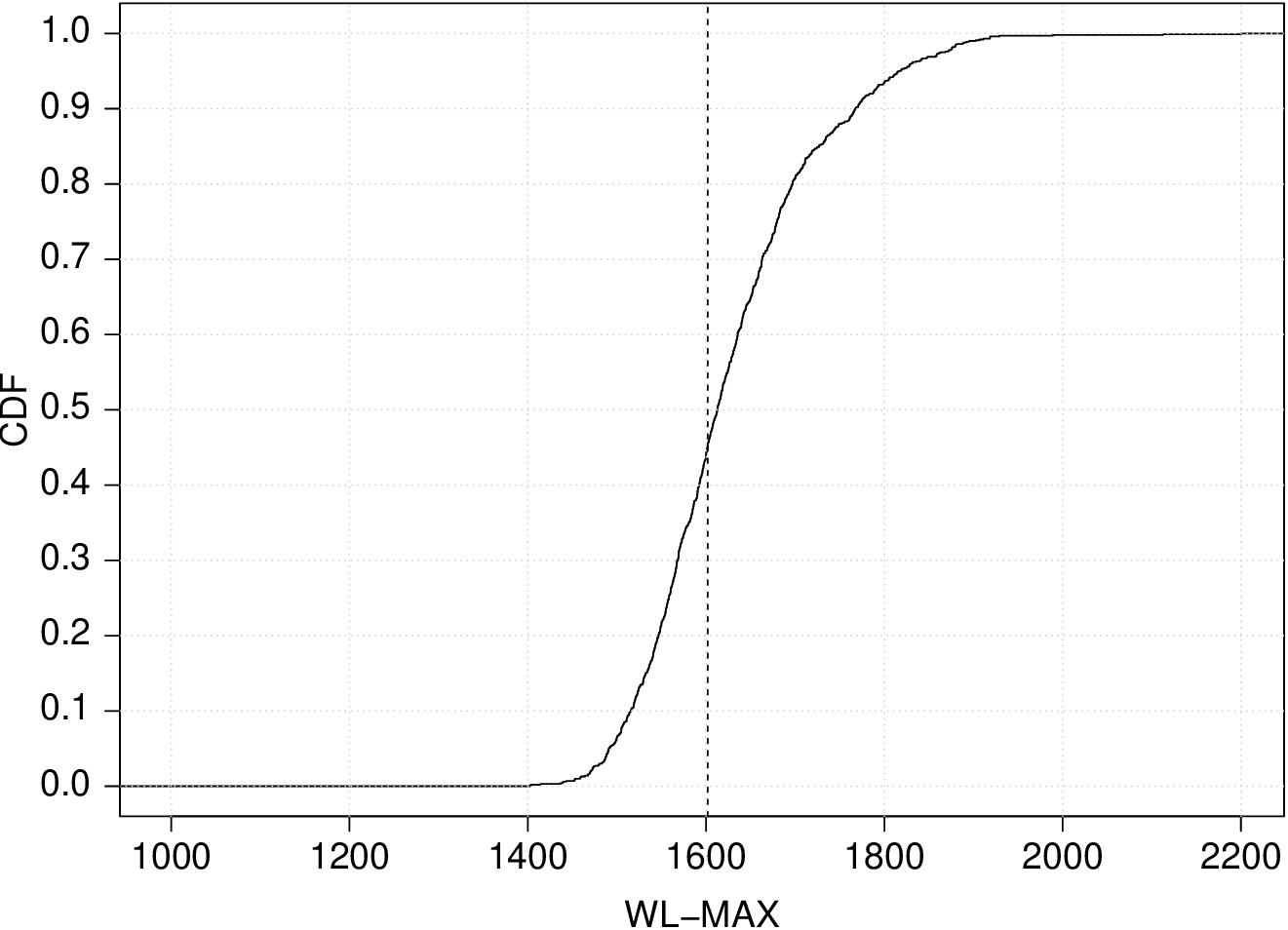} \\
\includegraphics[width=1.55in,angle=-90]{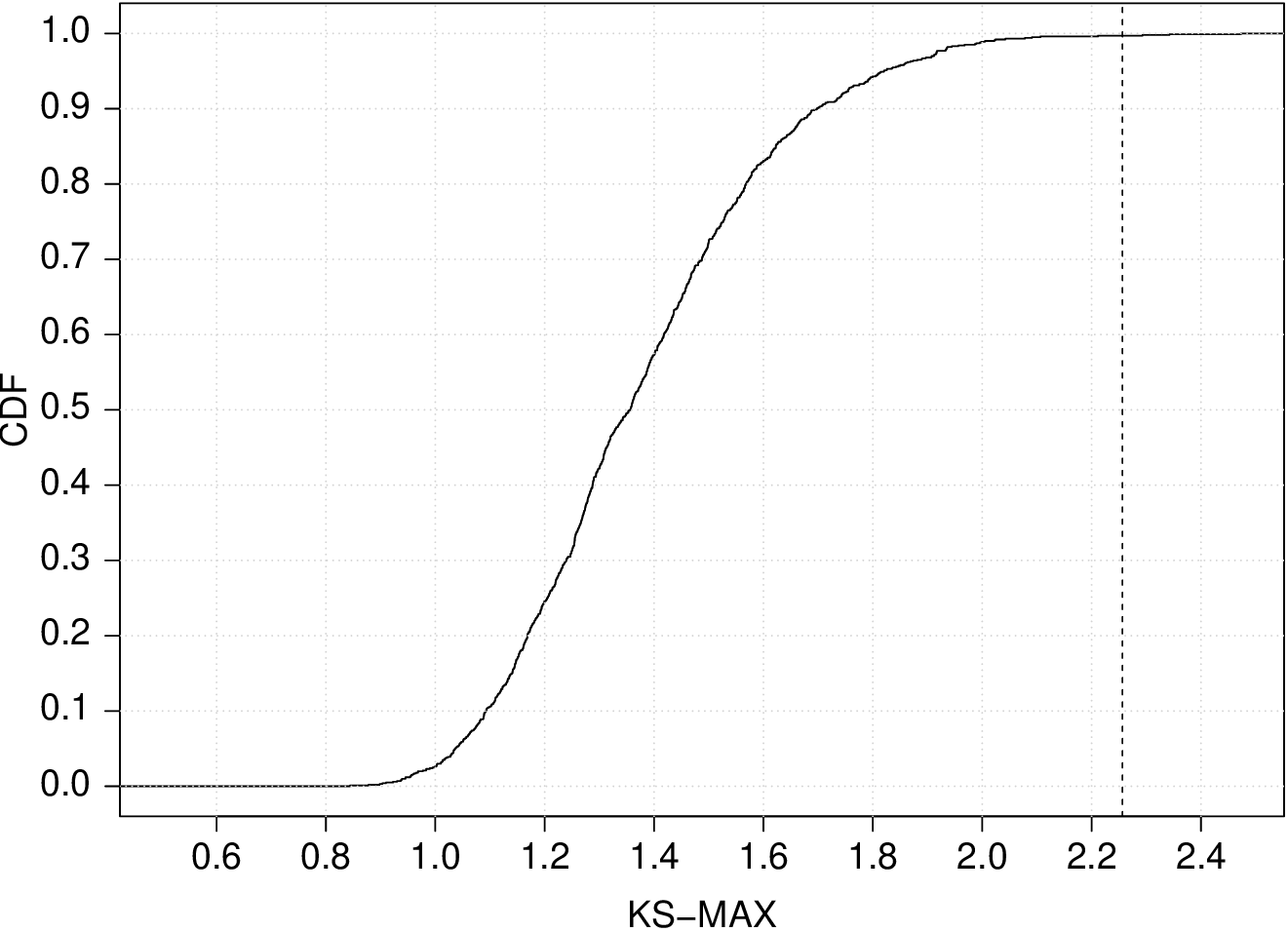}  \hspace{0.1in}
\includegraphics[width=1.55in,angle=-90]{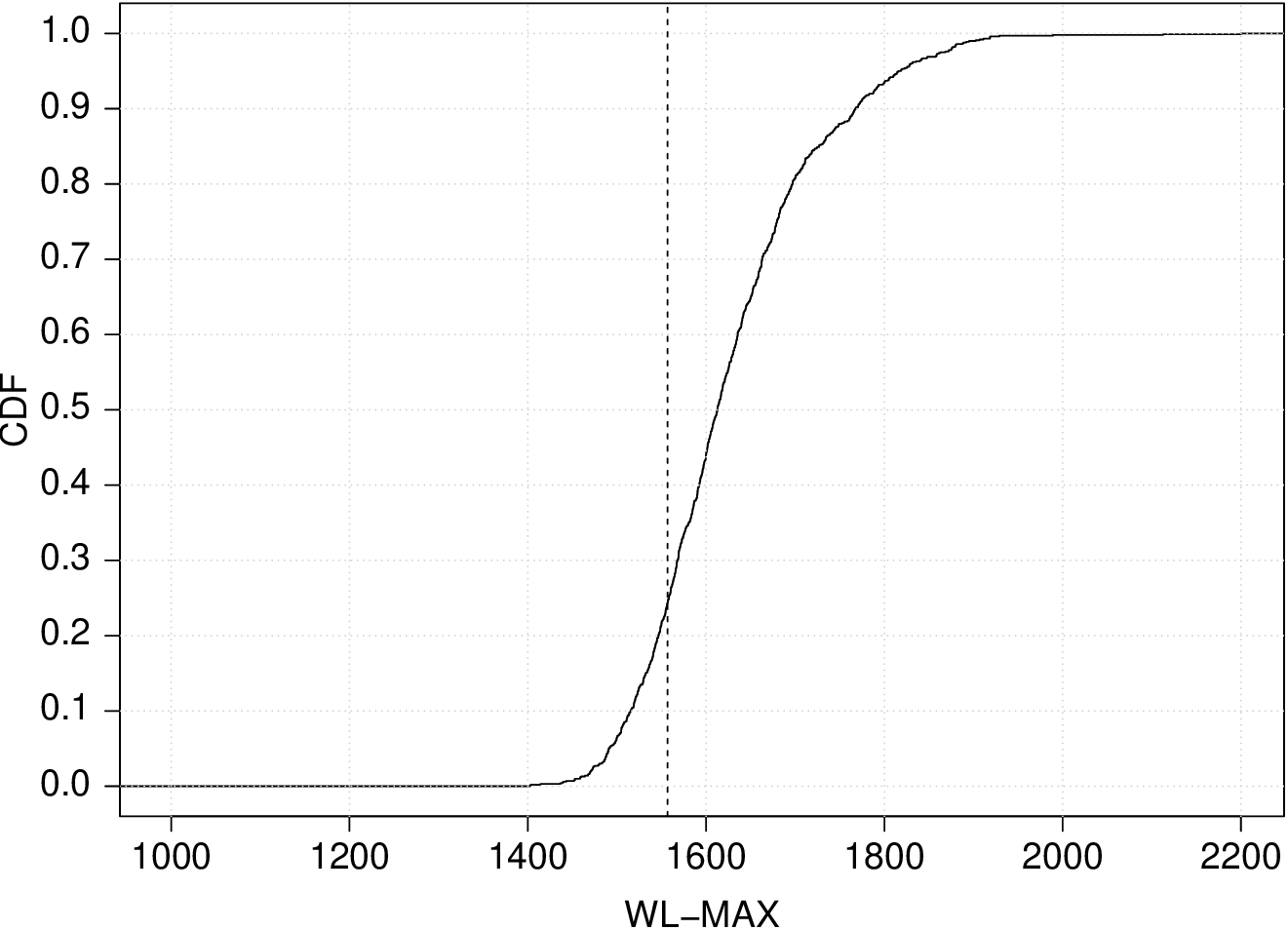} 
\caption{Simulated cumulative distribution functions of $KS_{\rm max}$ (left) and $WL_{\rm max}$ (right) for Gaussian white noise, superimposed with the corresponding metrics (vertical line) observed from the GARCH residuals (top row) and  the GJR-GARCH residuals (bottom row) of series 2008--2012.}
\label{fig:null}
\centering
\centerline{\footnotesize \hspace{0.0in}1992--1996\hspace{1.6in}1998--2002\hspace{1.6in}2008--2012\vspace{-0.3in}}
\includegraphics[width=1.55in,angle=-90]{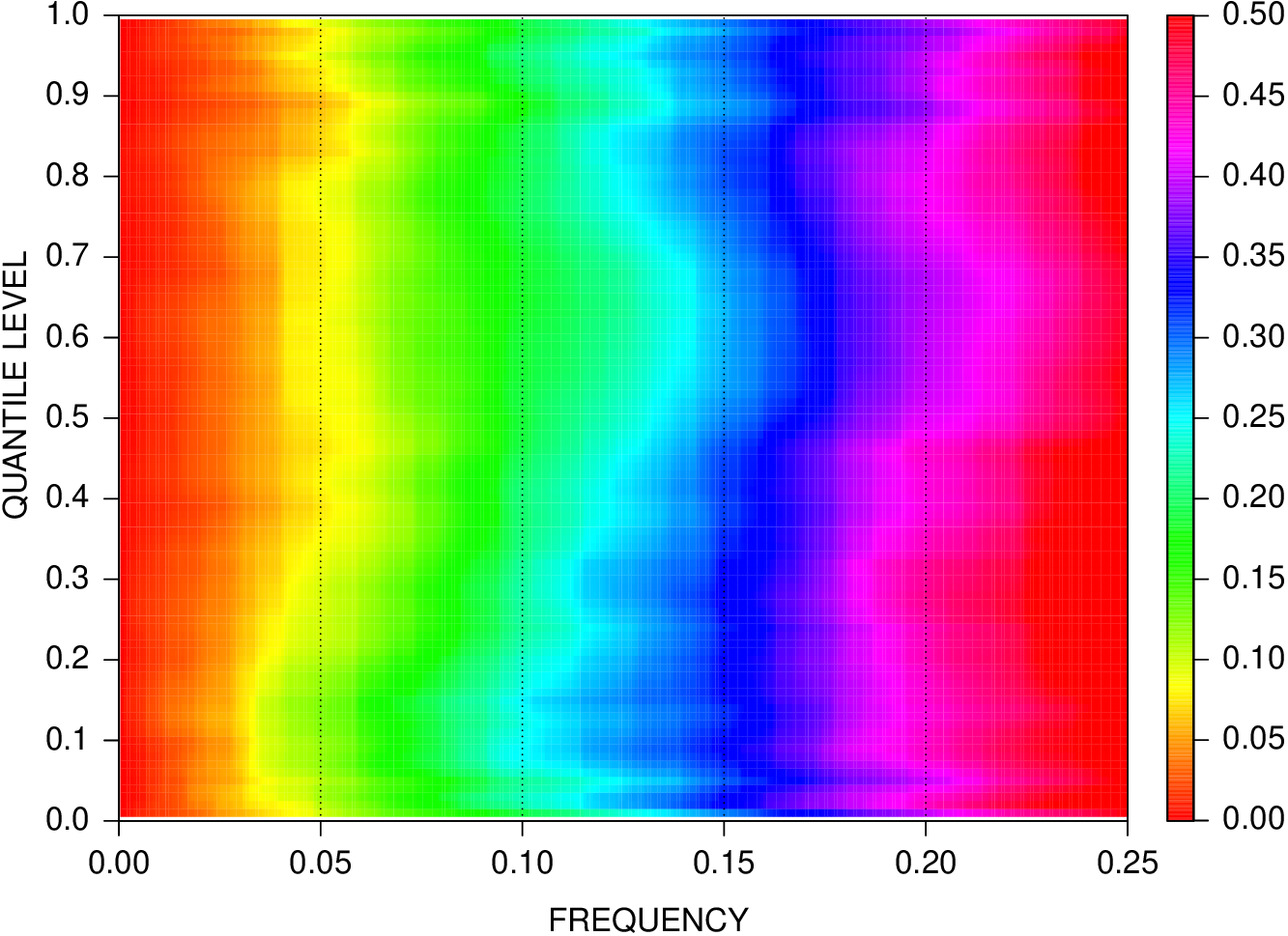} \hfill
\includegraphics[width=1.55in,angle=-90]{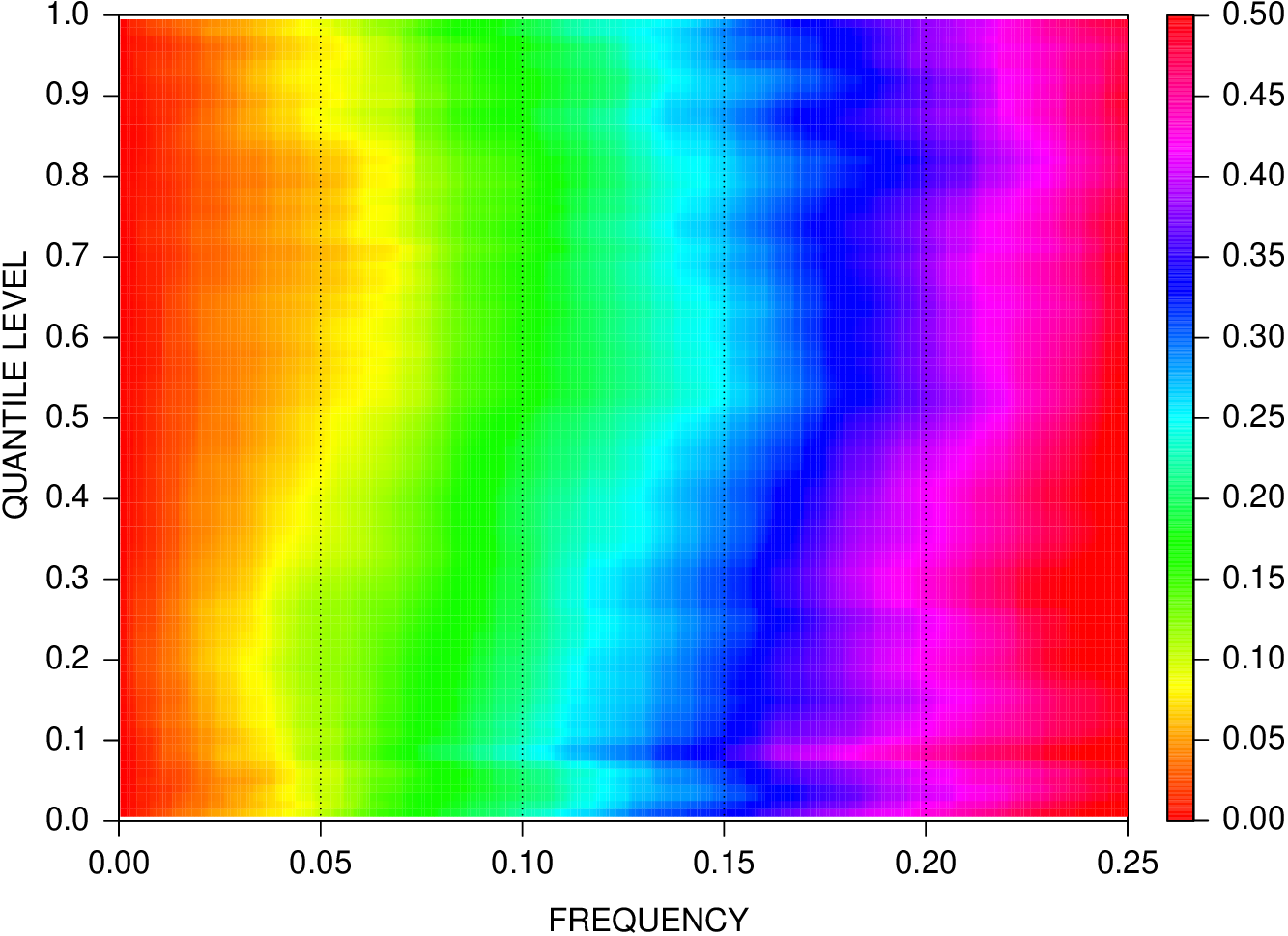} \hfill
\includegraphics[width=1.55in,angle=-90]{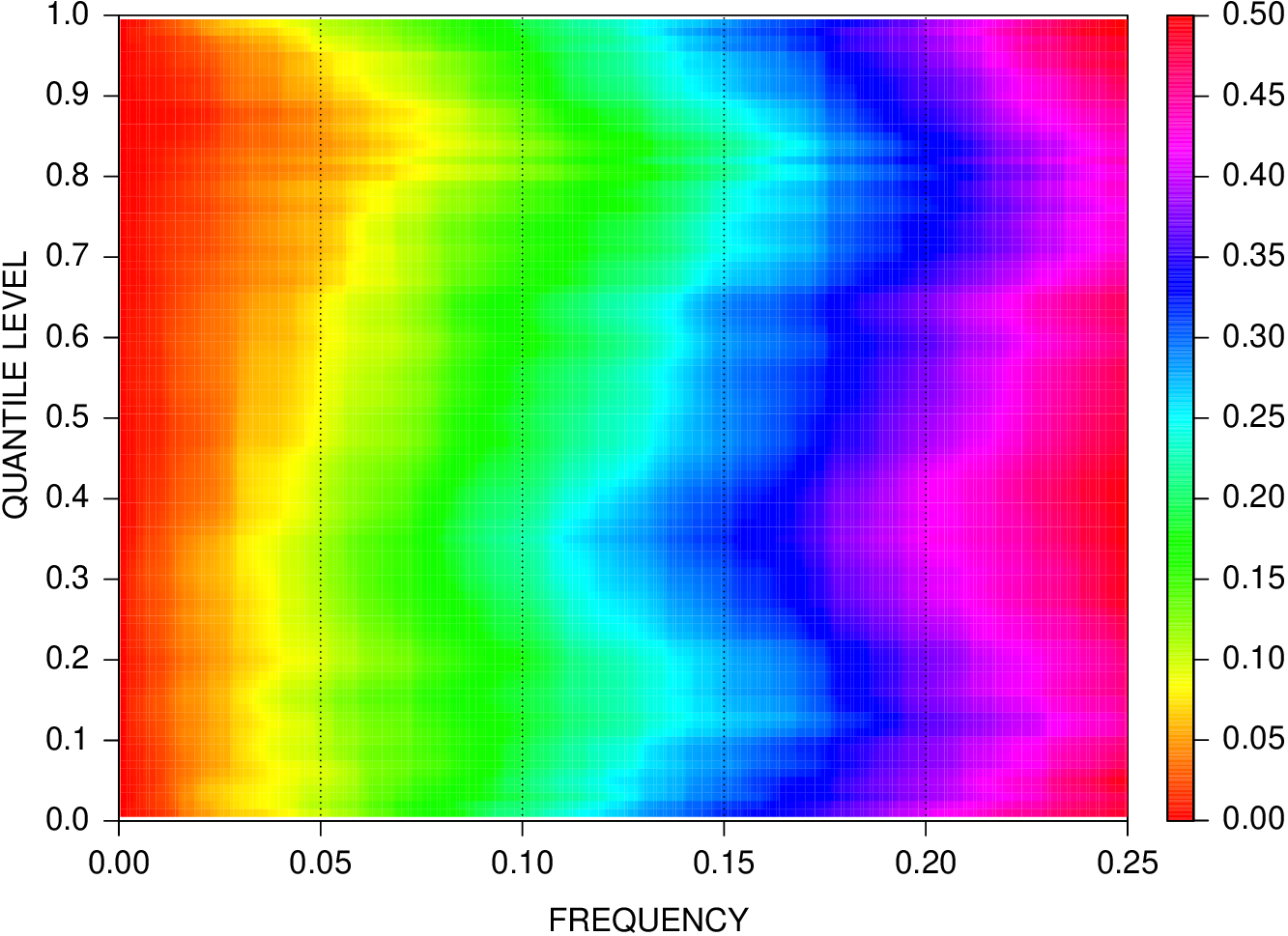} \\
\includegraphics[width=1.55in,angle=-90]{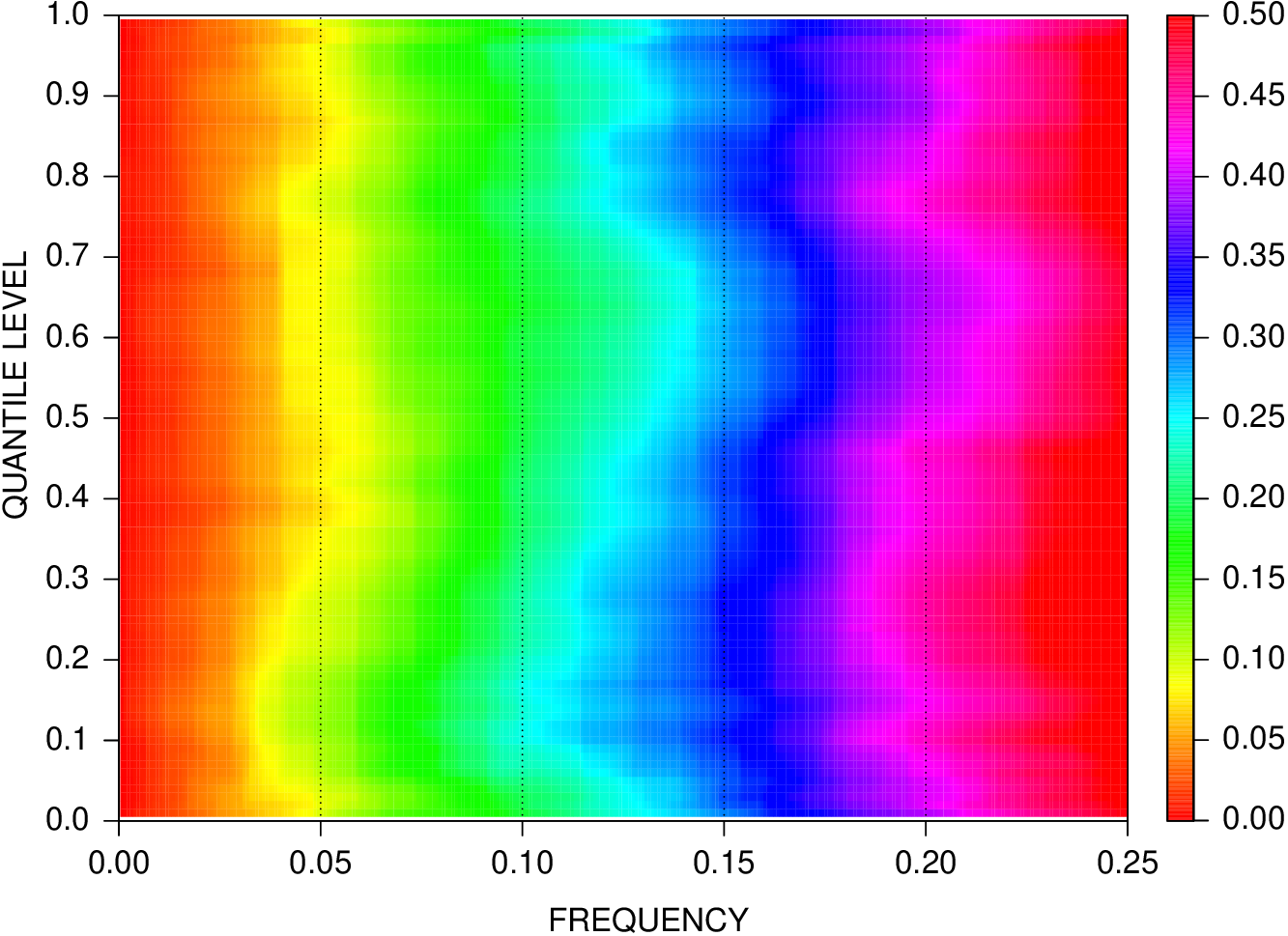} \hfill
\includegraphics[width=1.55in,angle=-90]{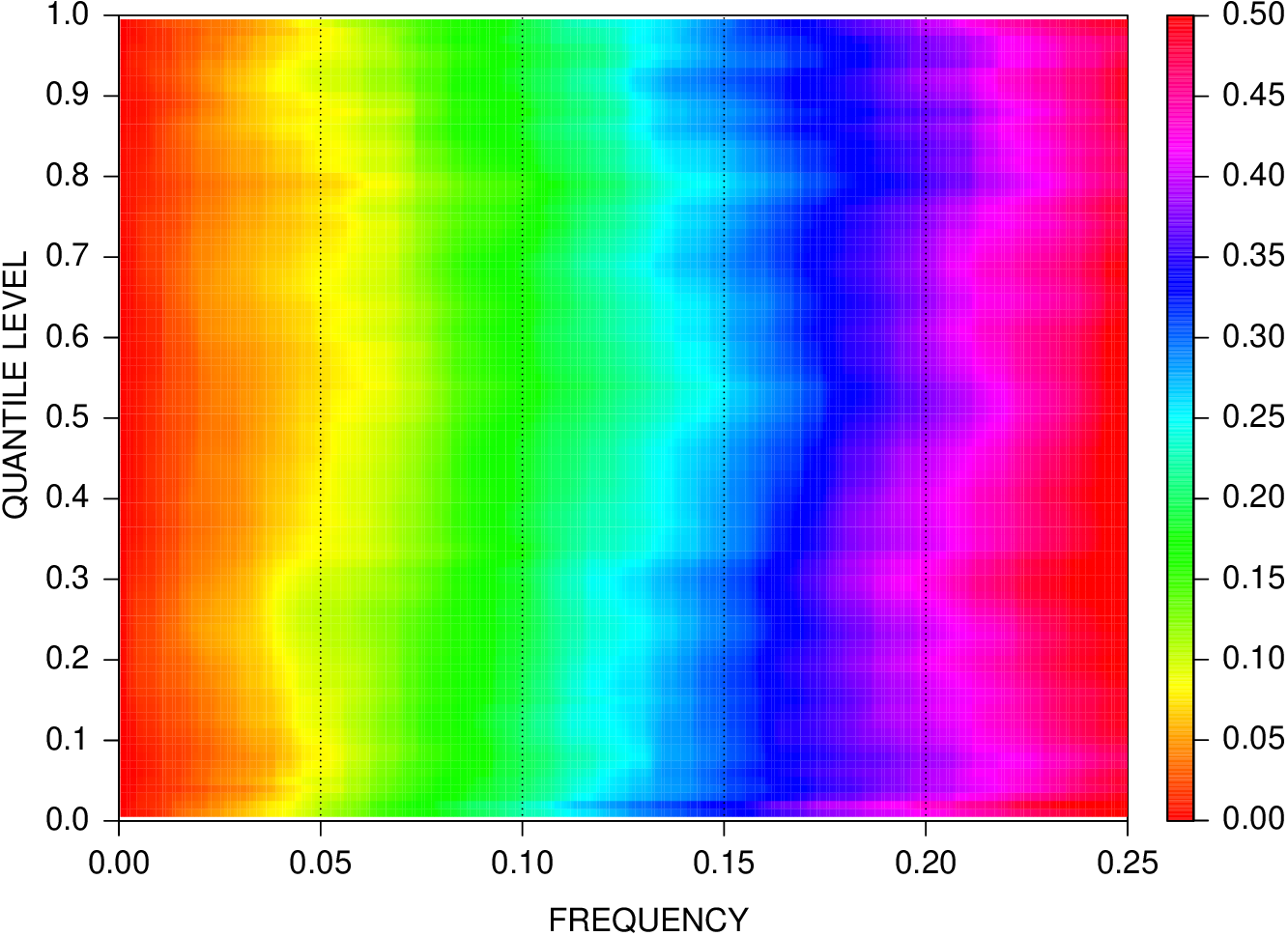} \hfill
\includegraphics[width=1.55in,angle=-90]{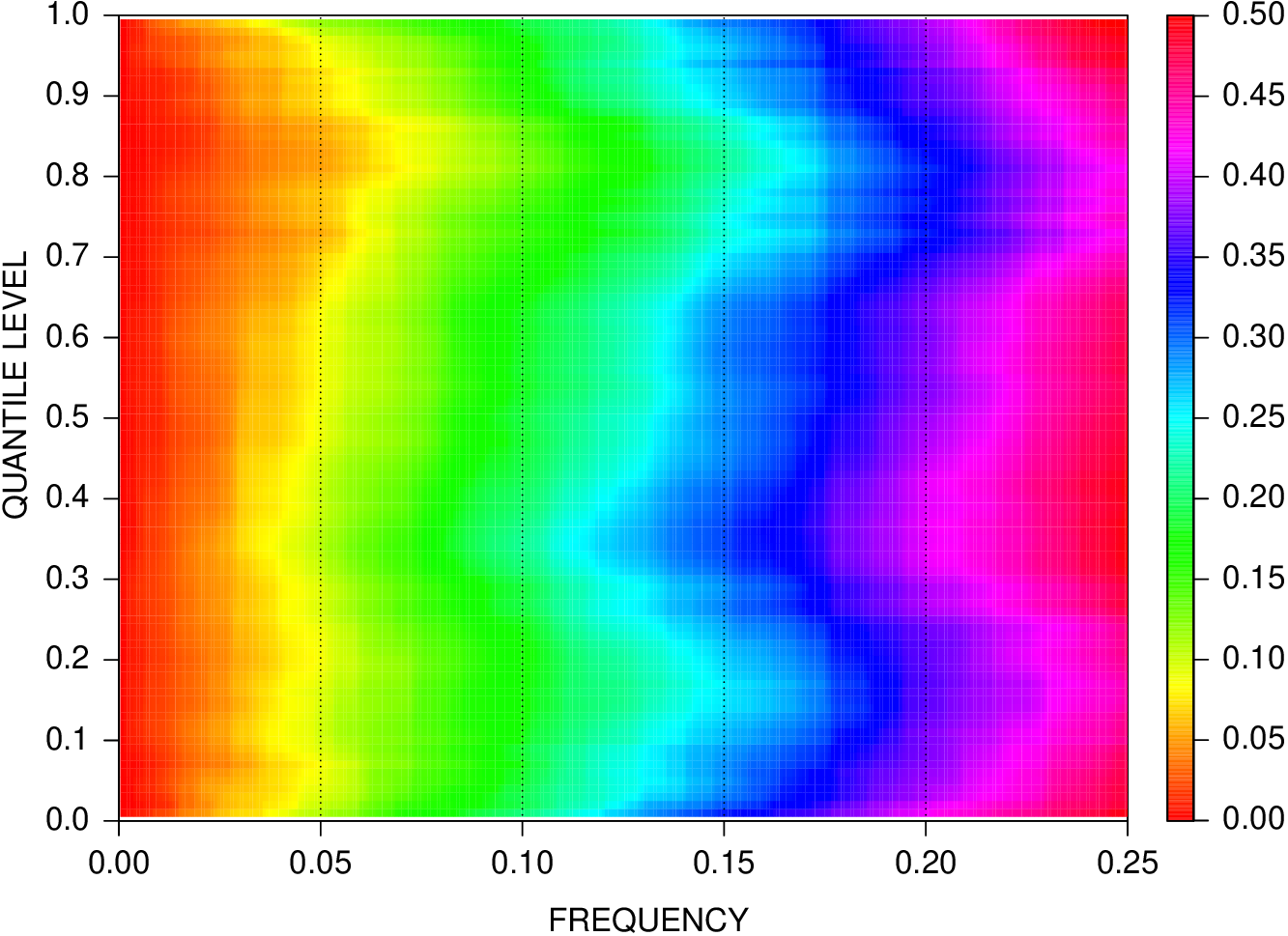} \\
\caption{Cumulative quantile periodograms of residuals from the GARCH model (top row) 
and the GJR-GARCH model (bottom row) for series 1992--1996 (left), 1998--2002 (middle), and 2008--2012 (right). Vertical lines indicate the expected behavior of cumulative quantile periodogram for white noise: uniform growth in frequency and constant in quantile level.}
\label{fig:residQFA}
\end{figure}

First consider the goodness of fit testing.
Table~\ref{tab:QFA} contains the results of the residual approach, where the $p$-values are calculated from simulated Gaussian white noise. As an example, Figure~\ref{fig:null} shows the cumulative distribution functions of $KS_{\rm max}$ and $WL_{\rm max}$ from the simulation
together with the corresponding metrics of the GARCH and GJR-GARCH residuals of series 2008--2012. 

Similarly to the standard tests in Table~\ref{tab:fit}, the QFA-based tests in Table~\ref{tab:QFA} suggest the possibility for a lack of fit in both models of series 2008--2012, as indicated by the small $p$-values of the KS metrics regardless of the aggregation method over the quantiles. The somewhat smallish $p$-values of $KS_{\rm mean}$ and $WL_{\rm mean}$ for the GARCH model of series 1998--2002 may also raise suspicions for a lack of fit.

Figure~\ref{fig:residQFA} facilitates a closer examination of the residuals by QFA. 
Visual inspection of the top panel confirms that certain deviations from white noise exist in the residuals of the GARCH models, especially for series 2008--2012, and the deviations occur at many quantile levels, which justifies the small $p$-values of $KS_{\rm mean}$ in Table~\ref{tab:QFA}. 
Comparing the top panel with the bottom one shows that the GJR-GARCH model apparently reduces 
the deviations in the residuals for series 1998--2002, which is manifested in a decrease of the spectral measures
and an increase of their $p$-values.

\begin{figure}[p]
\centering
\centerline{\footnotesize \hspace{0.0in}1992--1996\hspace{1.6in}1998--2002\hspace{1.6in}2008--2012\vspace{-0.3in}}
\includegraphics[width=1.55in,angle=-90]{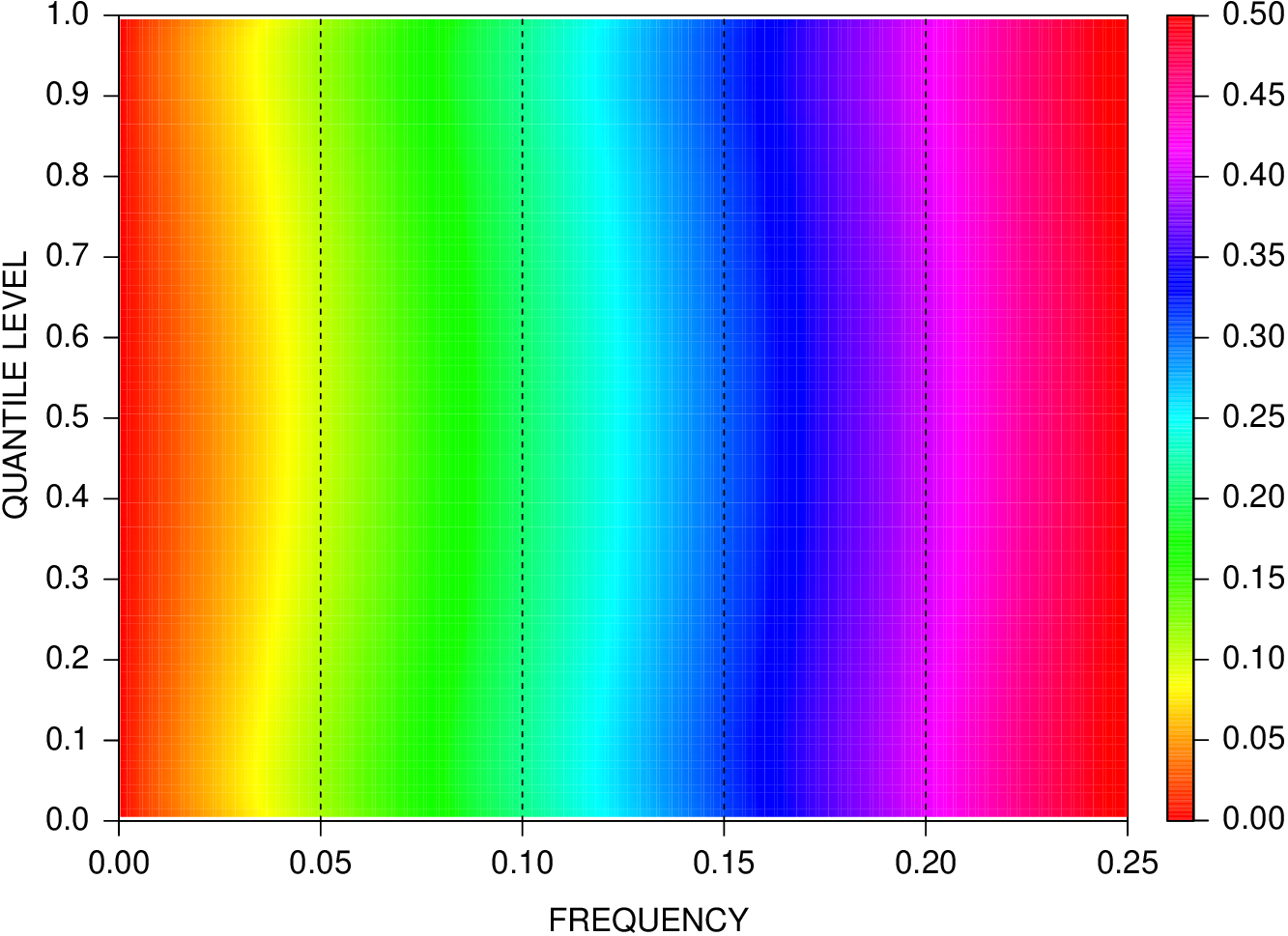} \hfill
\includegraphics[width=1.55in,angle=-90]{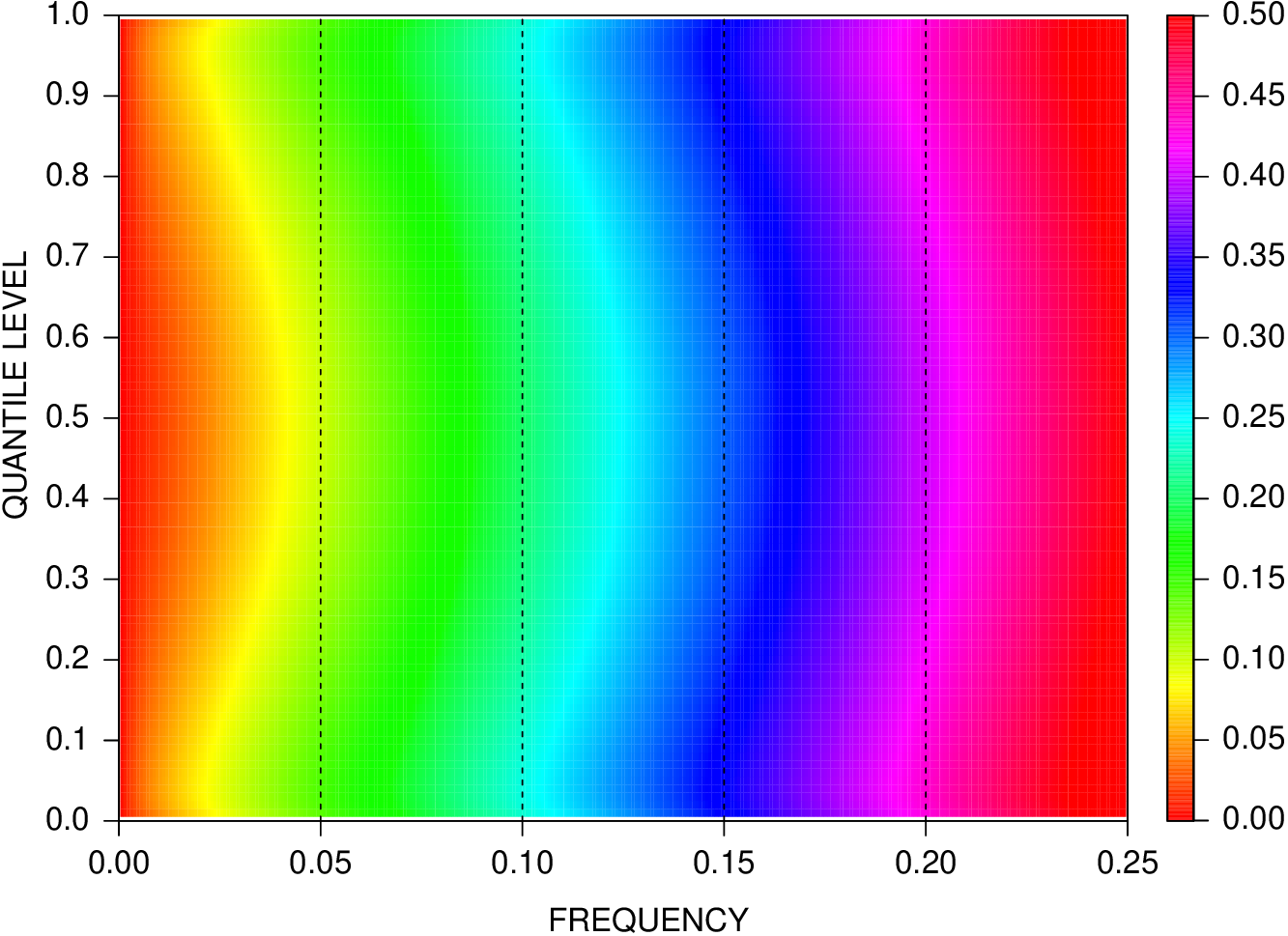} \hfill
\includegraphics[width=1.55in,angle=-90]{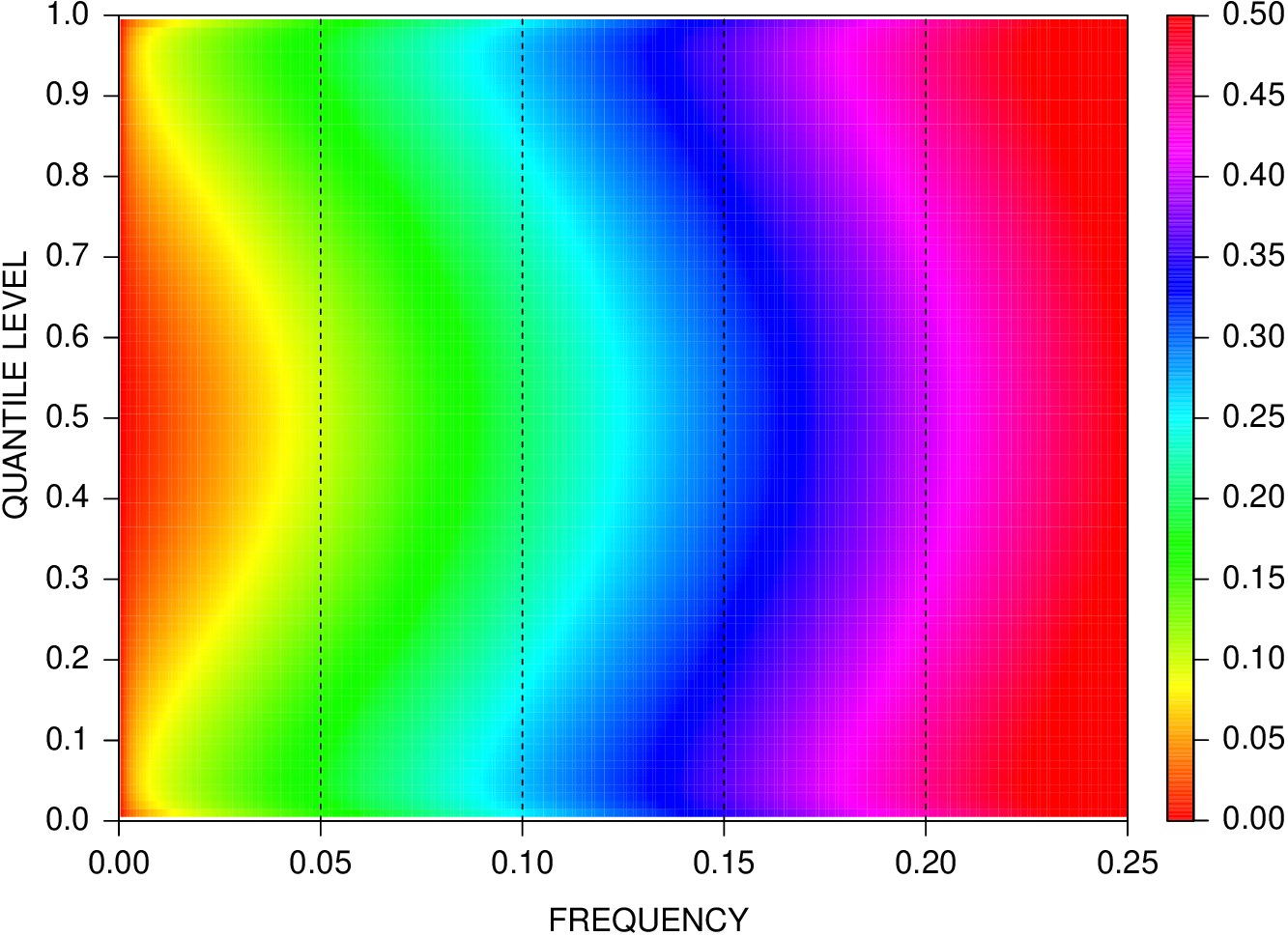} \\
\includegraphics[width=1.55in,angle=-90]{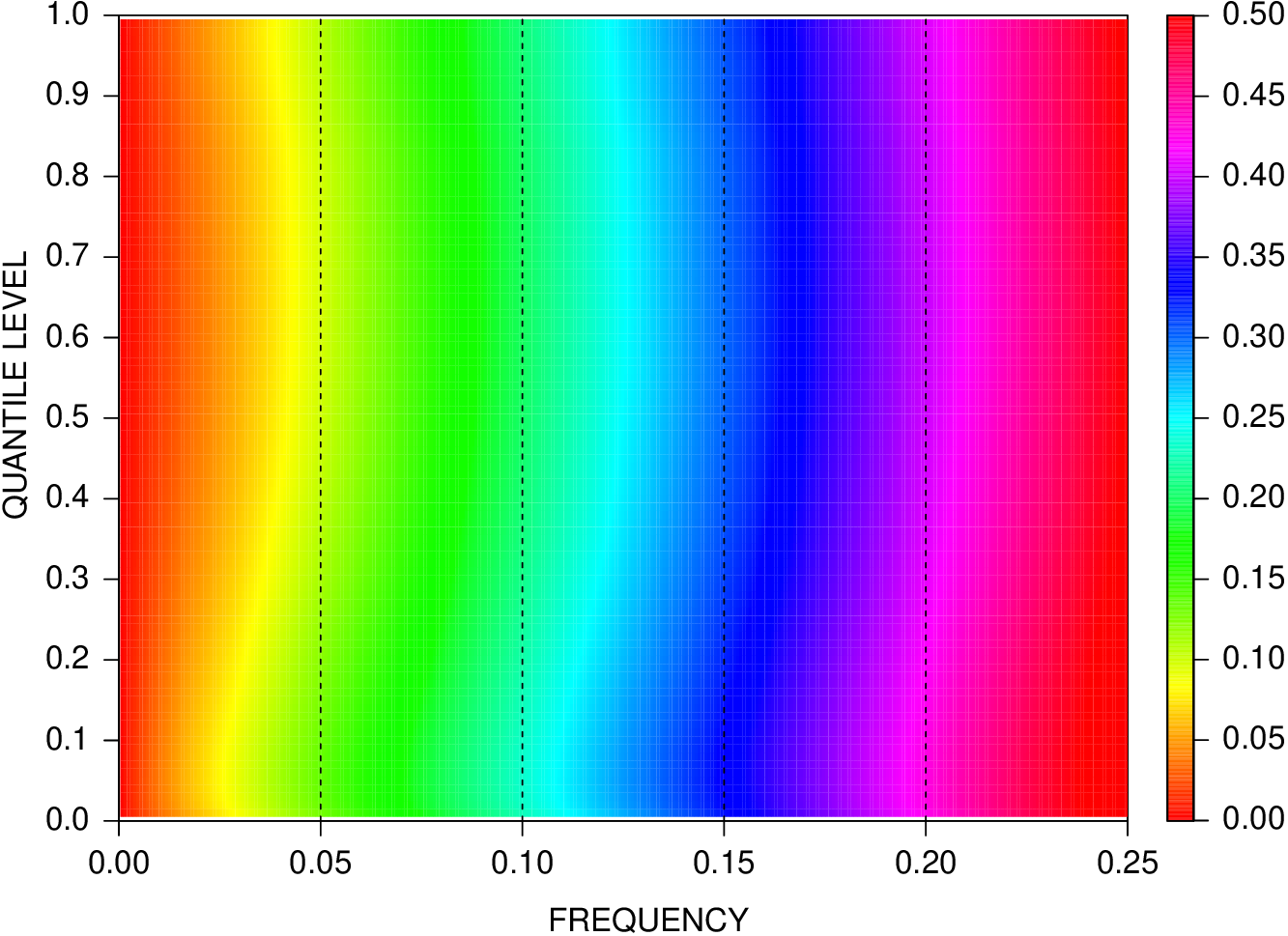} \hfill
\includegraphics[width=1.55in,angle=-90]{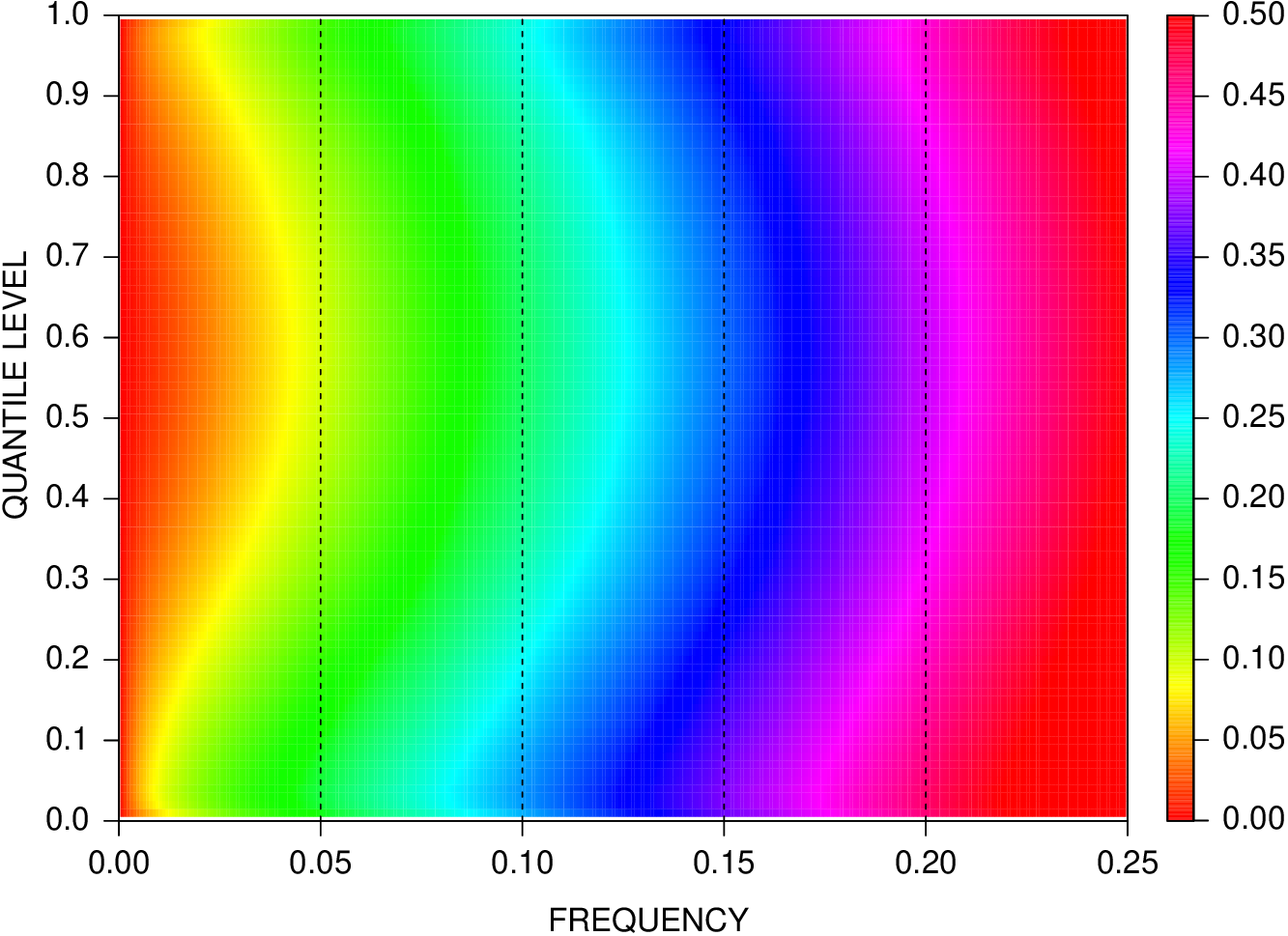} \hfill
\includegraphics[width=1.55in,angle=-90]{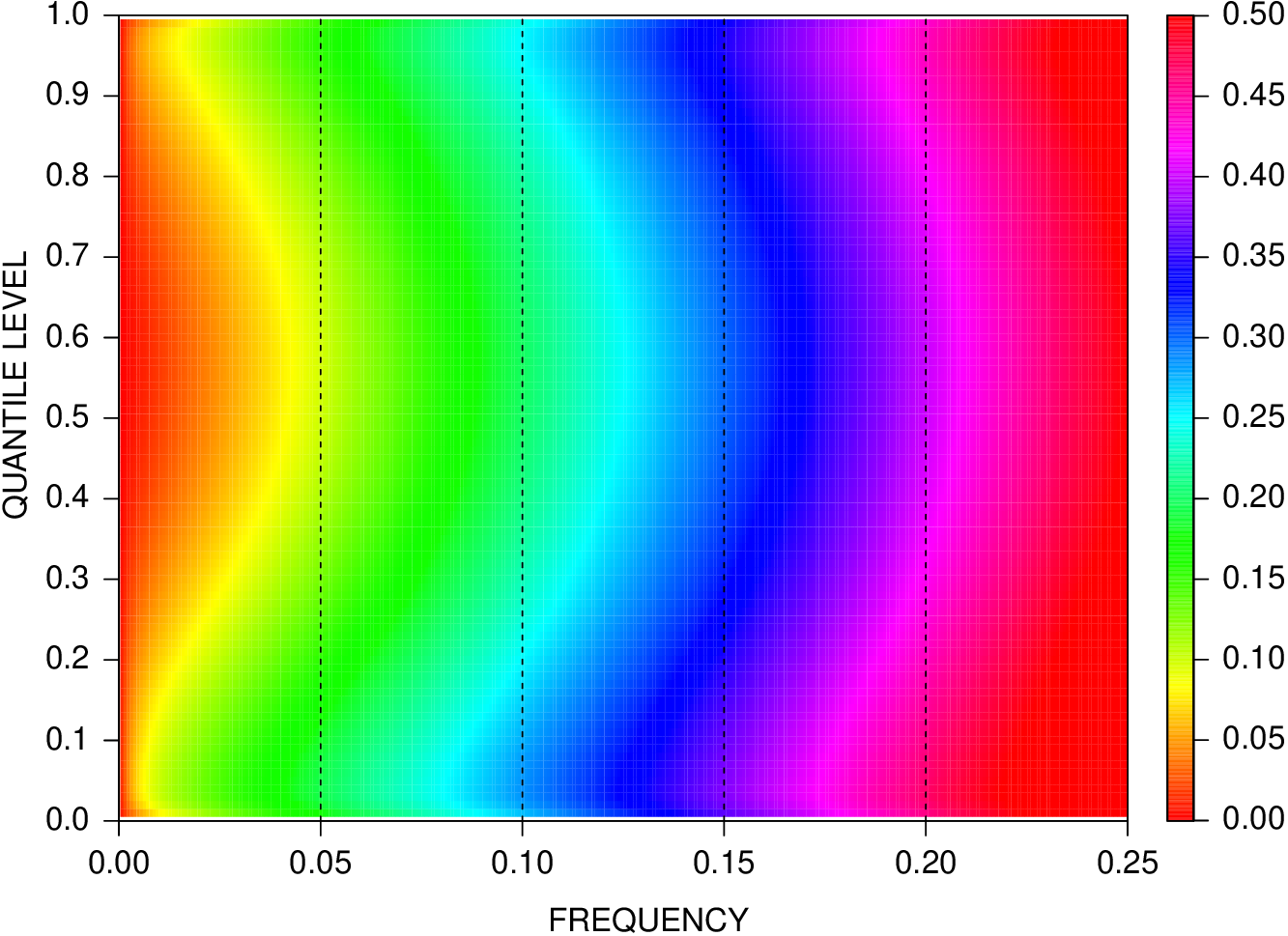} \\
\caption{Simulated cumulative quantile spectra of the GARCH model (top row) 
and the GJR-GARCH model (bottom row) for series 1992--1996 (left), 1998--2002 (middle), and 2008--2012 (right). }
\label{fig:garch}
\centering
\vspace{0.15in}
\centerline{\footnotesize \hspace{-1in}$\al = 0.1$ \hspace{1.75in}$\al=0.5$\hspace{1.8in}$\al=0.9$ \vspace{-0.55in}}
\includegraphics[width=1.55in,angle=-90]{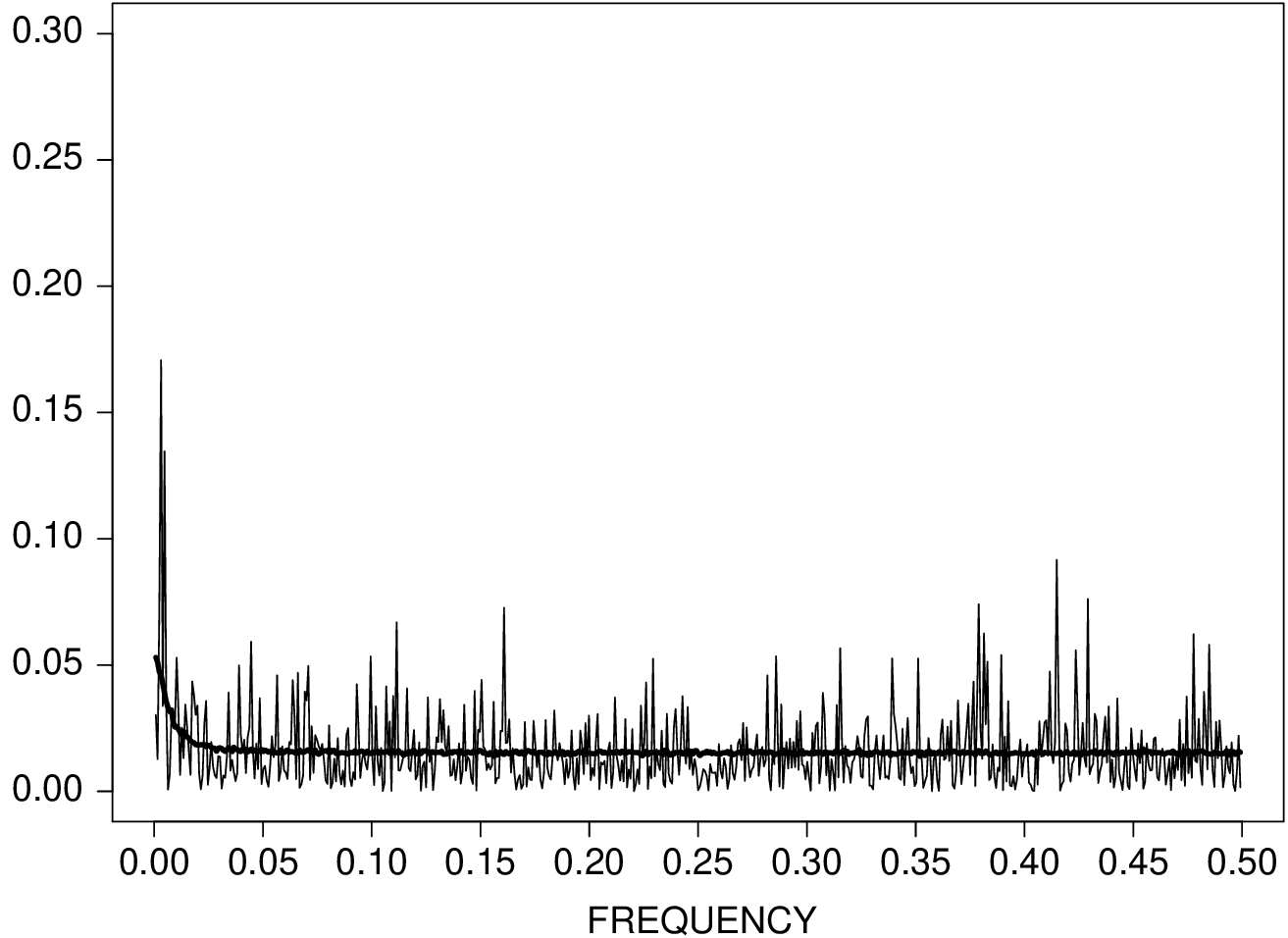} \hfill
\includegraphics[width=1.55in,angle=-90]{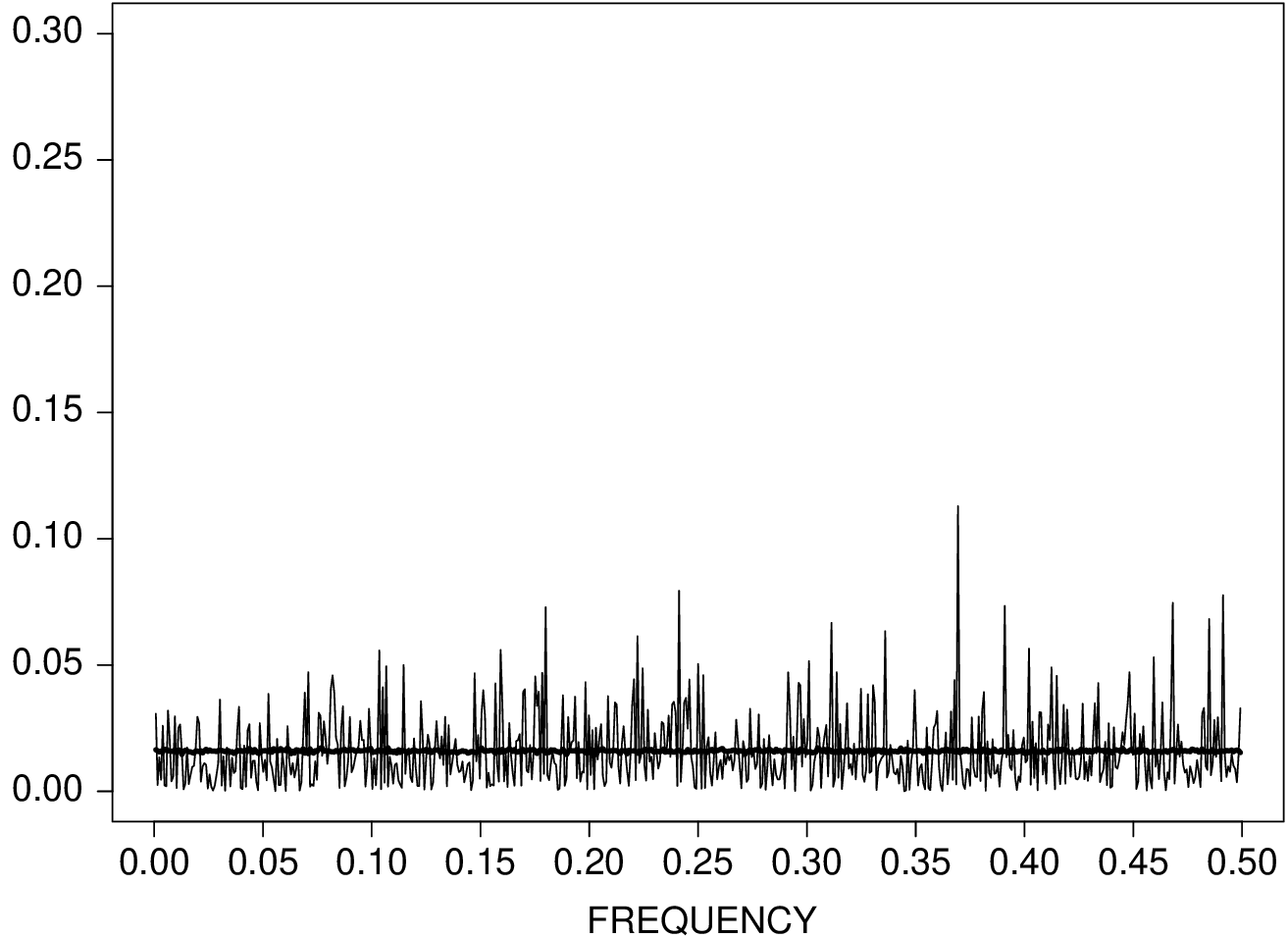} \hfill
\includegraphics[width=1.55in,angle=-90]{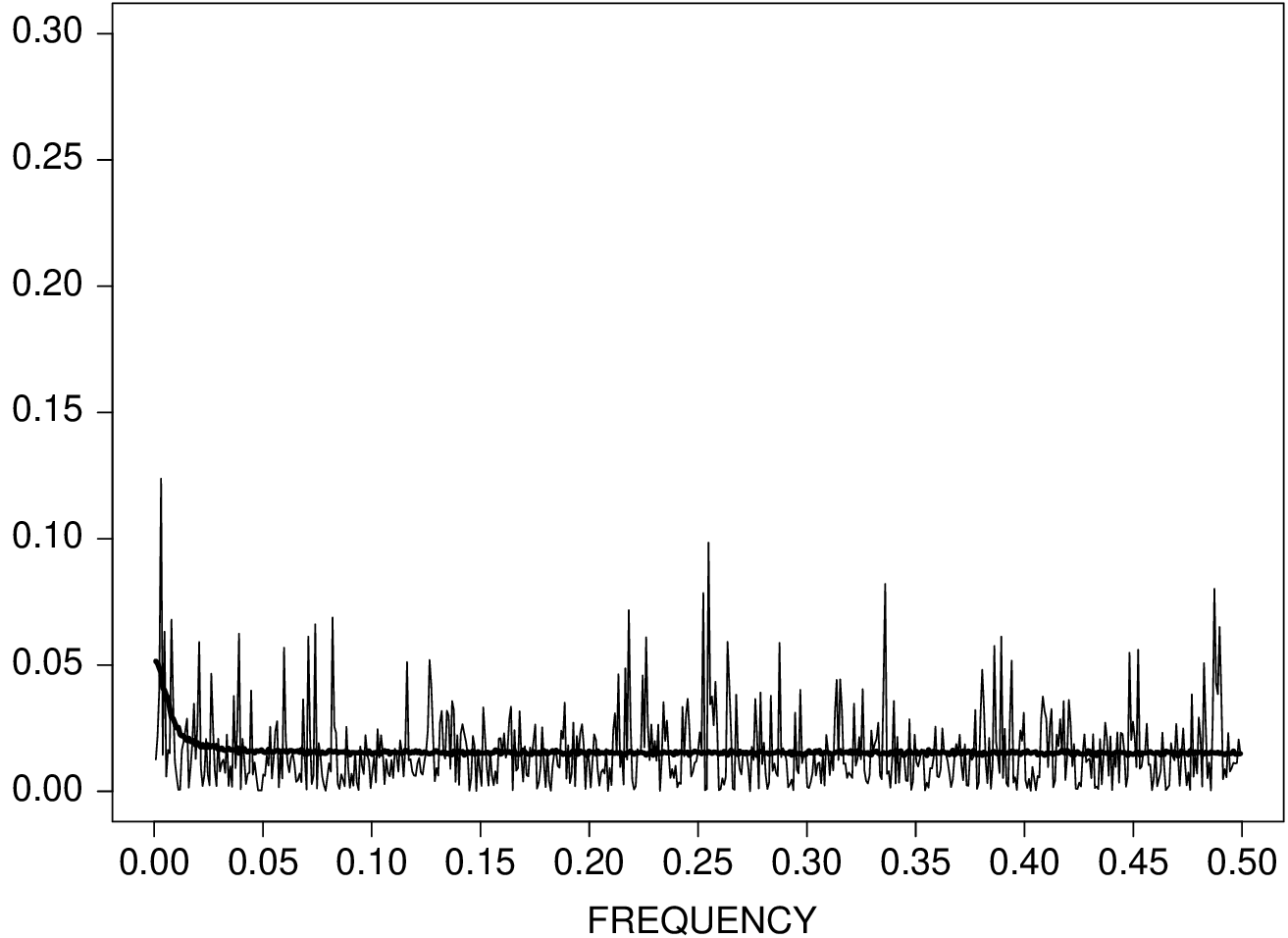} \\
\vspace{0.15in}
\centerline{\footnotesize \hspace{-1in}$\al = 0.1$ \hspace{1.75in}$\al=0.5$\hspace{1.8in}$\al=0.9$ \vspace{-0.55in}}
\includegraphics[width=1.55in,angle=-90]{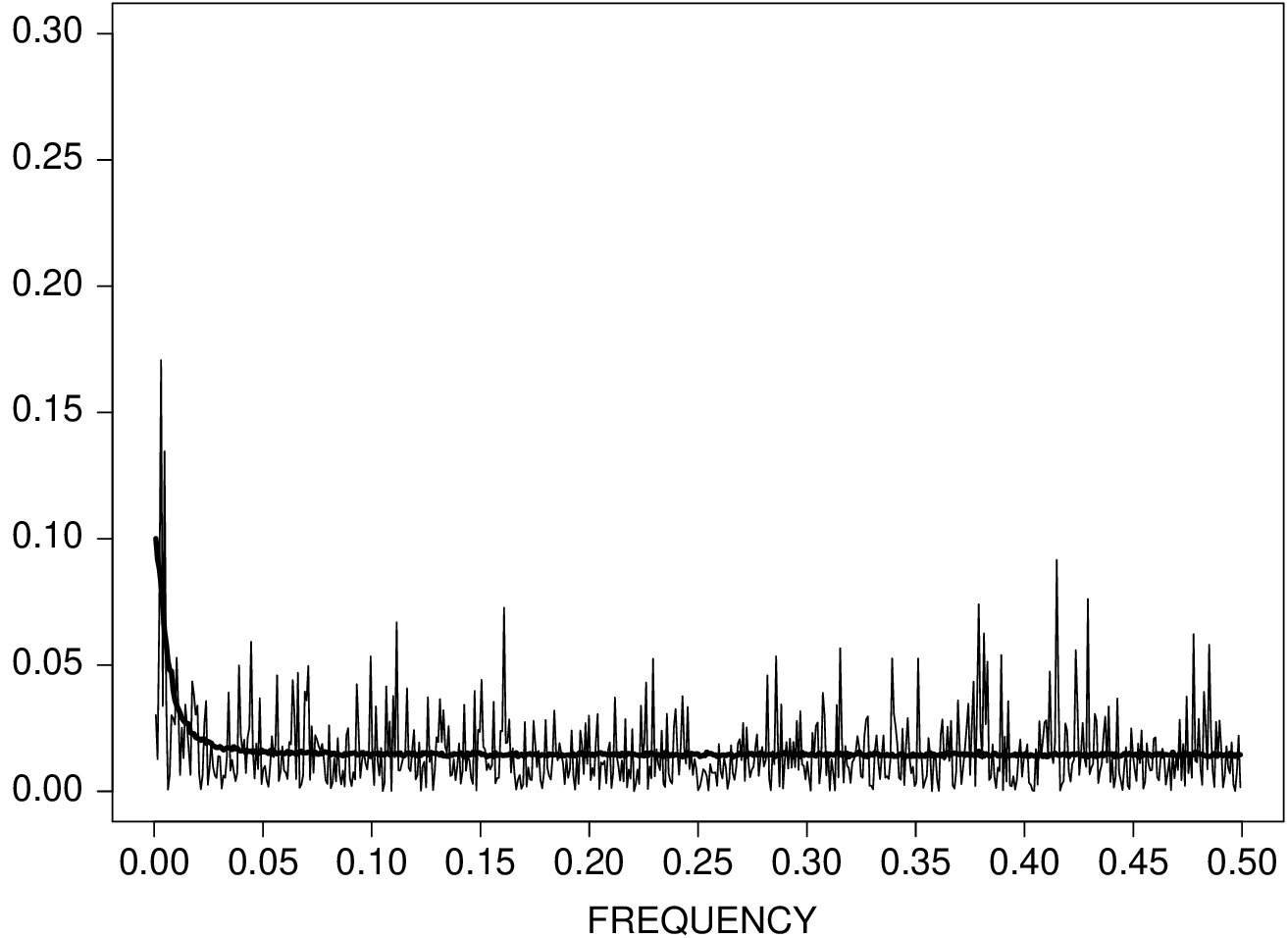} \hfill
\includegraphics[width=1.55in,angle=-90]{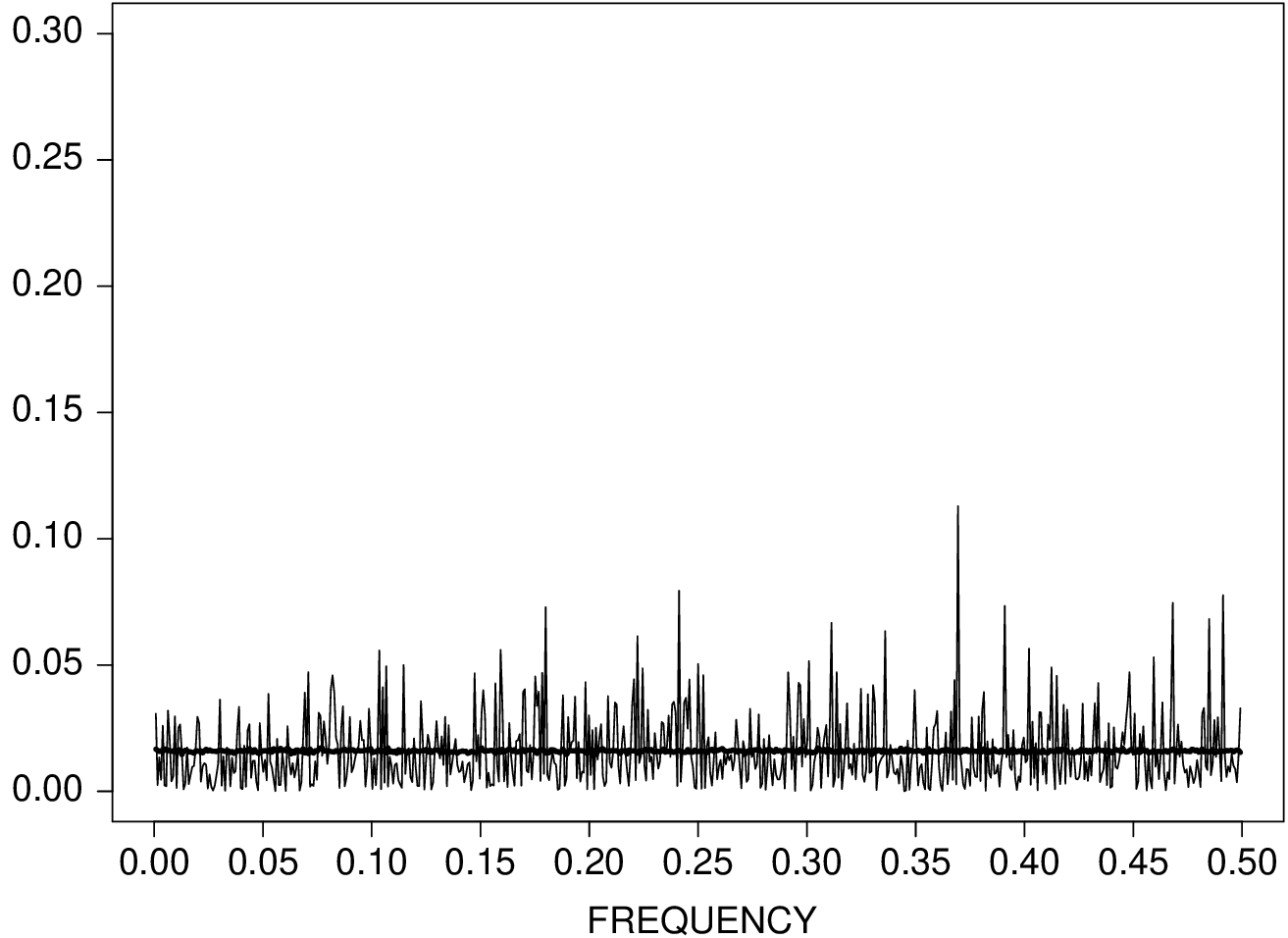} \hfill
\includegraphics[width=1.55in,angle=-90]{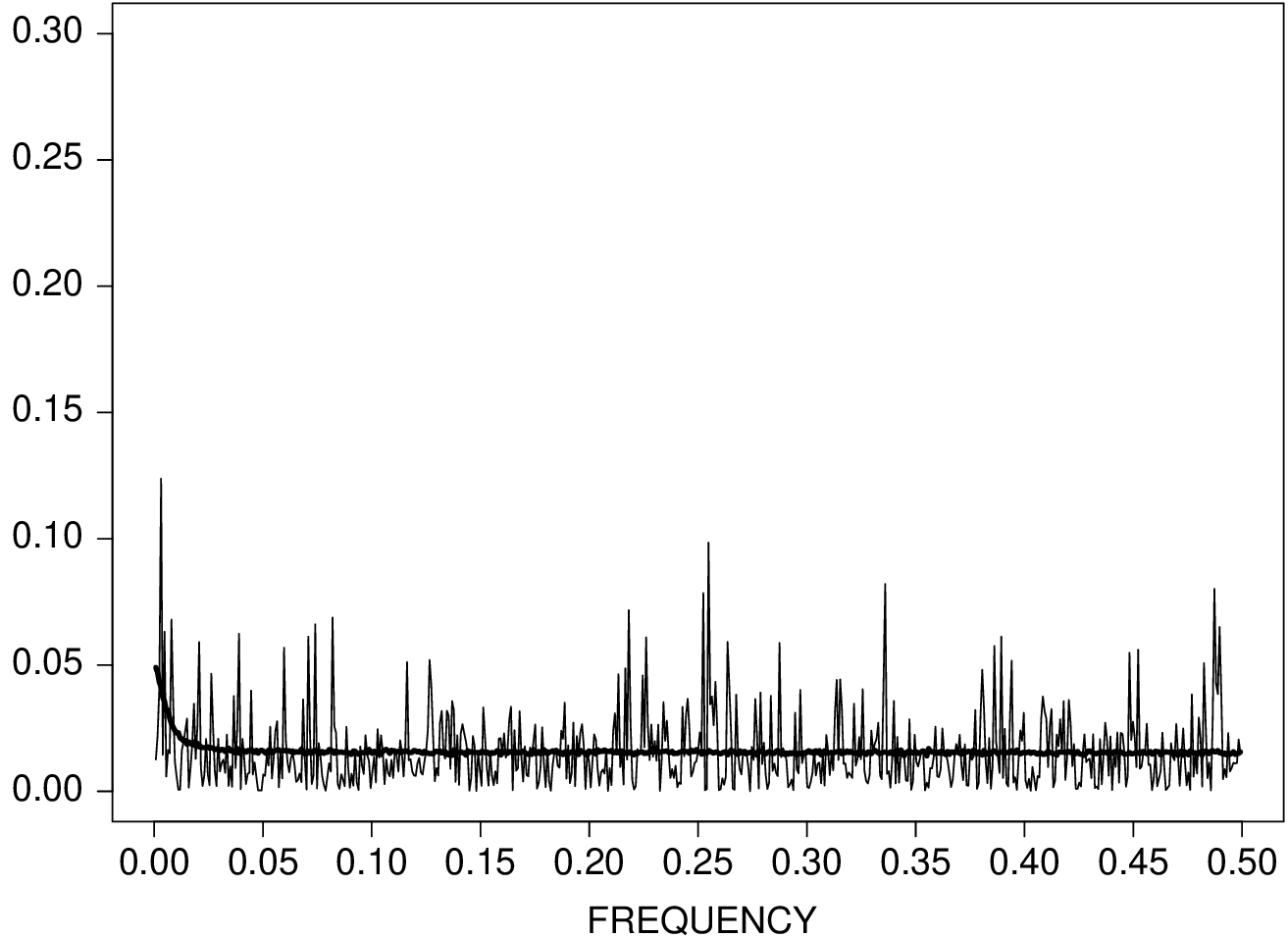} 
\caption{Quantile periodogrms of series 1998--2002 superimposed with the quantile spectra (thicker line) of the GARCH model (top row) and the GJR-GARCH model (bottom row) at level 0.1 (left), 0.5 (middle), and 0.9 (right).}
\label{fig:qperfit}
\end{figure}

In the direct approach, we first obtain the expected 
spectra $\tilde{q}(\om,\al)$ and $Q(\om,\al)$ in (\ref{KSmax})--(\ref{WLmean}) from the ensemble average of the quantile periodograms of  simulated time series according to (\ref{model}) using the $\tt garchSim$ function with estimated parameters and random samples from the standard Gaussian distribution. Figures~\ref{fig:garch} depict the resulting cumulative quantile spectra of the GARCH and GJR-GARCH models. As can be seen, the GARCH models produce symmetric quantile spectra with respect to the quantile level, whereas the GJR-GARCH models are able to generate some of the asymmetric characteristics observed in the quantile periodograms of the SPX series (Figure~\ref{fig:QFA}).

The improvement achieved by GJR-GARCH over GARCH for series 1998--2002 can be further appreciated by inspecting the graphs in Figure~\ref{fig:qperfit}, where the quantile periodograms of the series at levels 0.1, 0.5, and 0.9 are shown together with the quantile spectra of the GARCH and GJR-GARCH models. Both models  appear nearly identical at levels 0.5 and 0.9; but GJR-GARCH yields a stronger low-frequency peak than GARCH does at level 0.1, matching the observed pattern more closely.

Table~\ref{tab:QFA2} contains the $p$-values of the tests for all model-series combinations based on parametric bootstrapping. 
It shows a lack of fit in both models of series 2008--2012 according to $KS_{\rm mean}$, 
which is consistent with the residual tests  in Table~\ref{tab:QFA}. It also suggests
a possible lack of fit in both models of series 1992--1996 and 1998--2002 due to the WL metrics.

\begin{table}
{\footnotesize
\begin{center} 
\caption{$p$-Vaules of QFA-Based Goodness of Fit Tests by Direct Approach} 
\label{tab:QFA2}
\begin{tabular}{l|l|cccc} \hline
 \multicolumn{1}{c|}{Model} & \multicolumn{1}{c|}{Series} &
$KS_{\max}$ &  $WL_{\max}$ 
& $KS_{\rm mean}$ & $WL_{\rm mean}$  \\  \hline
GARCH                  &  1992--1996 & 0.604 &  0.075
                                                 & 0.390 & {\bf 0.022}  \\
                            &  1998--2002 & 0.432 & {\bf 0.027}  
                                                 & 0.252 & 0.360  \\
                            &  2008--2012 & 0.111 & 0.700 
                                                 & {\bf 0.031} & 0.846 \\
GJR-GARCH            & 1992--1996 & 0.916 & 0.059 
                                                 & 0.690 & {\bf 0.031}  \\
                            &  1998--2002 & 0.625 & {\bf 0.029} 
                                                 & 0.405 & 0.407 \\
                            &  2008--2012 & 0.209 & 0.670 
                                                 & {\bf 0.039} & 0.818 \\
\hline
\end{tabular} 
\end{center}
Note. Results are based on 1000 parametric bootstrap runs using fitted models driven by Gaussian white noise.
}
\end{table}

To further localize the lack of fit in the quantile-frequency framework, 
one can perform  drill-down diagnostic tests by steering the focus of QFA on different 
quantile and frequency regions through the choice of $\Om$ and $\cA$ in (\ref{KSmax})--({\ref{WLmean}). 
For example, motivated by the asymmetric dynamics of the time series across quantiles, 
one may zoom in on the middle quantiles, lower quantiles, or higher quantiles, respectively, to identify possible failures of a model in capturing the behavior of small-to-moderate returns, large negative returns, or large positive returns. Table~\ref{tab:QFA3} contains the results of such an analysis that help illuminate the results in Table~\ref{tab:QFA2} (which are based on aggregated metrics across all quantiles).

\begin{table}[p]
{\footnotesize
\begin{center} 
\caption{$p$-Vaules of QFA-Based Goodness of Fit Tests by Direct Approach: Drill Down} 
\label{tab:QFA3}
\begin{tabular}{l|l|cccc} \hline
& &  \multicolumn{4}{c}{Middle Quantiles} \\[-0.1in]
 \multicolumn{1}{c|}{Model} & \multicolumn{1}{c|}{Series} &
$KS_{\max}$ & $WL_{\max}$ 
& $KS_{\rm mean}$ & $WL_{\rm mean}$  \\  \hline
GARCH                  &  1992--1996 & 0.485 &  {\bf 0.023}  
                                                 & 0.304  & {\bf 0.001} \\
                            &  1998--2002 & 0.254 & 0.110 
                                                 & 0.209 & 0.780  \\
                            &  2008--2012 & {\bf 0.029} & 0.816
                                                 & 0.228 & 0.931 \\
GJR-GARCH            & 1992--1996 & 0.519 & {\bf 0.014}
                                                 & 0.417 & {\bf 0.001}  \\
                            &  1998--2002 & 0.418 & 0.110 
                                                 & 0.273 & 0.777 \\
                            &  2008--2012 & 0.099 & 0.783
                                                 & 0.289 & 0.935  \\
\end{tabular} 
\begin{tabular}{l|l|cccc} \hline
& &  \multicolumn{4}{c}{Lower Quantiles} \\[-0.1in]
 \multicolumn{1}{c|}{Model} & \multicolumn{1}{c|}{Series} &
$KS_{\max}$ & $WL_{\max}$
& $KS_{\rm mean}$ & $WL_{\rm mean}$  \\  \hline
GARCH                  &  1992--1996 & 0.318 &  0.575 
                                                 & 0.209 & 0.595 \\
                            &  1998--2002 & 0.920 & {\bf 0.012} 
                                                 & 0.701 & 0.235 \\
                            &  2008--2012 & 0.284 & 0.817 
                                                 & 0.180 & 0.415 \\
GJR-GARCH            & 1992--1996 & 0.913 & 0.605 
                                                 & 0.730 & 0.592 \\
                            &  1998--2002 & 0.887 & {\bf 0.008}
                                                 & 0.521 & 0.246 \\
                            &  2008--2012 & 0.257 & 0.839 
                                                 & 0.141 & 0.429 \\
\end{tabular} 
\begin{tabular}{l|l|cccc} \hline
& &  \multicolumn{4}{c}{Upper Quantiles} \\[-0.1in]
 \multicolumn{1}{c|}{Model} & \multicolumn{1}{c|}{Series} &
$KS_{\max}$ & $WL_{\max}$ 
& $KS_{\rm mean}$ & $WL_{\rm mean}$  \\  \hline
GARCH                  &  1992--1996 & 0.805 &  0.684  
                                                 & 0.797 & 0.516 \\
                            &  1998--2002 & 0.235 & 0.173 
                                                 & 0.159 & 0.092  \\
                            &  2008--2012 & 0.070 & 0.322 
                                                 & {\bf 0.011} & 0.665 \\
GJR-GARCH            & 1992--1996 & 0.573 & 0.659 
                                                 & 0.747 & 0.541  \\
                            &  1998--2002 & 0.254 & 0.197  
                                                 & 0.429 & 0.131 \\
                            &  2008--2012 & 0.068 & 0.317 
                                                 &  {\bf 0.010} &  0.675 \\
\hline
\end{tabular} 
\end{center}
Note. Middle quantiles include $0.31,0.32,\dots,0.69$; lower quantiles include $0.05,0.06,\dots,0.30$; 
and upper quantiles include $0.70,0.71,\dots,0.95$. 
See footnote of Table~\ref{tab:QFA2} for additional comments.
}
\end{table}

The drill-down diagnostics in Table~\ref{tab:QFA3} show that the middle quantile region is where a lack of fit may occur in the GARCH and GJR-GARCH models of series 1992--1996. In contrast, the lower quantile region exhibits some lack of fit symptoms in the models of series 1998--2002, and the upper quantile region suggests a possible lack of fit in the models of series 2008--2012.

\subsection{Discriminant Testing}

Table~\ref{tab:QFA4} contains the results of discriminant testing based on the method outlined in Section~3. In this application, we would like to determine whether the SPX return series from the three periods may be regarded as the product of a common underlying stochastic mechanism or different mechanisms. 
The method depends on how we define the common stochastic mechanism. The results in Table~\ref{tab:QFA4} are based on three GJR-GARCH models which are chosen over the GARCH models because of their greater flexibility. Each model is fit to one series and then checked against the remaining two series using the spectral measures (\ref{KSmax})--(\ref{WLmean}). 

Different models being used in this experiment means that the test results are not necessarily reciprocal. 
To be precise, we say ``series A differs from series B'' if the tests reject the null hypothesis that series A is generated from the model trained on series B. It is not the same as saying ``series B differs from series A'' because latter means the model trained on series A does not fit series B.

\begin{table}[t]
{\footnotesize
\begin{center} 
\caption{$p$-Vaules of QFA-Based Discriminant Tests Using GJR-GARCH Models}
\label{tab:QFA4}
\begin{tabular}{l|l|cccc} \hline
\multicolumn{1}{c|}{Series} & 
\multicolumn{1}{c|}{Model} & 
$KS_{\max}$ & $WL_{\max}$ &  $KS_{\rm mean}$ & $WL_{\rm mean}$ \\ \hline
1992--1996
&   1998--2002 & 0.191 & 0.055 & 0.182 & {\bf 0.035}  \\
&   2008--2012 & 0.169 & 0.075 & 0.092 & {\bf 0.035}  \\
1998--2002
&   1992--1996 & 0.794 & {\bf 0.027} &  0.526 & 0.299 \\
&   2008--2012 & 0.471 & 0.055 & 0.284 & 0.447  \\ 
2008--2012
&   1992--1996  & {\bf 0.001} &  0.421 & {\bf 0.009} & 0.370 \\ 
&   1998--2002  & 0.234 &  0.618 & {\bf 0.031} & 0.736  \\ 
\hline
\end{tabular} 
\end{center}
Note. Results are based on 1000 parametric bootstrap runs using fitted models driven by Gaussian white noise.
}
\end{table}

\begin{table}[p]
{\footnotesize
\begin{center} 
\caption{$p$-Vaules of QFA-Based Discriminant Tests  Using GJR-GARCH Models: Drill Down} 
\label{tab:QFA5}
\begin{tabular}{l|l|cccc} \hline
& &  \multicolumn{4}{c}{Middle Quantiles} \\[-0.1in]
 \multicolumn{1}{c|}{Series} & \multicolumn{1}{c|}{Model} &
$KS_{\max}$ & $WL_{\max}$ 
& $KS_{\rm mean}$ & $WL_{\rm mean}$ \\  \hline
1992--1996
&   1998--2002 & 0.420 & {\bf 0.016} &  0.275 & {\bf 0.000} \\
&   2008--2012 & 0.365 & {\bf 0.016} &  0.233 & {\bf 0.000} \\
1998--2002
&   1992--1996 & 0.383 &  0.132 &  0.308 &  0.763 \\
&   2008--2012 & 0.409 &  0.135 &  0.244 &  0.769  \\                                        
2008--2012
&   1992--1996  & 0.160 & 0.726 & 0.320 & 0.929  \\ 
&   1998--2002  & 0.127 & 0.762 & 0.304 & 0.935 \\ 
\end{tabular} 
\begin{tabular}{l|l|cccc} \hline
& &  \multicolumn{4}{c}{Lower Quantiles} \\[-0.1in]
 \multicolumn{1}{c|}{Series} & \multicolumn{1}{c|}{Model} &
$KS_{\max}$ & $WL_{\max}$ 
& $KS_{\rm mean}$ & $WL_{\rm mean}$  \\  \hline
1992--1996
&   1998--2002 & 0.159 & 0.545 & 0.110 & 0.486  \\
&   2008--2012 & 0.143 & 0.576 & 0.084 & 0.448 \\
1998--2002
&   1992--1996 & 0.882 & {\bf 0.017} & 0.609 & 0.209 \\
&   2008--2012 & 0.834 & {\bf 0.014} & 0.530 & 0.272 \\                                        
2008--2012
&   1992--1996 & {\bf 0.001} & 0.732 & {\bf 0.005} & 0.086 \\ 
&   1998--2002 & 0.254 & 0.821 & 0.112 & 0.339 \\ 
\end{tabular} 
\begin{tabular}{l|l|cccc} \hline
& &  \multicolumn{4}{c}{Upper Quantiles} \\[-0.1in]
 \multicolumn{1}{c|}{Series} & \multicolumn{1}{c|}{Model} &
$KS_{\max}$ & $WL_{\max}$ 
& $KS_{\rm mean}$ & $WL_{\rm mean}$  \\  \hline
1992--1996
&   1998--2002 & 0.337 &  0.681 & 0.492 &  0.482  \\
&   2008--2012 & 0.156 &  0.679 & 0.172 &  0.448  \\
1998--2002
&   1992--1996 & 0.805 & 0.164 & 0.705 & 0.106 \\
&   2008--2012 & 0.185 & 0.173 & 0.215 & 0.121 \\
2008--2012
&   1992--1996  & {\bf 0.001} & 0.137 & {\bf 0.004} &  0.165  \\ 
&   1998--2002  & 0.056 & 0.228 & {\bf 0.008} &  0.524  \\ 
\hline
\end{tabular} 
\end{center}
Note.  Middle quantiles include $0.31,0.32,\dots,0.69$; lower quantiles include $0.05,0.06,\dots,0.30$;
and upper quantiles include $0.70,0.71,\dots,0.95$. See footnote of Table~\ref{tab:QFA4} for additional comments.
}
\end{table}

The results in Table~\ref{tab:QFA4} suggest that (a) series 1992--1996 may differ from series 1998--2002 
and 2008--2012 and (b) series 2008--2012 may differ from series 1992--1996 and 1998--2002. The results do not suggest that series 1998--2002 may differ from series 1992--1996 based on $WL_{\rm max}$ because the same metric in Table~\ref{tab:QFA2} indicates a lack of fit in the model for series 1992--1996.

More can be said about these series based on the drill-down diagnostic tests in Table~\ref{tab:QFA5} together with the corresponding goodness of fit tests in Table~\ref{tab:QFA3}. 
These results further indicate that it is mainly at middle quantiles where series 1992--1998 differs from series 1998--2002 and  2008--2012. They also suggest that series 2008-2012 differs from series 1992--1996 
mainly at lower and upper quantiles rather than middle quantiles, whereas it differs from series 1998--2002 only at upper quantiles.

\subsection{Additional Experiments}

The supplementary material provides the results of some additional experiments.
A summary is given in the following, where tables and figures in the supplementary material
are referred to by using the prefix {\it cf.}

\subsubsection{Quantile Levels Near 0 And 1}

In the experiments so far, we have limited the quantile levels to the range between 0.05 
and 0.95. This range is extended to 0.01 and 0.99 in an experiment reported in the supplementary material.  
The results show that the inclusion of these quantile levels weakens 
some of the tests in terms of the number of highlighted cases. Specifically, 4 out of 6 cases in Table~\ref{tab:QFA2} are no longer highlighted for the goodness of fit testing ({\it cf.}\ Table~2), and so are 2 out of 6 cases in Table~\ref{tab:QFA4} for the discriminant testing ({\it cf.}\ Table~4).
The highlighted cases in Tables~\ref{tab:QFA}, \ref{tab:QFA3}, and \ref{tab:QFA5}  remain unchanged 
({\it cf.}\ Tables~1, 3, and 5).

\subsubsection{Non-Gaussian Distributions}

Besides Gaussian white noise, non-Gaussian distributions, especially those that are able to 
accommodate asymmetric and heavy-tailed behaviors, are also of interest in financial 
and econometric applications. The supplementary material provides the results of experiments based on the so-called skewed Student's $t$-distribution (Fern\'{a}ndez and Steel 1998), the probability 
density function of which takes the form of $(2/(\xi+1/\xi)) p_\nu(x/\xi)$ 
for $x \ge 0$ and $(2/(\xi+1/\xi)) p_\nu(\xi x)$ for $ x < 0$, where $p_\nu(\cdot)$ is the
probability density function of the ordinary $t$-distribution with $\nu > 0$ degrees of freedom 
($\nu$ is also known as the shape parameter)
and where $\xi> 0$ is the skewness parameter (left-skewed if $\xi < 1$ and right-skewed if $\xi > 1$).
These parameters are estimated together with the remaining parameters in the GARCH and GJR-GARCH models by the method of maximum likelihood using the {\tt garchFit} function in 
the {\tt fGarch} package (Wuertz, 2017) with the option {\tt cond.dist} = ``{\tt sstd}''. 

The estimated parameters as well as the corresponding results of goodness of fit testing and discriminant testing are provided in the supplementary material ({\it cf.}\ Tables~6--11). Also included are the plots of simulated cumulative quantile spectra of the corresponding GARCH and GJR-GARCH models ({\it cf.}\ Figure~1). 

For goodness of fit testing, the results of the residual approach
({\it cf.}\ Table 7) highlight an additional case in comparison with Table~\ref{tab:QFA}
regarding the GARCH model of series 1998--2002.
The results of the direct approach ({\it cf.}\ Table~8) no longer highlight the GJR-GARCH models of series 1998--2002 
and 2008--2012 as in Table~\ref{tab:QFA2}, but the drill-down analysis ({\it cf.}\ Table~9) 
still highlights them as in Table~\ref{tab:QFA3}. 

For discriminant testing, the results ({\it cf.}\ Table~10) still suggest that series 1992--1996 may differ from series 2008--2012 as the results in Table~\ref{tab:QFA4} but with the additional 
support by $KS_{\rm mean}$ and the corresponding drill-down analysis on the upper quantiles ({\it cf.}\ Table~11). The case of series 2008--2012 versus 
the model of series 1998--2002 is no longer highlighted when using all quantiles ({\it cf.}\ Table~10) but  remains highlighted as in Table~\ref{tab:QFA5} when focused on the upper quantiles  ({\it cf.}\ Table~11).

\subsubsection{Non-Crossing Quantile Regression}

It is well known that when computed at each quantile level independently the quantile regression solution may experience the so-called quantile crossing problem, i.e., the quantile regression fit at a lower quantile level may take a larger value than its counterpart at a higher quantile level (Koenker, 2005, p.\ 55). 
When quantile regression is used as a means of estimating quantile functions, the quantile crossing problem 
is a serious one and needs to be dealt with, for which a number of techniques have been 
proposed (e.g., He, 1997; Wu and Liu, 2009; Bondell, Reich and Wang, 2010; Lian, Meng and Fan, 2015).
For the purpose of spectral analysis, the quantile periodograms employ the regression coefficients 
and the minimized values of the objective function rather than the quantile curves. Therefore, 
except for extraneous variability, the quantile crossing problem does not have a detrimental 
effect on the quantile periodograms and QFA. 

One way to impose the non-crossing requirement on 
the quantile periodograms for a given set of quantile levels $\al_\ell$
($\ell=1,\dots,m$) is to consider the following joint optimization problem with 
non-crossing constraints for fixed frequency $\om \in (0,\pi)$ and the corresponding regressor $\bz_t(\om) := [1,\cos(\om t)$, $\sin(\om t)]^T$:
\eqn
\{ \hat{\bmbeta}_{n,\ell}(\om): \ell=1,\dots,m \} := 
 \arg\min_{\substack{\bmbeta_\ell \in \bbR^3, \, \ell=1,\dots,m \\ 
 \bz_t^T(\om) (\bmbeta_{\ell+1} - \bmbeta_\ell) \ge 0 \\ t=1,\dots,n; \, \ell=1,\dots,m-1}}
\sum_{\ell=1}^m \sum_{t=1}^n \rho_{\al_{\ell}}(X_t - \bz_t^T(\om) \bmbeta_\ell).
 \label{rqnc}
\eqqn
The solution of (\ref{rqnc}) guarantees the non-crossing requirement
\eqn
\bz_t^T(\om) \hat{\bmbeta}_{n,\ell+1}(\om) 
\ge \bz_t^T(\om) \hat{\bmbeta}_{n,\ell}(\om), \quad \mbox{for all $t=1,\dots,n$ and $\ell=1,\dots,m-1$}.
\label{nc}
\eqqn
Although it remains solvable as a linear program, the problem (\ref{rqnc}) is computationally demanding due to 
the large number of variables and constraints. We consider a suboptimal but more practical method similar to the stepwise technique proposed in Wu and Liu (2009).

The method begins with an unconstrained solution $\hat{\bmbeta}_{n,\ell}(\om) := [\hat{\lam}_n(\om,\al_\ell),
\hat{A}_n(\om,\al_\ell), \hat{B}_n(\om,\al_\ell)]^T$ at a middle quantile indexed by $\ell_0$, 
say $\al_{\ell_0} = 0.5$, and proceeds to obtain the remaining solutions as follows: For $\ell = \ell_0 + 1,\dots,m$, the solutions are given recursively by 
\eqn
 \hat{\bmbeta}_{n,\ell}(\om) := 
 \arg\min_{\substack{\bmbeta \in \bbR^3 \\ 
 \bz_t^T(\om) \bmbeta \ge a_t(\om) \\ t=1,\dots,n}}
 \sum_{t=1}^n \rho_{\al_{\ell}}(X_t - \bz_t^T(\om) \bmbeta),
 \label{rqnc1}
 \eqqn
where 
 $a_t(\om) :=\bz_t^T(\om) \hat{\bmbeta}_{n,\ell-1}(\om)$. For $\ell = \ell_0 - 1,\dots,1$, 
 the solutions are given recursively by
\eqn
 \hat{\bmbeta}_{n,\ell}(\om) := 
 \arg\min_{\substack{\bmbeta \in \bbR^3 \\ 
 \bz_t^T(\om) \bmbeta \le b_t(\om) \\ t=1,\dots,n}}
 \sum_{t=1}^n \rho_{\al_{\ell}}(X_t - \bz_t^T(\om) \bmbeta),
 \label{rqnc2}
 \eqqn
where $b_t(\om) :=\bz_t^T(\om) \hat{\bmbeta}_{n,\ell+1}(\om)$. The constrained quantile regression problems (\ref{rqnc1}) and (\ref{rqnc2}) can be solved numerically by using the {\tt rq.fit.fnc}  function
in the {\tt quantreg} package (Koenker, 2005). 

Note that our method defined by (\ref{rqnc1}) and (\ref{rqnc2}) differs from the method of Wu and Liu (2009) 
in the sense that we employ less stringent constraints. The latter is aimed at a more general problem of quantile regression  and therefore has to allow each entry in the regressor to take any values in a finite interval  independently (see also Bondell, Reich and Wang, 2010). Our method solves a
specific problem of trigonometric quantile regression and only requires the regressor to take the possible values of the trigonometric series, which is sufficient to guarantee the non-crossing condition (\ref{nc}).

The experiments using the non-crossing quantile regression method show that the resulting 
quantile periodograms of the SPX series and the resulting quantile spectra of the GARCH and GJR-GARCH models
({\it cf.}\ Figures~2 and 3) are visually indistinguishable from their counterparts in Figures~\ref{fig:QFA}
and \ref{fig:garch} obtained by unconstrained quantile regression. Furthermore, for diagnostic checks, 
the non-crossing quantile regression method does not produce meaningfully different results 
({\it cf.}\ Tables~12--16) from those in Tables~\ref{tab:QFA}--\ref{tab:QFA5}.

\section{Concluding Remarks}

In this paper, we introduced some spectral measures for diagnostic checks of time series models and for discriminant analysis of time series. The metrics are based on the recently proposed quantile periodograms. Derived from trigonometric quantile regression, the quantile periodograms describe the oscillatory behavior 
of time series around different quantiles and thereby provide a way to analyze the serial dependence 
of time series in the quantile-frequency domain.  

The motivating application of the SPX  daily log returns demonstrates that the proposed quantile-frequency analysis (QFA) method offers a richer view of serial dependence than the traditional autocorrelation function and periodogram do by enabling the spectral measures to identify potential lack of fit of time series models in different quantile-frequency regions. This capability may be leveraged to develop better models 
and help better understand the behavior of regime changes in financial markets.

For future research, we would like to investigate the asymptotic distributions of the QFA-based spectral measures for white noise as well as for suitably broad classes of random processes.

\section*{References}

{\footnotesize

\begin{description} 

\item
Ang, A. and Timmermann, A. (2012) Regime changes and financial markets. 
{\it Annual Review of Financial Economics}, 4, 313--337.
  
\item
Bollerslev, T. (1986) Generalized autoregressive conditional
heteroskedasticity. {\it Journal of Econometrics}, 31, 307--327.

\item
Bondell, H., Reich, B. and Wang, H. (2010) 
Noncrossing quantile regression curve estimation.
{\it Biometrika}, 97, 825--838.

\item
B$\ddot{\rm u}$hlmann, P. (2002) Bootstraps for time  series. 
{\it Statistical Science}, 17, 52--72.



\item 
Dette, H., Hallin, M., Kley, T. and Volgushev, S. (2015) Of copulas,
quantiles, ranks and spectra: an $L_1$-approach to spectral analysis. 
{\it Bernoulli}, 21, 781--831.

\item 
Ding, Z., Granger, C.  and Engle, R.  (1993) A long
memory property of stock market returns and a new model.
{\it Journal of Empirical Finance}, 1, 83--106.

\item
Engle, R.  (1982) Autoregressive conditional heteroskedasticity 
with estimates of the variance of United Kingdom inflation. {\it Econometrica}, 50, 987--1007.

\item
Efron, B. and  Tibshirani, R. (1993) {\it An introduction to the bootstrap}. Boca Raton, FL: Chapman \& Hall/CRC. 

\item
Fajardo, F. A., Reisen, V. A., L\'{e}vy-Leduc, C. and Taqqu, M. S. (2018)
M-periodogram for the analysis of long-range-dependent time series.
{\it Statistics: A Journal of Theoretical and Applied Statistics}, 52, 665--683.

\item 
Fama, E. F. (1965) The behaviour of stock market prices. {\it Journal of Business},38, 34--105.

\item 
Fern\'{a}ndez. C. and Steel, M. (1998) On Bayesian modeling of fat tails and skewness.
{\it Journal of the American Statistical Association}, 93, 359--371.


\item
Glosten, L., Jagannathan, R. and Runkle, D. (1993) On the relation between the expected value and the volatility of the nominal excess return on stocks. {\it Journal of Finance}, 48, 1779--1801.

\item
Hagemann, A. (2011) Robust spectral analysis. {\tt arXiv:1111.1965}.

\item
Hansen, P. and Lunde, A. (2005) A forecast comparison of volatility models: does anything beat a GARCH(1,1)? {\it Journal of Applied Econometrics}, 20, 873--889.

\item
He, X. (1997) Quantile curves without crossing. {\it American Statistician}, 51, 186--192.

\item
Hecke, R., Volgushev, S. and Dette, H. (2018) Fourier analysis of serial dependence
measures. {\it Journal of Time Series Analysis}, 39, 75--89.

\item
Higgins, M. L. and Bera, A. K. (1992) A class of nonlinear ARCH models. {\it International Economic Review}, 33, 137--158.

\item
Hong, Y. (2000) Generalized spectral tests for serial dependence. {\it Journal of the Royal Statistical Society Series B}, 62, 557--574. 


\item
Huang, H., Ombao, H. and Stoffer, D. (2004) Discrimination and classification of nonstationary time series using the SLEX model. {\it Journal of the American Statistical Association}, 99, 763--774.

\item
Jordanger, L. and Tj{\o}stheim, D. (2017) Nonlinear spectral analysis via the local
Gaussian correlation. {\tt arXiv:1708.02166}.

\item
Kakizawa, Y., Shumway, R. and Tanaguchi, M. (1998) Discrimination and clustering for multivariate time series. {\it  Journal of the American Statistical Association}, 93, 328--340.

\item 
Koenker, R. and Bassett, G. (1978) Regression quantiles.
{\it Econometrica}, 46, 33--50.

\item
Koenker, R. (2005) {\it Quantile Regression}. Cambridge, UK: Cambridge University Press.

\item
Kullback, S. and Leibler, R. (1951) On information and sufficiency. {\it Annals of Mathematical Statistics}, 22, 79--86.
 
\item
Lee, J. and Subba Rao, S. (2011) The quantile spectral density
and comparison based tests for nonlinear time series. {\tt arXiv:1112.2759}.


\item
Li, T. H. (2008) Laplace periodogram for time series analysis. 
{\it Journal of the American Statistical Association}, 103, 757--768.

\item
Li, T. H. (2012a) Quantile periodograms. {\it Journal of the American Statistical Association}, 107, 765--776. Complete version available at {\tt domino.research.ibm.com/library/cyberdig.nsf} as IBM Research Report RC25199.

\item
Li, T. H. (2012b)  On robust spectral analysis by least absolute
deviations. {\it Journal of Time Series Analysis},  33, 298--303.

\item
Li, T. H. (2013) {\it Time Series with Mixed Spectra}. Boca Raton, FL: CRC Press.

\item
Li, T. H. (2014) Quantile periodogram and time-dependent variance. 
{\it Journal of Time Series Analysis}, 35, 322-–340.

\item
Li, T. H. (2019) From zero crossings to quantile-frequency analysis of time
series with an application to nondestructive evaluation. 
{\it Applied Stochastic Models for Business and Industry}, DOI: 10.1002/asmb.2499.

\item
Lian, H., Meng, J. and Fan, Z. (2015)
Simultaneous estimation of linear conditional quantiles with penalized splines.
{\it Journal of Multivariate Analysis}, 141, 1--21.

\item
Lim, Y. and Oh, H. S. (2015) Composite quantile periodogram for spectral analysis.
{\it Journal of Time Series Analysis}, 37, 195--221.

\item
Ljung, G. M. and Box, G. E. P.  (1978) On a measure of a lack of fit in time series models. {\it Biometrika}, 65, 297--303. 

\item
Paparoditis, E. and Politis, D. (1999) 
The local bootstrap for periodogram statistics. 
{\it Journal of Time Series Analysis}, 20, 193--222.

\item
Portnoy, S. and Koenker, R. (1997) The Gaussian hare and the Laplacian tortoise: 
computability of squared-error versus absolute error estimators. 
{\it Statistical Science}, 12, 279--300.

\item
Priestley, M. B. (1981) {\it Spectral Analysis and Time Series}. San Diego, CA: Academic Press.

\item 
Skaug, H.  and  Tj{\o}stheim, D. (1993) A nonparametric test of serial independence based on the empirical distribution function. {\it Biometrika}, 80, 59--602.

\item
Schwert, G. W. (1990)  Stock volatility and the crash of '87. {\it Review of Financial Studies}, 3, 77--109.
 
\item
Taylor, S. J. (1986) {\it Modelling Financial Time Series}. New York: Wiley.


\item
Weston, S. (2019) Using the foreach package. https://cran.r-project.org/web/packages/foreach/vignettes/foreach.pdf. 

\item
Whittle, P. (1962) Gaussian estimation in stationary time series.
{\it Bulletin of the International Statistical Institute}, 39, 105--129.

\item
Wuertz, D. (2017) Rmetrics -- Autoregressive conditional heteroskedastic modelling.
{\tt CRAN.R-project.org/package=fGarch}.

\item
Zakoian, J. M. (1994) Threshold heteroskedastic models. {\it Journal of Economic Dynamics and Control}, 18, 931--955.

\end{description}}

\vfill
\noindent
{\it Supporting Information}

\noindent
Supplementary material: ``Quantile-frequency analysis and spectral measures for diagnostic checks of time series with nonlinear 
dynamics: Additional experiments.''

\end{document}